\documentclass[aps, twocolumn,prx, showpacs,preprintnumbers,amsmath,amssymb,superscriptaddress]{revtex4-2}
\usepackage{amsmath,amsfonts,amssymb,graphics,graphicx,epsfig,color,times,indentfirst,layout,hyperref,multirow,bm,subcaption}
\usepackage{hyperref, soul}
\usepackage{ulem}
\hypersetup{linktocpage,,colorlinks=true}
\usepackage{threeparttable}
\usepackage{graphicx}
\usepackage{dcolumn}
\usepackage{bm}
\usepackage{color}

\usepackage{tikz}
\usepackage{tikz-feynhand}
\usetikzlibrary{shapes.misc}
\usetikzlibrary{decorations.pathmorphing}
\usepackage{caption}
\usepackage{subcaption}
\usepackage{comment}
\usepackage{array}
\newcolumntype{C}[1]{>{\centering\arraybackslash}m{#1}}

\tikzset{middlearrow/.style={
        decoration={markings,
            mark= at position 0.5 with {\arrow{#1}} ,
        },
        postaction={decorate}
    }
}

\begin{document}

\title{
Prethermalization and  transient dynamics of the Multi-Channel Kondo systems under generic quantum quenches: Insights form Large-$N$ Schwinger-Keldysh approach}

\author{Iksu Jang}
\email{jisjhm@gmail.com}
\affiliation{Department of Physics, National Tsing Hua University, Hsinchu 30013, Taiwan}

\author{Po-Yao Chang}
\email{pychang@phys.nthu.edu.tw}
\affiliation{Department of Physics, National Tsing Hua University, Hsinchu 30013, Taiwan}

 \date{\today}
\begin{abstract}
Understanding out-of-equilibrium many-body quantum systems is an essential task in contemporary physics. While advanced numerical methods have been developed, capturing the universal dynamics of generic many-body quantum systems still remains a significant challenge. 
In this study, we focus on the multi-channel Kondo impurity (MCKI) model, an intriguing theoretical model hosting an over-screened Kondo state with non-Fermi liquid characteristics, which serves as a theoretical platform for investigating universal properties in many-body quantum dynamics. 
Utilizing the large-$N$ Schwinger-Keldysh approach, we systematically investigate both transient dynamics and long-time quasi-equilibrium properties in the MCKI model following a sudden change of the Kondo coupling. 
Our investigations encompass two distinct initial states: the over-screened Kondo state and the high-temperature Fermi liquid state. For the over-screened Kondo state initial condition, we observe oscillations 
in various physical observables, including spin-spin correlations and the Kondo order parameter. These oscillations signify the quantum revival of the entangled state that characterizes the over-screened Kondo state. 
On the other hand, in the case of the high-temperature Fermi liquid initial state, the absence of oscillations can be attributed to the de-phasing mechanism. 
Furthermore, we discover that the system reaches a quasi-equilibrium on an $\mathcal{O}(1)$ timescale. This quasi-equilibrium manifests as incoherent thermalization between the impurity and the conduction electrons,
in which we observe the non-vanishing effective temperature difference between the Abrikosov fermion representing the impurity spin and the composite boson formed by the Abrikosov fermion and the conduction electrons at the impurity site.
This quasi-equilibrium can be interpreted as prethermalization. Incorporating the $1/N$ correction, we demonstrate that the system attains complete thermalization on an $\mathcal{O}(N)$ timescale.
Additionally, we discuss the quantum cooling effect and quantum Boltzmann equations. Our comprehensive study establishes a foundation for investigating quantum many-body systems
using large-N quantum field theory treatment, while our findings reveal several universal properties of quantum dynamics and provide a different perspective on prethermalization.

\end{abstract}

\maketitle

\tableofcontents

\section{Introduction}
\label{sec:intro}

The multi-channel Kondo impurity (MCKI) model~\cite{Schlottmann}, a variant of the original Kondo model~\cite{Kondo} which considers the conduction electrons of more than one channel coupled to the impurity spin,
has attracted significant attention in the field of condensed matter physics.
With its unique features, including the emergence of non-Fermi liquid phase~\cite{Seaman,Andraka,Ralph,Emery,Affleck:1990iv,Ludwig:1994nf,Affleck:1990zd,Gan,Andrei1,Andrei2,RMPNRG,Parcollet} and fractionalized excitations~\cite{Emery,Landau,Gabay,Papaj}, 
the MCKI models have been intensively studied for decades using several theoretical methods. In particular, the two-channel Kondo (2CK) model~\cite{VONDELFT19981,VONDELFT1999175,Kirchner2020}, which serves as a minimal model of the MCKI, has been investigated theoretically and proposed for experimental realizations.~\cite{Potok2007,Zhu2016,GUPTA2022114547}. 

While the equilibrium properties of the MCKI models have been extensively studied over the past few decades, recent research efforts have shifted toward their non-equilibrium dynamics. Exploring the non-equilibrium properties of the MCKI models provide valuable insight into extracting universal properties of generic many-body systems far from equilibrium. The presence of non-Fermi liquid phases in the MCKI models adds further intrigue due to its relevance to the quantum chaos~\cite{Dora2017,Han2021}. Moreover, the existence of fractionalized excitations in the MCKI models~\cite{Lotem} has motivated recent investigation into its non-equilibrium properties, particularly in the context of qubit state manipulation in quantum computing physics.

To study the non-equilibrium physics in the MCKI models, several theoretical methods have been employed. 
One prominent approach is the time-dependent numerical renormalization group (TDNRG) method~\cite{Pletyukhov}, which, in principle, covers the entire time scale~\cite {Anders2005,Anders2006}. However, it usually requires substantial computing resources. 
Alternatively, several analytical methods have been utilized, including the renormalization group method~\cite{Lin2020}, bosonization with the Emery-Kivelson (EK) mapping~\cite{Ratiani2010,Heyl2010,Dalum2020,Dalum20202}, conformal field theory (CFT) approach~\cite{Karki2020}, 
and large-$N$ methods employing the $SU(N)$ MCKI model~\cite{Kirchner2009,Ribeiro2013}. 
However, most of these analytical techniques mainly focus on the universal properties of non-equilibrium steady states (NESS), particularly non-equilibrium transport through the MCKI, rather than transient non-equilibrium phenomena. 
Additionally, a holographic approach~\cite{Erdmenger2017}, based on the AdS/CFT conjecture, is used to study the quantum quench dynamics in the Kondo model. 

This overview of numerical and theoretical methods illuminates the diverse approaches with their advantage and disadvantage in exploring the non-equilibrium physics of the MCKI models.
The numerical methods tend to focus on obtaining short-time properties, while the analytical techniques aim to understand long-time behaviors.

 To comprehensively and controllably cover the non-equilibrium properties, ranging from transient to long-time equilibration, we investigate the MCKI model by subjecting it to a sudden change of the Kondo coupling. 
We employ the $SU(N)$ MCKI model~\cite{Parcollet} and Schwinger-Keldysh formalism with the large-$N$ method.
Before delving into the specifics, let us briefly discuss the equilibrium states of the MCKI model. For temperature below $T_K$ (Kondo temperature), denoted as 
$T< T_K$, the low energy state of the $SU(N)$ MCKI model exhibits a non-Fermi liquid state known as the over-screened Kondo state.
On the other hand, for high temperature $T>T_K$, the system is characterized by a Fermi liquid state.
While the previous studies~\cite{Kirchner2009,Ribeiro2013} on non-equilibrium phenomena using the $SU(N)$ MCKI model focus on the properties of NESS assuming equilibration, our work explicitly investigates the transient dynamics and the quasi-equilibrium properties resulting from the sudden quench of the Kondo coupling.
We consider different initial states, including the non-Fermi liquid state and Fermi liquid state, as well as different quench parameters within the MCKI model. 

\begin{table}
\begin{center}
    \begin{tabular}{| C{2cm} || C{3cm}| m{3cm} |}
    \hline 
    Initial state & Transient dynamics of physical quantities & Decay rates of spin-spin correlators \\ 
    \hline
    \hline
    Non-Fermi Liquids & oscillations & 
     insensitive to the quench protocol \\ \hline
    Fermi Liquids & no oscillations & 
     depend on the quench protocol  \\
    \hline
    \end{tabular}
\end{center}
\caption{A summary of the transient dynamics in the MCKI model.}
\label{T1}
\end{table}

\begin{figure}[h]
\centering
\begin{subfigure}{0.45\textwidth}
\includegraphics[scale=0.24]{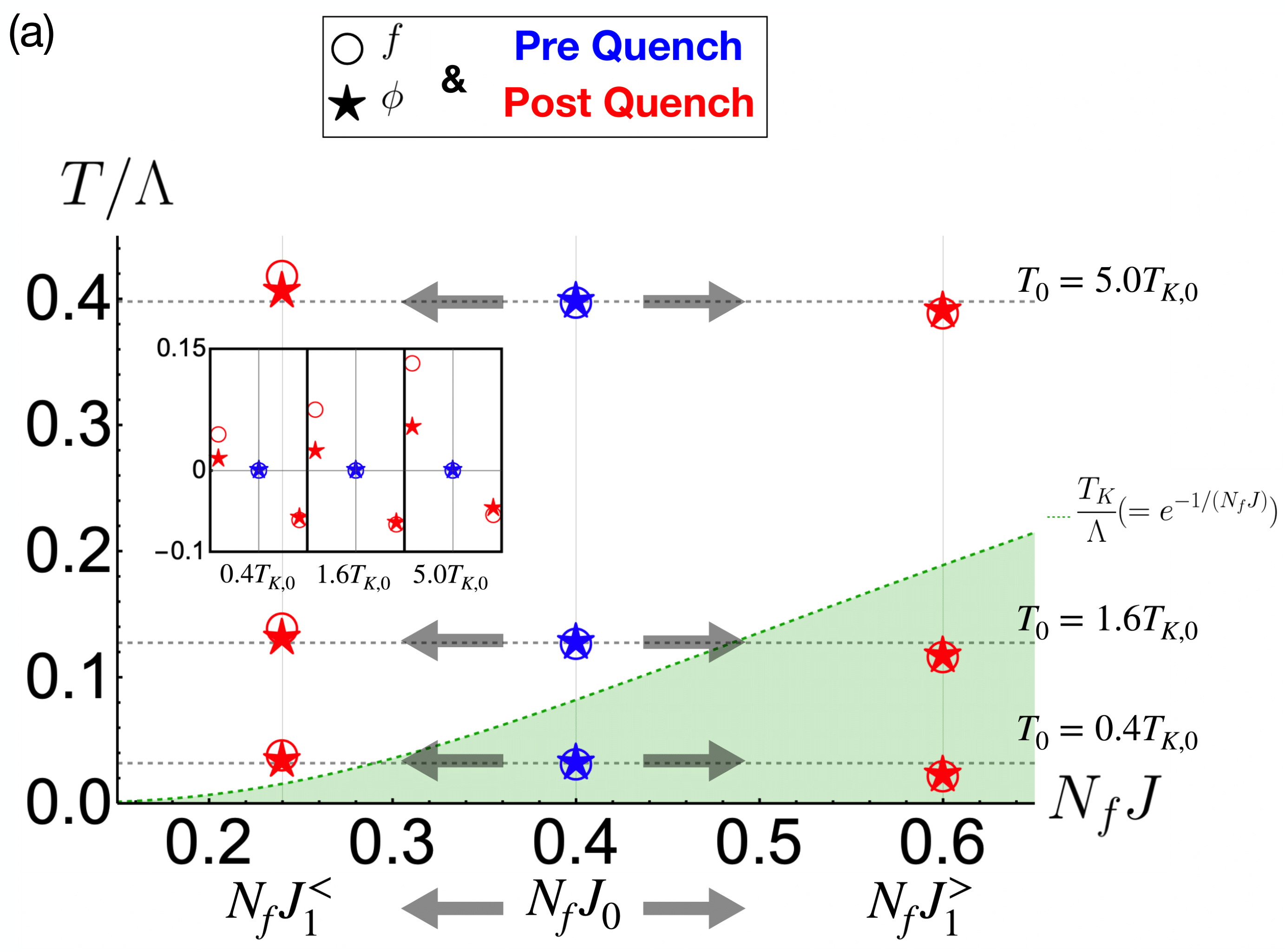}
\end{subfigure}
\hfill
\begin{subfigure}{0.45\textwidth}
\begin{tikzpicture}
\node at (-3,.8){(b)};
\draw[-,very thick] (-2.8,-0.1) -- (-2.8,0.1)  node[above] {$0$};
\node at (-1.8,-0.3){Transient};
\node at (-1.8,-0.65){dynamics};
\draw[->,very thick] (-2.8,0) -- (4,0) node[right] {$t$};
\draw[-,very thick] (-1,-0.1) -- (-1,0.1);
\node at (-1,0.3){$\mathcal{O}(1)$};
\node at (.5,-0.5){Quasi-equilibrium};
\draw[-,very thick] (2,-0.1) -- (2,0.1);
\node at (2,0.3){$\mathcal{O}(N)$};
\node at (3.,-0.5){Thermalization};
\end{tikzpicture}
\end{subfigure}
\caption{(a) Phase diagram of pre- and post-quench states. $T$, $\Lambda$, $N_f$, and $J$ are temperature, cut-off of conduction electron, conduction electron’s density of states near the Fermi energy, and Kondo coupling constant respectively. The circle ($\circ$) and star ($\star$) denote the $f$-fermion and $\phi$-boson respectively. The blue-colored ones correspond to the pre-quench states with $J=J_0$ while the red-colored ones denote the post-quench states with $J=J_1^<(<J_0)$ and $J_1^>(>J_0)$. The green dotted line shows the Kondo temperature as a function of the Kondo coupling constant $J$. Here the results of the two quench protocols: (i) $J_0 (=10)\rightarrow J_{1}^<(=6)$ and (ii) $J_0 (=10)\rightarrow J_{1}^>(=15)$ of the initial states with $J_0=10$ and $T_0=0.4T_{K,0},1.6T_{K,0}$ and $5.0T_{K,0} $ (denoted by the horizontal dashed lines) are depicted. $T_{K,0}$ is the Kondo temperature with $J=J_0$. The Inset figure is a zoomed-in picture of the temperature changes.
(b) Different time-scales for the $SU(N)$ MCKI model under a sudden quench.}
\label{fig:PhaseDiagram}
\end{figure}
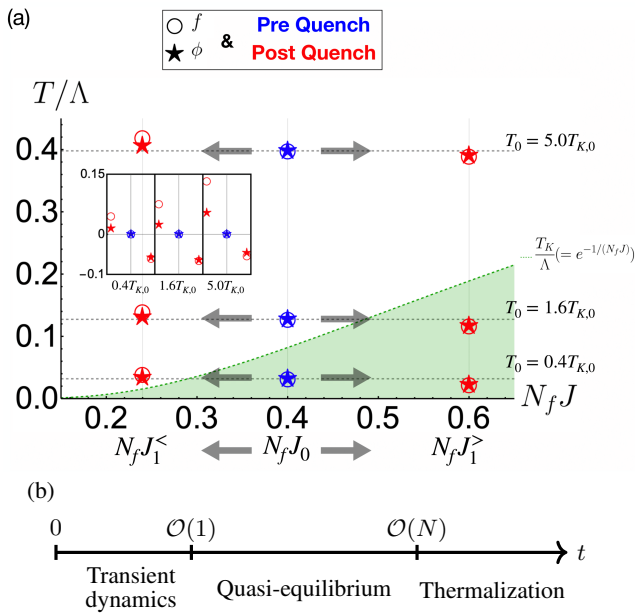

To investigate the non-equilibrium properties on the transient timescale, we examine various physical quantities including the spin-spin correlation functions, the Kondo order parameter, and the spectral functions. 
Analyzing the spin-spin correlation functions, we observe that 
the exponential decay rates of the spin-spin correlation function remain unchanged when the initial states are the over-screened Kondo states with larger post-quench coupling constant. However, when the initial states are high-temperature Fermi liquid states, the exponential decay rates become sensitive to the quench protocols. Furthermore, we discover that all physical quantities exhibit oscillatory properties with a universal frequency~\cite{Bayat}, which is determined by the energy cut-off of the conduction electrons, in the case of initial states being the over-screened Kondo states.
These oscillations can be interpreted as the quantum revival of the entangled state that characterizes the over-screened Kondo state~\cite{Michailidis}. 
On the other hand, the absence of oscillations in the high-temperature Fermi liquid state can be attributed to its de-phasing property. 
The transient properties are summarized in Table~\ref{T1}.

To analyze the long-time quasi-equilibrium properties in the MCKI model, we calculate the effective temperatures and quasi-equilibration times of the two objects: 
the Abrikosov fermion ($f$) representing the impurity spin, and the composite boson ($\phi$) formed by the Abrikosov fermion and the conduction electrons at the impurity site.
Fig.~\ref{fig:PhaseDiagram}(a) presents the results of quantum quenches with different quench protocols for different initial states. As the Kondo coupling constant decreases, we observe an increase in the final temperatures of both the fermion and the boson. Conversely, when the Kondo coupling is increased during the quenching process, we observe a quantum cooling effect. This effect results in decreased effective temperatures for both the fermion and the boson, which can be attributed to the formation of a stronger over-screened Kondo state with lower impurity entropy. This quantum cooling effect is reminiscent of algorithmic cooling~\cite{QuantumCooling}, which is a technique employed in quantum information processing to decrease the entropy of a quantum system.

 Furthermore, we observe that the difference between the final temperatures of the fermion and boson becomes smaller as the final state approaches the over-screened state, which corresponds to a non-Fermi liquid phase, as shown in Fig.~\ref{fig:PhaseDiagram}(a).
However, we find that the final effective temperatures of the fermion and boson are not the same for all the cases we investigated. This incoherent thermalization between the impurity spin and the composite boson
resembles the mixed Sachdev-Ye-Kitaev (SYK) models~\cite{Eberlein2017,Gu2017,Bhattacharya2019,Haque2019,Haldar2020,Zhang2020,Kuhlenkamp2020,Samui2021,Cheipesh2021,Chunxiao2021,Kulkarni2022,Zanoci2022,Zanoci20222,Sohal_2022,Larzul2022,AncelLarzul},
which includes both SYK$_2$ and SYK$_4$ terms, and exhibits prethermalization properties~\cite{Larzul2022}.
To investigate the possibility of full thermalization, we take into account the feedback effect and compute the $1/N$ corrections to the conduction electrons.
We observe that this feedback effect results in coherent temperature changes of the conduction electrons along with the impurity. 
We anticipate the full thermalization can happen on an $\mathcal{O}(N)$ timescale,
which is estimated based on the relaxation rate of the conduction electrons, roughly of the order $\mathcal{O}(N)$.
We conclude the quench dynamics of the MCKI model has the following properties [see Fig.~\ref{fig:PhaseDiagram}(b)]: (1) On an $\mathcal{O}(1)$ timescale, the system approaches a quasi-equilibrium state. Prior to reaching the quasi-equilibrium state, the transient dynamics strongly depend on the initial states. (2)
On an $\mathcal{O}(N)$ timescale, the system will be fully thermalized.

On top of the numerical analysis, we discuss the dependence of the equilibration on the final state of the system phenomenologically using the quantum Boltzmann equations.

This paper is organized as follows: In Sec.~\ref{sec:Model}, we introduce the $SU(N)$ MCKI model and discuss the Abrikosov fermion representation of the quantum impurity. We also provide details of the Schwinger-Keldysh formalism in the MCKI model with the large-$N$ treatment. 
Furthermore, we present Green’s function representations of the physical quantities. Readers who are familiar with the subject can directly go to the results in Sec.~\ref{sec:NumericalResultsAnalysis}. 
 In Sec.~\ref{sec:NumericalResultsAnalysis}, we first discuss the equilibrium properties briefly. Then, we present the analysis of the non-equilibrium properties of various physical quantities, including spin-spin correlation functions, the Kondo order parameter, and the spectral functions on the transient timescale. 
We then examine the spectral functions in quasi-equilibrium and analyze the effective final temperatures and the quasi-equilibration times for different quench protocols.
By considering $1/N$ correction, we find that the time for the full thermalization is $\mathcal{O}(N)$. In Sec.~\ref{sec:QBEapproach}, we discuss the phenomenological description of thermal quasi-equilibration using the quantum Boltzmann approach. Finally, we provide a conclusion summarizing our findings in Sec.~\ref{sec:Conclusion}.

\section{Multi-Channel Kondo Impurity (MCKI) Model  \& Schwinger-Keldysh formalism}
\label{sec:Model}
The model for the Multi-Channel Kondo Impurity system with a time-dependent Kondo coupling is given as follows~\cite{Affleck:1990iv,Ludwig:1994nf,Affleck:1990zd,Parcollet}:
\begin{align}
&H_{MCKI}(t)=\sum_{\mathbf{p}}\sum_{i=1}^K\sum_{\alpha=1}^N\epsilon(\vec{p})c^\dagger_{i,\alpha}(p)c_{i,\alpha}(p)\nonumber\\
&+J_K(t)\sum_{A=1}^{N^2-1}S^A\sum_{i,\alpha,\beta}c_{i,\alpha}^\dagger(r=0) t_{\alpha\beta}^Ac_{i,\beta}(r=0)\label{eq:KondoHamiltonian}
\end{align}
where $K$ and $N$ are a channel number and a spin number respectively, and $J_K(t)$ is a time-dependent Kondo coupling. $S^A$ is a $A-$component of the spin of an impurity in $SU(N)$-representation and $t^A$ is one of generators of the $SU(N)$ group. 

Here we use a fundamental representation of the $SU(N)$ group which satisfies the following properties:
\begin{subequations}
\label{eq:SUNGroupProperty}
\begin{gather}
tr[t^At^B]=2\delta_{AB}\\
\sum_{A=1}^{N^2-1}t_{\alpha\beta}^At_{\gamma\eta}^A=2\Big(\delta_{\alpha\eta}\delta_{\beta\gamma}-\frac{1}{N}\delta_{\alpha\beta}\delta_{\gamma\eta}\Big)
\end{gather}
\end{subequations}

Following~\cite{Parcollet}, we use the Abrikosov fermion representation to express the impurity spin as follows:
\begin{align}
S^A=\frac{1}{2}\sum_{\alpha,\beta=1}^N f_{\alpha}^\dagger t_{\alpha\beta}^A f_{\beta},\quad \sum_{\alpha=1}^Nf_{\alpha}^\dagger f_{\alpha}=Q,
\label{eq:ImpuritySpinAR}
\end{align}
where $Q$ is a total number of $f$-fermions.

Then, $H_{MCKI}$ is re-expressed as follows:
\begin{align}
&H_{MCKI}(t)=\sum_p \sum_{i=1}^K\sum_{\alpha=1}^N \epsilon(p)c_{i,\alpha}^\dagger(p)c_{i,\alpha}(p)\nonumber\\
&+J_K(t)\sum_{i=1}^K\Big(\sum_{\alpha,\beta} f_{\alpha}^\dagger f_{\beta}c_{i,\beta}^\dagger c_{i,\alpha}-\frac{Q}{N}\sum_{\gamma}c_{i,\gamma}^\dagger c_{i,\gamma}\Big)
\end{align}
where $c_{i,\alpha}=c_{i,\alpha}(r=0)$.

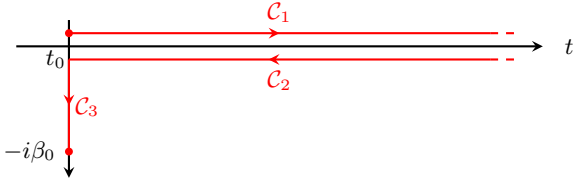
\begin{figure}[h]
\centering
\begin{tikzpicture}[scale=0.35,>=stealth]
	\draw[thick,->] (0,0)--(20,0);
	\draw[thick,->] (2,1)--(2,-5);
	\node at (1.5,-0.5) {$t_0$};
	\fill[red] (2,0.5) circle (0.15);
	\draw[thick,dashed,red] (18,0.5)--(19,0.5);
	\draw[thick,dashed,red] (18,-0.5)--(19,-0.5);
	\node at (21,0) {$t$};
	\draw[middlearrow={>},thick,draw=red] (2,0.5)--(18,0.5);
	\draw[middlearrow={<},thick,draw=red] (2,-0.5)--(18,-0.5);
	\draw[middlearrow={>},thick,draw=red] (2,-0.5)--(2,-4);
	\fill[red]  (2,-4) circle (0.15);
	\node at (.5,-4) {$-i\beta_0$};
	\node at (10,1.3) {$\textcolor{red}{\mathcal{C}_1}$};	
	\node at (10,-1.3) {$\textcolor{red}{\mathcal{C}_2}$};	
	\node at (2.7,-2.3) {$\textcolor{red}{\mathcal{C}_3}$};	
\end{tikzpicture}
\caption{Keldysh contour $\mathcal{C}=\mathcal{C}_1\cup\mathcal{C}_2\cup \mathcal{C}_3$.  The initial state with the temperature $1/\beta_0$ is prepared at $t=t_0$.}\label{fig:KeldyshContour}
\end{figure}

To derive saddle point equations, we start at a path integral formulation of the MCKI Hamiltonian and integrate out the bulk conduction electrons $c_{i,\alpha}(z,r\neq 0)$.
The resulting partition function and the action are given as follows:
\begin{align}
&Z=tr[\rho(t)]=\int\mathcal{D}(c^\dagger, c)\int \mathcal{D} (f^\dagger, f )\int \mathcal{D}\lambda e^{iS_{MCKI}},
\end{align}

\begin{align}
&S_{MCKI,r=0}=\int_{\mathcal{C}}dz\Bigg[\sum_{i=1}^K\sum_{\alpha=1}^N \int_{\mathcal{C}}dz' c_{i,\alpha}^\dagger(z)G^{-1}_{c,0}(z,z’)c_{i,\alpha}(z’)\nonumber\\
&+\sum_{\alpha=1}^N f_\alpha^\dagger(z)[i\partial_z-\lambda(z)]f_{\alpha}(z)+\lambda(z)Q\nonumber\\
&-J_{K}(z)\sum_{i=1}^K\sum_{\alpha,\beta=1}^Nc_{i,\alpha}^\dagger(z)c_{i,\beta}(z)\Big(f_{\beta}^\dagger(z)f_{\alpha}(z)-\frac{Q}{N}\delta_{\alpha\beta}\Big)\Bigg]\label{eq:MCKImpurityAction}
\end{align}
where the Keldysh contour $\mathcal{C}$ is depicted in Fig.~\ref{fig:KeldyshContour}, and $c_{i,\alpha}(z)=c_{i,\alpha}(r=0,z )=\frac{1}{\sqrt{N_{site}}} \sum_p c_{i,\alpha}(p,z)$ is a conduction electron operator at the impurity site $r=0$. 
Here Lagrangian multiplier $\lambda(z)$ is introduced to make the Hilbert space of $f$-electrons restricted for every time indices $z$, and 
\begin{gather}
G_{c,0}(z,z’)=\frac{1}{N_{site}}\sum_p [i\partial_z-\epsilon(p,z)]^{-1}_{z,z’}\label{eq:Gc0}
\end{gather}
is a free Green's function of conduction electron at the impurity site. Now all degrees of freedom in the $S_{MCKI,r=0}$ are defined on the impurity site only. 

\subsection{Large-$N$: Saddle point equations}
Now we will consider the large-$N$ limit following~\cite{Parcollet}, with $K=N\gamma$, $J_K(z)=\frac{J(z)}{N}$,
and  $Q=q_0N$. The $N\rightarrow \infty$ limit is taken for fixed values of the $\gamma$, $J(z)$ and $q_0$. 

To derive an effective action for the saddle point approximations, we first introduce a new boson field $B_i(z)$ to the action (Eq.~\eqref{eq:MCKImpurityAction}) using the Hubbard-Stratonovich transformation. 
Then, the effective action after intergrading out the $c$ field is
\begin{align}
&S_{eff} = \int_{\mathcal{C}}dz\Bigg\{\sum_{\alpha=1}^N\int_{\mathcal{C}}dz'  f_{\alpha}^\dagger(z)G_{f,0}^{-1}(z,z')f_{\alpha}(z’)\nonumber\\
&+N\lambda(z)q_0-\frac{1}{J(z)}\sum_{i=1}^K B_i^{\dagger}(z)B_i(z)\nonumber\\
&-\sum_{i=1}^K\sum_{\alpha=1}^N\frac{1}{N}\int_\mathcal{C}dz' B_i^\dagger(z)f_{\alpha}^\dagger(z)G_{c}(z,z')B_i(z')f_{\alpha}(z')\Bigg\} 
\label{eq:LargeNEffectiveAction1}
\end{align}
where 
\begin{subequations}
\begin{align}
G^{-1}_{c}(z,z’)&=G_{c,0}^{-1}(z,z’)+\frac{q_0J(z)}{N}\delta_{\mathcal{C}}(z-z’)\nonumber\\
&\approx G_{c,0}^{-1}(z,z’),\label{eq:Gc}\\
G^{-1}_{f,0}(z,z’)&=[i\partial_z-\lambda(z)]\delta(z-z’).
\end{align}
\end{subequations}
Note that the $\frac{1}{N}$ correction to the $G_{c}^{-1}(z,z’)$ is ignored and the $G_{c,0}(z,z’)$ (Eq.~\eqref{eq:Gc0}) is used in the follow-up self-consistent calculations.

We can further decouple the interaction vertex by introducing bi-local fields $Q(z,z’)$, $\bar{Q}(z,z’)$, $\Sigma_f(z,z’)$ and $\Sigma_B(z,z’)$ and the resulting effective action is 
\begin{align}
 &S_{eff}=N \int_{\mathcal{C}}dz\int_{\mathcal{C}}dz'\Bigg\{q_0\delta(z-z')\lambda(z)\nonumber\\
&-\gamma\bar{Q}(z',z)Q(z',z)G_c(z,z')-i\Sigma_f(z,z’)Q(z’,z)\nonumber\\
&-i\gamma\Sigma_B(z,z’)\bar{Q}(z',z)\Bigg\}-itr\ln[G_{f,0}^{-1}(z,z’)-\Sigma_f(z,z’)]\nonumber\\
&+i\gamma tr\ln[J^{-1}(z)\delta(z-z’)-\Sigma_B(z,z’)].\label{eq:SeffN}
\end{align}

Since the effective action $S_{eff}$ (Eq.~\eqref{eq:SeffN}) is $\mathcal{O}(N)$ order, the saddle point equations become exact in the $N\rightarrow \infty$ limit. The resulting saddle point equations are given as follows:
\begin{subequations}
\begin{gather}
\Sigma_f(z,z’)=i\gamma G_B(z’,z)G_c(z,z’)\\
\Sigma_B(z,z’)=iG_f(z’,z)G_c(z,z’)\\
G_f(z,z’)=[(G_{f,0}^{-1}-\Sigma_f)^{-1}]_{z,z’}\label{eq:SaddlePointEqGf}\\
G_B(z,z’)=[(-J^{-1}(z)+\Sigma_B)^{-1}]_{z,z’}\label{eq:SaddlePointEqGB}\\
q_0=-iG_f(z,z+0^+)\label{eq:numberConstraint}
\end{gather}
\label{eq:SKEquations}
\end{subequations}
where the fields $Q(z,z’)$ and $\bar{Q}(z,z’)$ are replaced with $G_f(z,z’)$ and $G_B(z,z’)$ based on the following definitions of the contour-ordered Keldysh Greens functions:
\begin{subequations}
\begin{gather}
G_f(z,z’)=-\frac{i}{N}\sum_{\alpha=1}^N\langle  \mathcal{T}_{\mathcal{C}}f_{\alpha}(z)f_{\alpha}
^\dagger(z’)\rangle,\\
G_B(z,z’)=-\frac{i}{K}\sum_{i=1}^K\langle \mathcal{T}_{\mathcal{C}}B_{i}(z)B_i^\dagger(z’)\rangle
\end{gather}
\end{subequations}
where $\mathcal{T}_{\mathcal{C}}$ is a contour-ordering operator.

Eqs.~\eqref{eq:SaddlePointEqGf} and~\eqref{eq:SaddlePointEqGB} can be transformed to the form of integro-differntial equations as follows:
\begin{subequations}
\label{eq:IntegroDiffSKEquations}
\begin{align}
&(i\partial_z-\lambda(z))G_f(z,z')-\int_{\mathcal{C}}d\bar{z}\Sigma_f(z,\bar{z})G_f(\bar{z},z’)\nonumber\\
&=\delta_{\mathcal{C}}(z-z'),\\
&(-i\partial_{z'}-\lambda(z'))G_f(z,z')-\int_{\mathcal{C}}d\bar{z}G_f(z,\bar{z})\Sigma_f(\bar{z},z’)\nonumber\\
&=\delta_{\mathcal{C}}(z-z'),\\
&-\frac{1}{J(z)}G_B(z,z')+\int_{\mathcal{C}} d\bar{z} \Sigma_B(z,\bar{z})G_B(\bar{z},z’)\nonumber\\
&=\delta_{\mathcal{C}}(z-z'),\label{eq:GBSKEquation}\\
&-\frac{1}{J(z')}G_B(z,z')+\int_{\mathcal{C}}d\bar{z}G_B(z,\bar{z})\Sigma_B(\bar{z},z’)\nonumber\\
&=\delta_{\mathcal{C}}(z’-z).
\label{eq:GBSKEquation}
\end{align}
\end{subequations} 
These integro-differential equations are used in deriving the Kadanoff-Baym equations in Sec.~\ref{subsec:KB}.

 We would like to comment on the difference between the MCKI model and the SYK model.
In the MCKI model, there is a constraint on the number of $f$-fermions which is expressed as Eq.~\eqref{eq:numberConstraint}. 
The number of the $f$-fermions is determined by $\lambda(z)$ and is a constant in the equilibrium.
Here, by using the Ward-identity, we find the $\lambda(z)$ is still a constant in the non-equilibrium setting 
and the details are given in Appendix~\ref{app:U1symmetry}.

\subsection{Kadanoff-Baym equations}
\label{subsec:KB}

Following the~\cite{Aoki}, the Kadanoff-Baym equations can be expressed in terms of the greater, lesser, retarded, advanced, and Keldysh Green’s functions using the following relations
\begin{subequations}
	\label{eq:defOfGRGAGgGl}
	\begin{align}
		G^>(t,t’)&=G^{21}(t,t'),\; G^<(t,t')=G^{12}(t,t'),\\
		G^R(t,t’)&=\frac{1}{2}\Big(G^{11}-G^{12}+G^{21}-G^{22}\Big)(t,t'),\\
		G^A(t,t’)&=\frac{1}{2}\Big(G^{11}+G^{12}-G^{21}-G^{22}\Big)(t,t’),\\
		G^K(t,t’)&=\frac{1}{2}\Big(G^{11}+G^{12}+G^{21}+G^{22}\Big)(t,t’)
	\end{align}
\end{subequations}
where $G^{ij}(t,t')=G(z,z')\Big|_{z\in \mathcal{C}_i, z'\in \mathcal{C}_j}$. 

It is well known that the retarded, advanced, and Keldysh Green’s functions can be re-expressed in terms of the greater and lesser Greens functions:
\begin{subequations}
	\label{eq:GRGAInGgGl}
\begin{align}
	G^R(t,t’)&=\Theta(t-t')\Big(G^>(t,t')-G^<(t,t')\Big),\\
	G^A(t,t’)&=\Theta(t'-t)\Big(G^<(t,t')-G^>(t,t')\Big),\\
	G^K(t,t’)&=\Big(G^<(t,t’)+G^>(t,t')\Big)
\end{align}
\end{subequations}
where $\Theta(t)$ is a step function with a value $1/2$ at $t=0$. The expressions for the retarded and advanced Greens functions are nothing but the conventional definitions of them.

However, the relations Eq.~\eqref{eq:GRGAInGgGl} do not apply to $G_B$ and $\Sigma_f(\propto G_B)$ since the value of $G_B(z,z')$ has a singularity at $z=z'$ reflected in the differential equation Eq.~\eqref{eq:GBSKEquation} while usual Green's functions of fermion and boson fields have a singular derivative at $z=z'$. Related to this fact, the fields $B_i$ and $B_i^\dagger$ also do not follow the usual boson commutation relation which is reflected in the absence of the term $B_i^\dagger(z)\partial_zB_i(z)$ in the action~\eqref{eq:LargeNEffectiveAction1} . As a result, it is difficult to figure out the properties of the field $B$ in the operator language. It causes a problem or confusion in using the relation Eq.~\eqref{eq:GRGAInGgGl} and finding a sum-rule for the $G_B$. To remedy this problem, it is necessary to find a relation between the $G_B$ and fermion operators $f$ and $c$. To do that, we introduce a source field $\eta_i(z)$ which couples to the boson field $B_i(z)$ in the partition function Eq.~\eqref{eq:LargeNEffectiveAction1} as follows:
\begin{align}
	&Z[\eta^\dagger,\eta]=\int\mathcal{D}(c^\dagger,c)\int\mathcal{D}(f^\dagger,f)\int\mathcal{D}\lambda \int\mathcal{D}(B^\dagger,B)\nonumber\\
	&\times \exp\Bigg[i\int_{\mathcal{C}}dz\Bigg\{\sum_{i=1}^K\sum_{\alpha=1}^N \int_{\mathcal{C}}dz' c_{i,\alpha}^\dagger(z)G_{c}^{-1}(z,z')c_{i,\alpha}(z’)\nonumber\\
	&+\sum_{\alpha=1}^N\int_{\mathcal{C}}dz' f_{\alpha}^\dagger(z)G_{f,0}^{-1}(z,z')f_{\alpha}(z)+N\lambda(z)q_0\nonumber\\
	& -\frac{1}{J(z)}\sum_{i=1}^KB_i^\dagger(z) B_i(z)-\frac{1}{\sqrt{N}}\sum_{i=1}^K\sum_{\alpha}^N\Big(B_i^\dagger(z)f_{\alpha}^\dagger (z)c_{i,\alpha}(z)\nonumber\\
	&+B_i(z)c_{i,\alpha}^\dagger(z)f_{\alpha}(z)\Big)+\sum_i\Big(\eta_i(z)B_i^\dagger(z)+B_i(z)\eta^\dagger(z)\Big) \Bigg\}\Bigg]
\end{align}

By calculating the $\frac{1}{Z[\eta^\dagger,\eta]}\frac{\delta^2 Z[\eta^\dagger,\eta]}{\delta \eta_i^\dagger(z’)\delta \eta_i(z)}\Big|_{\eta^\dagger=\eta=0}$, we find
\begin{align}
	G_B(z,z')&=-J(z)\delta_{\mathcal{C}}(z-z’)+J(z)J(z')G_{\phi}(z,z’),
	\label{eq:RelationGBGphi}
\end{align}
where a new boson field $\phi$ and a Green’s function are defined as follows:
\begin{subequations}
\label{eq:defOfGphi}
	\begin{gather}
		\phi_i(z)=\frac{1}{\sqrt{N}}\sum_{\alpha=1}^N f_\alpha^\dagger(z)c_{i,\alpha}(z),\label{eq:DefOfPhi}\\
		G_\phi(z,z')=-\frac{i}{K}\sum_{i=1}^K\langle 	\mathcal{T}_{\mathcal{C}}\phi_i(z)\phi_i^\dagger(z')\rangle.
	\end{gather}
\end{subequations}

Eq.~\eqref{eq:RelationGBGphi} shows the singularity at $z=z’$ of the $G_B(z,z’)$ explicitly. By using $G_\phi(z,z’)$ instead of the $G_B(z,z’)$, we can avoid the singularity issue and Eq.~\eqref{eq:GRGAInGgGl} is applicable to the $G_\phi(z,z’)$. Additionally, we can easily find a sum-rule for the boson field $\phi$ from Eq.~\eqref{eq:DefOfPhi}. For details of the sum-rules, see Appendix.~\ref{Appendix:SumRules}

Using Eq.~\eqref{eq:RelationGBGphi}, the self-consistent saddle point equations (Eqs.~\eqref{eq:SKEquations} and~\eqref{eq:IntegroDiffSKEquations}) can be re-expressed in terms of the $G_\phi(z,z’)$ as follows:
\begin{widetext}
\begin{subequations}
	\label{eq:newSKEquations}
	\begin{align}
		&[i\partial_z-\lambda_0+i\gamma J(z)G_c(z,z)]G_f(z,z’)
	           -\int_{\mathcal{C}}d\bar{z}\tilde{\Sigma}_f(z,\bar{z})G_f(\bar{z},z')=\delta_{\mathcal{C}}(z-z’),\label{eq:IntegroDiffGfEq1}\\
		&[-i\partial_{z'}-\lambda_0+i\gamma J(z’)G_c(z’,z’)]G_f(z,z’)
		   -\int_{\mathcal{C}}d\bar{z}G_f(z,\bar{z})\tilde{\Sigma}_f(\bar{z},z')=\delta_{\mathcal{C}}(z-z’),\label{eq:IntegroDiffGfEq2}\\
		&G_\phi(z,z')+\Sigma_B(z,z’)
		   =\int_\mathcal{C} d\bar{z}\Sigma_B(z,\bar{z})J(\bar{z})G_\phi(\bar{z},z’), \quad
		   G_\phi(z,z')+\Sigma_B(z,z’)
		   =\int_\mathcal{C} d\bar{z}G_\phi(z,\bar{z})J(\bar{z})\Sigma_B(\bar{z},z')
	\end{align}
\end{subequations}
where
\begin{subequations}
\begin{align}
	   &\tilde{\Sigma}_f(z,z')=i\gamma J(z)J(z') G_\phi(z',z)G_c(z,z'),\quad
		\Sigma_B(z,z')=iG_f(z',z)G_c(z,z'),\\
           & q_0=-iG_f(z,z+0^+)=-iG_f^<(z,z)
\end{align}
\end{subequations}
and the relation
\begin{gather}
\Sigma_f(z,z’)=-i\gamma  J(z)G_c(z,z’)\delta_{\mathcal{C}}(z-z’)+\tilde{\Sigma}_f(z,z’)
\end{gather}
has been used and $\lambda(z)$ is set to be the constant value $\lambda_0$. 

Now let us re-write the above Eqs.~\eqref{eq:newSKEquations} in terms of the greater, lesser, retarded, and advanced Green’s functions using Eqs.~\eqref{eq:defOfGRGAGgGl} and~\eqref{eq:GRGAInGgGl}. Then the resulting Kadanoff-Baym equations are given by 
\begin{subequations}
\begin{align}
&\Big[i\partial_t+i\gamma J(t)G_c^<(t,t)-
\lambda_0\Big]G_f^{R}(t,t’)-\int d\bar{t}\tilde{\Sigma}_f^{R}(t,\bar{t})G_f^{R}(\bar{t},t’)=\delta(t-t’),\\
&\Big[-i\partial_{t'}+i\gamma J(t')G_c^<(t',t')-\lambda_0\Big]G_f^R(t,t’)-\int d\bar{t} G_f^R(t,\bar{t})\tilde{\Sigma}_f^R(\bar{t},t’)=\delta(t-t’),\\
&\Big[i\partial_t-\lambda_0+i\gamma J(t)G_c^<(t,t)\Big]G_f^>(t,t’)-\int d\bar{t}\Big[\tilde{\Sigma}_f^>(t,\bar{t})G_f^{A}(\bar{t},t')+\tilde{\Sigma}_f^{R}(t,\bar{t})G_f^>(\bar{t},t')\Big]=0\label{eq:GfKBEq1},\\
&\Big[i\partial_t-\lambda_0+i\gamma J(t)G_c^<(t,t)\Big]G_f^<(t,t’)-\int d\bar{t}\Big[\tilde{\Sigma}_f^<(t,\bar{t})G_f^{A}(\bar{t},t')+\tilde{\Sigma}_f^{R}(t,\bar{t})G_f^<(\bar{t},t')\Big]=0,\label{eq:GfKBEq2}\\
&G_\phi^{R}(t,t')+\Sigma_B^{R}(t,t’)=\int d\bar{t}\Sigma_B^{R}(t,\bar{t})J(\bar{t})G_{\phi}^{R}(\bar{t},t’),\\
&G_\phi^{>}(t,t')+\Sigma_B^>(t,t’)=\int d\bar{t}\Big[\Sigma_B^>(t,\bar{t})J(\bar{t})G_\phi^A(\bar{t},t’)+\Sigma_B^R(t,\bar{t})J(\bar{t})G_\phi^>(\bar{t},t')\Big],\\
&G_\phi^{<}(t,t')+\Sigma_B^<(t,t’)=\int d\bar{t}\Big[\Sigma_B^<(t,\bar{t})J(\bar{t})G_\phi^A(\bar{t},t’)+\Sigma_B^R(t,\bar{t})J(\bar{t})G_\phi^<(\bar{t},t')\Big]
\end{align}
\label{eq:KBequation}
\end{subequations}
\end{widetext}
where
\begin{subequations}
\begin{gather}
\tilde{\Sigma}_f^{>/<}(t,t')=i\gamma J(t)J(t') G_{\phi}^{</>}(t',t)G_c^{>/<}(t,t’),\label{eq:Sigmaf} \\
\Sigma_B^{>/<}(t,t')=iG_{f}^{</>}(t',t)G_c^{>/<}(t,t'),\\
G_f^<(t,t)=iq_0,\; G_f^{>}(t,t)=-i(1-q_0).\label{eq:GfConstraint}
\end{gather}
\end{subequations}
 
The solutions of the Kadanoff-Baym equations~(\ref{eq:KBequation}) are the essential ingredients for computing all the physical observables.

\subsection{Physical observables}
We derive expressions of physical observables in terms of the Greens functions here. 

\subsubsection{Spin-spin correlation functions}
First, we consider a spin-spin correlation function defined as follows:
\begin{gather}
\Theta(t-t’) \frac{\langle \vec{S}_{imp}(t)\cdot\vec{S}_{imp}(t’)\rangle}{N^2-1} =C_{imp}(t,t’)-\frac{i}{2}\chi_{imp}(t,t’)\label{eq:SpinSpinCorrel}
\end{gather}
where $C_{imp}(t,t’)$ and $\chi_{imp}(t,t’)$ are given by
\begin{subequations}
\begin{gather}
C_{imp}(t,t’)=\frac{1}{2}\frac{1}{N^2-1}\sum_{A=1}^{N^2-1}\langle \{S^A_{imp}(t),S^A_{imp}(t’)\}\rangle,\\
\chi_{imp}(t,t)=i\frac{\Theta(t-t')}{N^2-1}\sum_{A=1}^{N^2-1}\langle[S_f^A(t),S_f^A(t')]\rangle.
\end{gather}
\end{subequations}
Note that the $\chi_{imp}$ is an impurity spin susceptibility.

To obtain the spin-spin correlation function in terms of Green’s function, we re-express the spin-spin correlation function using the Abrikosov representation (Eq.~\eqref{eq:ImpuritySpinAR}) and Keldysh formalism as follows:
\begin{align}
&\Theta(t-t’) \frac{\langle \vec{S}_{imp}(t)\cdot\vec{S}_{imp}(t’)\rangle}{N^2-1} =\frac{\Theta(t-t’)}{4(N^2-1)}\sum_{A=1}^{N^2-1}\nonumber\\
&\times \Big\langle\mathcal{T}_{\mathcal{C}} f^\dagger_\alpha(z)t^A_{\alpha\beta}f_\beta(z) f^\dagger_\gamma(z’)t^A_{\gamma\eta}f_\eta(z’)\Big\rangle\Big|_{Re z=t,\; Re z’=t’}\nonumber\\
&\equiv \Theta(t-t’) \mathcal{F}(z,z’)\Big|_{Re z=t,\; Rez’=t’}
\end{align}
where $z$ and $z’$ are points on Keldysh contour (Fig.~\ref{fig:KeldyshContour}).

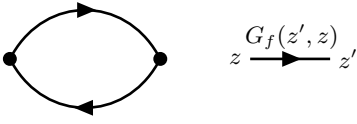
\begin{figure}[t]
\centering
\begin{tikzpicture}[baseline=-0.1cm]
	\begin{feynhand}
		\vertex[dot] (a0) at (0,0) {}; \vertex[dot] (b0) at (2,0){}; 
		\propag[fer] (a0) to [out=60,in=120,looseness=1.2](b0);
		\propag[fer] (b0) to [out=240,in=-60,looseness=1.2](a0);
	\end{feynhand}
	~
	\begin{feynhand}
\vertex (a)  at (3,0) {$z$}; \vertex (b) at (4.5,0) {$z'$};
\propag[fer] (a) to (b);
\node at (7.5/2,0.3) {$G_f(z’,z)$};
\end{feynhand}
\end{tikzpicture}
\caption{Feynman diagram of the spin-spin correlation functions with the order $\mathcal{O}(N^0)$.}
\label{fig:ZerothOrderSpinSpinCorrel}
\end{figure}

Using the path integral formalism based on the Feynman rules (Appendix.~\ref{Appendix:SelfConsFeynmanRules}), we calculate the value of $\mathcal{F}(z,z’)$ of the order $\mathcal{O}(N^0)$ given in Fig.~\ref{fig:ZerothOrderSpinSpinCorrel}. The diagrams with more than one-loop are ignored since they are suppressed in the $N\rightarrow \infty$ limit. The resulting $\mathcal{F}(z,z’)$ is given as follows:
\begin{align}
	\mathcal{F}(z,z’)&=\frac{1}{4}\frac{1}{N^2-1}\sum_{A=1}^{N^2-1}t^A_{\alpha\beta}t^A_{\gamma\eta}\Big\langle f_\alpha^\dagger(z)f_\beta(z)f_\gamma^\dagger(z')f_\eta(z’)\Big\rangle\nonumber\\
	&=\frac{1}{2}G_f(z,z')G_f(z',z)
\end{align}
where Eq.~\eqref{eq:SUNGroupProperty} and $tr[t^A]=0$ are used. 

As a result, the spin-spin correlation function of $\mathcal{O}(N^0))$ order is expressed in terms of the Green’s functions as follows:

\begin{align}
	\Theta(t-t’) \frac{\langle \vec{S}_{imp}(t)\cdot\vec{S}_{imp}(t’)\rangle}{N^2-1}& =\frac{\Theta(t-t’)}{2}G_f^>(t,t’)G_f^<(t’,t)
\label{eq:SpinCorrel}
\end{align} 
Then, $C_{imp}(t,t’)$ and $\chi_{imp}(t,t’)$ are given by 
\begin{subequations}
\begin{align}
C_{imp}(t,t’)&=\frac{1}{4}\Theta(t-t’)[G_f^>(t,t’)G_f^<(t’,t)\nonumber\\
&+G_f^<(t,t’)G_f^>(t’,t)],\\
\chi_{imp}(t,t’)&=i\frac{1}{2}\Theta(t-t')[G_f^>(t,t')G_f^<(t’,t)\nonumber\\
&-G_f^<(t,t')G_f^>(t’,t)]\label{eq:SpinSuscep}.
\end{align}
\end{subequations}

\subsubsection{Kondo order parameter}
In addition to the spin-spin correlation function, we also consider a Kondo order parameter  defined as follows:
\begin{align}
\mathcal{O}_{Kondo}(t)&=\frac{1}{K}\sum_{i}\Big\langle\vec{S}_{imp}(t)\cdot\vec{s}_{c,i}(t)\Big\rangle,
\end{align}
where $\vec{s}_{c,i}(t)=\sum_{\alpha,\beta=1}^Nc_{i,\alpha}^\dagger(t)t_{\alpha\beta}^A c_{i,\beta}(t)$.

Considering the $J(z)$ field in the partition function $Z$ (Eq.~\eqref{eq:LargeNEffectiveAction1}) as a source field, the $\mathcal{O}_{Kondo}(t)$ is given by 
\begin{align}
\mathcal{O}_{Kondo}(t)&=i\gamma \frac{\delta \ln Z[J(z)]}{\delta J(z)}\Big|_{Re z=t}\nonumber\\
&=iN\Big[q_0G_c^<(t,t)-\frac{1}{J^2(t)}G_B^<(t,t)\Big]\nonumber\\
&=iN\Big[q_0 G_{c,0}^<(t,t)-G_\phi^<(t,t)\Big]\label{eq:KondoOrder}
\end{align}
where the relation Eq.~\eqref{eq:RelationGBGphi} is used and $\frac{1}{N}$-correction in Eq.~\eqref{eq:Gc} is considered.

\section{Numerical results \& Analysis}\label{sec:NumericalResultsAnalysis}
\begin{figure*}[!htbp]
\begin{center}
\includegraphics[scale=0.4]{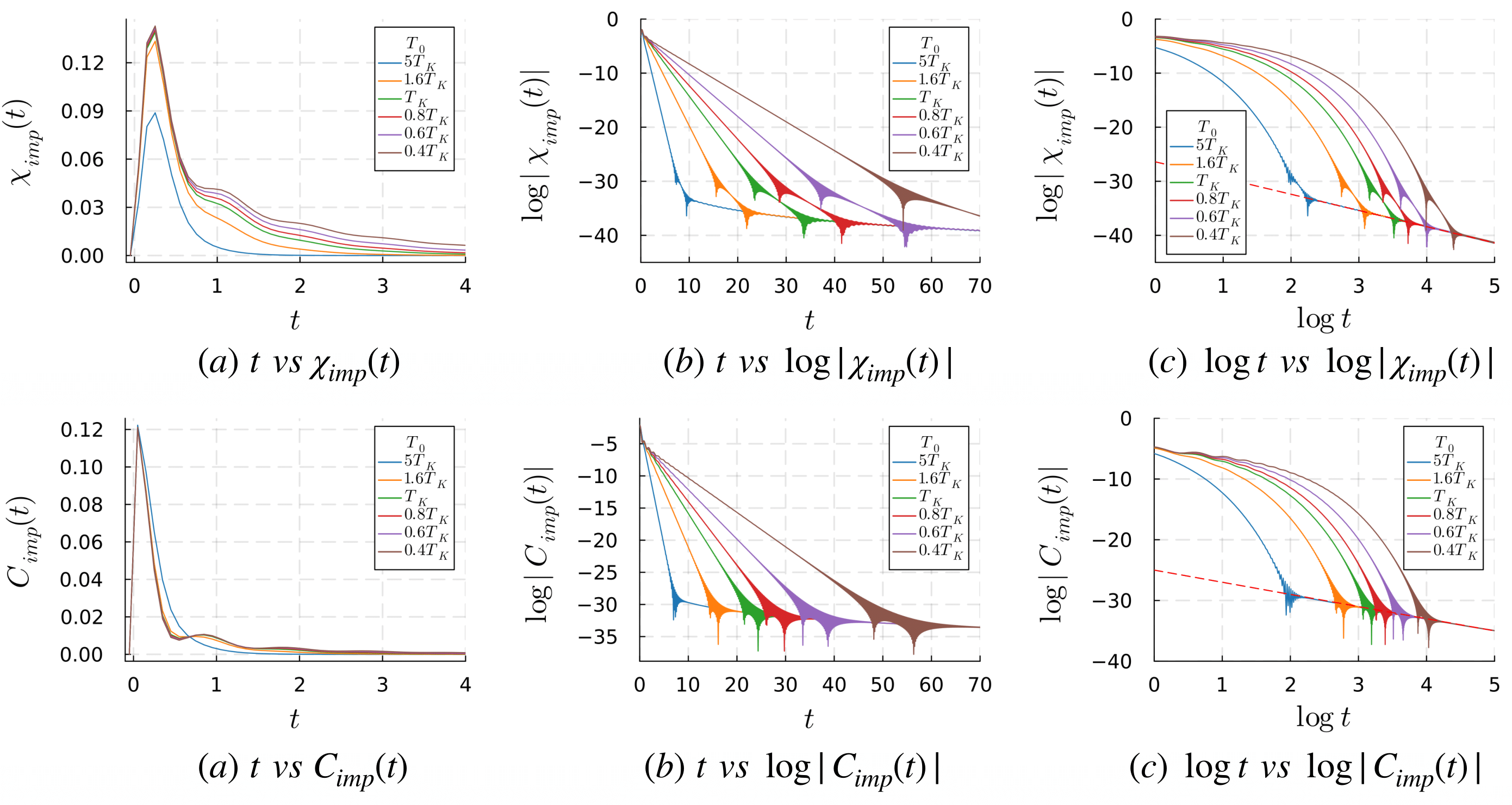}
\caption{$\chi_{imp}(t)$ (top) and $C_{imp}(t)$ (bottom) of initial states with different temperatures ($5T_K$, $1.6T_K$, $T_K$, $0.8T_K$, $0.6T_K$ and $0.4T_K$). The slopes (Red dashed lines) of the fitting lines in the log-log plots (Fig. (c)) are given by $-3$ and $-2$ for $\chi_{imp}(t)$ (top) and $C_{imp}(t)$ (bottom) respectively.}
\label{fig:CimpAndChiimpInEquil}
\end{center}
\end{figure*}

\begin{figure}[htbp]
\centering
\includegraphics[scale=0.06]{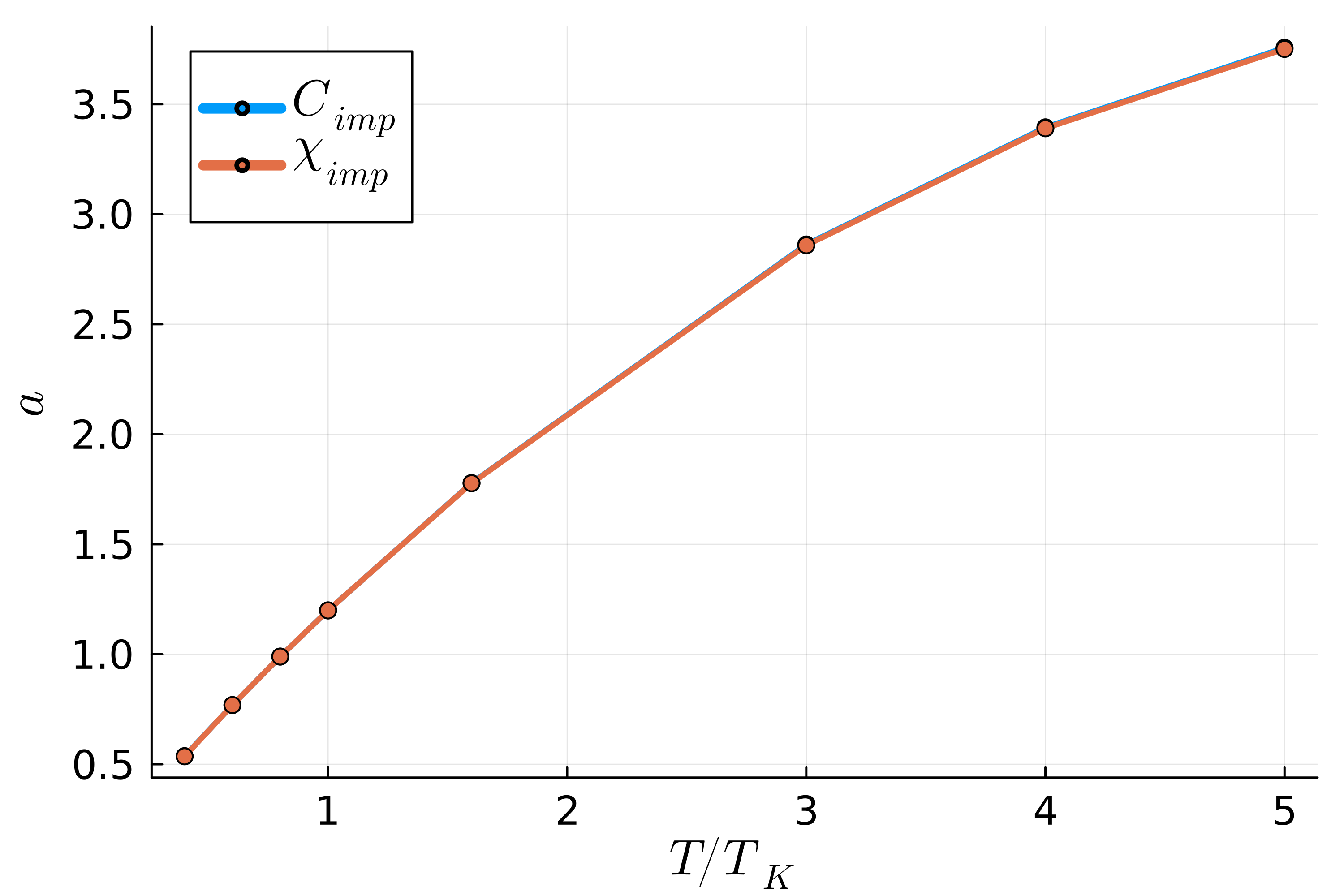}
\caption{Exponential decay rates of the $C_{imp}$ and $\chi_{imp}$ ($C_{imp}(t),\chi_{imp}(t)\sim e^{-at}$) as a function of the temperature in equilibrium.}
\label{fig:InitialStateDecayRate}
\end{figure}

To understand the dynamics of the Multi-Channel Kondo system under the quench protocol quantitatively, we have investigated an evolution of several physical quantities such as spin-spin correlation functions (Eq.~\eqref{eq:SpinSpinCorrel}), the Kondo-order parameter (Eq.~\eqref{eq:KondoOrder}), and the spectral functions. 
Additionally, we have calculated an effective temperature and a thermalization time. To examine how initial states affect quench dynamics, we have considered four different initial states with temperatures lower and higher than the Kondo temperature $T_K$. 

\subsection{Initial States \& Equilibrium properties}
\subsubsection{Initial States}
Before going to the results of the quench dynamics, we first discuss the properties of the initial states with different given temperatures. To get self-consistent solutions of the initial states with real-time arguments numerically, we followed~\cite{Eberlein2017}. For details, see Appendix~\ref{Appendix:PreparationOfInitialState}. 

From the Poor man’s RG~\cite{coleman_2015}, the Kondo temperature $T_K$ is given by $T_K=\Lambda e^{-\frac{1}{J_0N_F}}$ where $\Lambda$ is the energy cut-off and $N_F(=\frac{\nu_c}{2\Lambda})$ is the density of states near the Fermi energy of the conduction electron. See Appendix.~\ref{Appendix:AnalyticFormOfConduction} for details about the conduction electron setting. Based on the Kondo temperature, we have considered several cases with temperatures given by $0.4T_K$, $T_K$, $1.6T_K$, and so on. 

For numerical simulations, we used a time lattice of $N_{site}=8000$ with $\Delta t=0.1$. Then $\omega_{max}=\frac{\pi}{\Delta t}\approx 31.4$ and $\Delta \omega=\frac{2\omega_{max}}{N_{site}}\approx 0.008$. For settings of conduction electrons, we use $\Lambda=\frac{\omega_{max}}{5}\approx 6.28$ and $\nu_c=0.5$. The Kondo coupling constant $J_0$ and the filling of the $f$-fermion $q_0$ are set to be $10$ and $0.5$ respectively. The resulting numerical value of the Kondo temperature is around 0.5. In the set-up of numerical values, the energy resolution value $\Delta \omega$ should be the smallest value than any energy scale such as Kondo temperature $T_K$ or temperature $T$ while $\omega_{max}$ should be the largest one to get convergent physical solutions. For the $\gamma$ value, we fix it to be $\gamma=1.0$. Since it is known that the Multi-Channel Kondo system using the Abrikosov fermion representation in the large $N$ limit shows the over-screened phase for any $\gamma$ values~\cite{Parcollet}, we expect that there is no qualitative difference in the dynamics depending on the value of $\gamma$.

\subsubsection{Equilibrium: spin-spin correlation functions ($C_{imp}(t)$ and $\chi_{imp}(t))$.}

We discuss the $C_{imp}(t)$ and $\chi_{imp}(t)$ for different initial states. Fig.~\ref{fig:CimpAndChiimpInEquil} shows the $C_{imp}(t)$ and $\chi_{imp}(t)$, their log plots and log-log plots respectively for different temperatures. From the log plots of $C_{imp}(t)$ (Fig.~\ref{fig:CimpAndChiimpInEquil}(b) bottom) and $\chi_{imp}(t)$ (Fig.~\ref{fig:CimpAndChiimpInEquil}(b) top), we can see that both $C_{imp}(t)$ and $\chi_{imp}(t)$ exponentially decay in the early time. Fig.~\ref{fig:InitialStateDecayRate} shows that the exponential decay rates of the $C_{imp}(t)$ and $\chi_{imp}(t)$ almost perfectly match each other. The exponential decay rate decreases rapidly as the temperature is reduced. In the long time limit, both $C_{imp}(t)$ and $\chi_{imp}(t)$ show the power-law decay as shown in Fig.~\ref{fig:CimpAndChiimpInEquil}(c). However, they show the different power law exponents. For $C_{imp}(t)$, it is $2$ while it is $3$ for $\chi_{imp}(t)$. As a result, in the long time limit, $\langle \vec{S}_{imp}(t)\cdot\vec{S}_{imp}(0)\rangle$ shows the $\frac{1}{|t|^2}$ behavior which is consistent with the previous study done with the 2CK model~\cite{Heyl2010}.

\subsubsection{Equilibrium: spectral functions}
\begin{figure}
\centering
\begin{subfigure}{0.45\textwidth}
\includegraphics[scale=0.036]{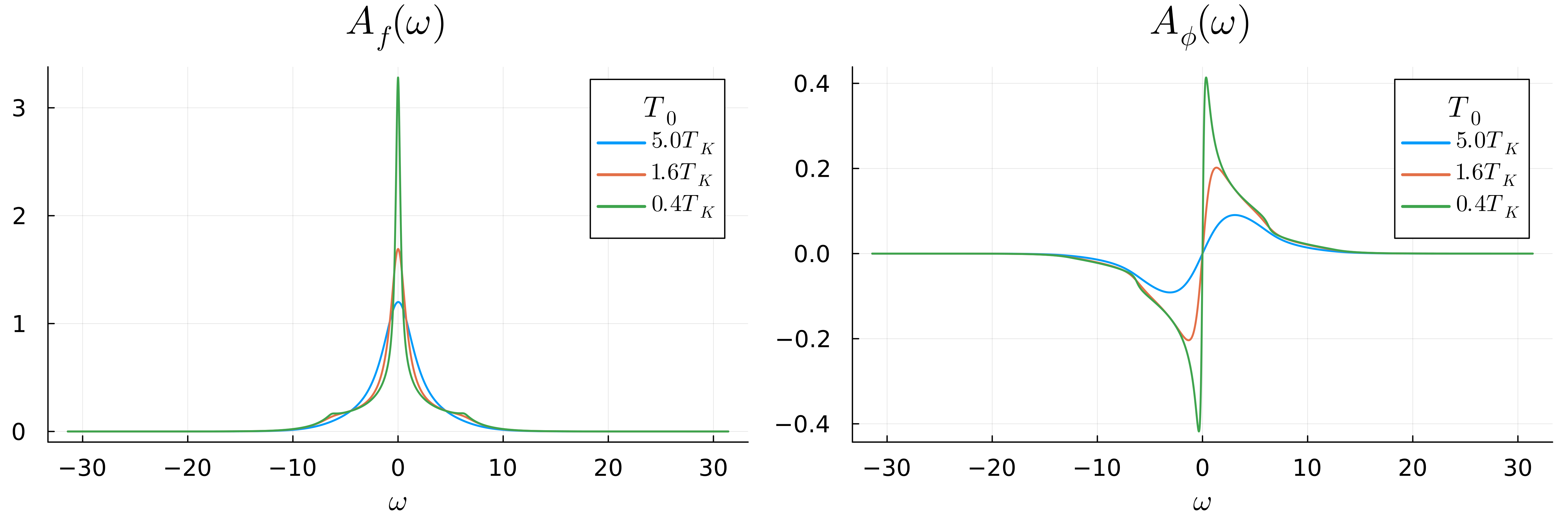}
\caption{Spectral functions of $f$-fermion and $\phi$-boson.}\label{fig:InitialStateSpectralFunctions}
\end{subfigure}
~
\begin{subfigure}{0.45\textwidth}
\includegraphics[scale=0.036]{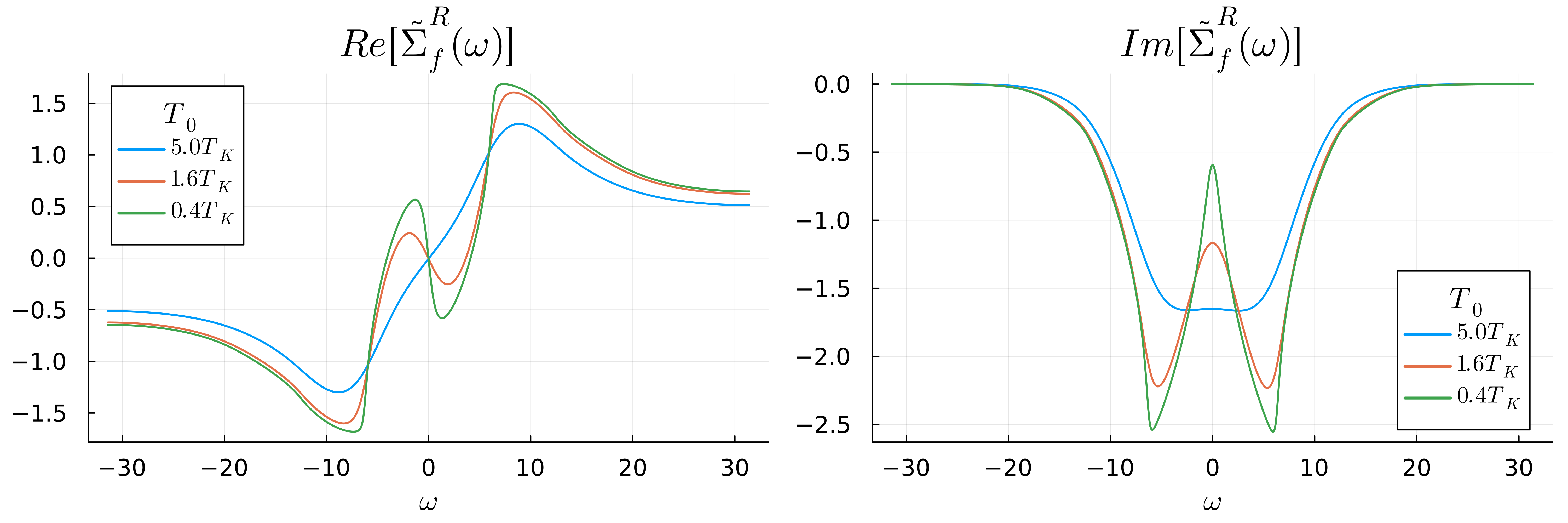}
\caption{Self-energies of the fermion $f$}\label{fig:InitialStateFermionSelfEnergy}
\end{subfigure}
~
\begin{subfigure}{0.45\textwidth}
\includegraphics[scale=0.036]{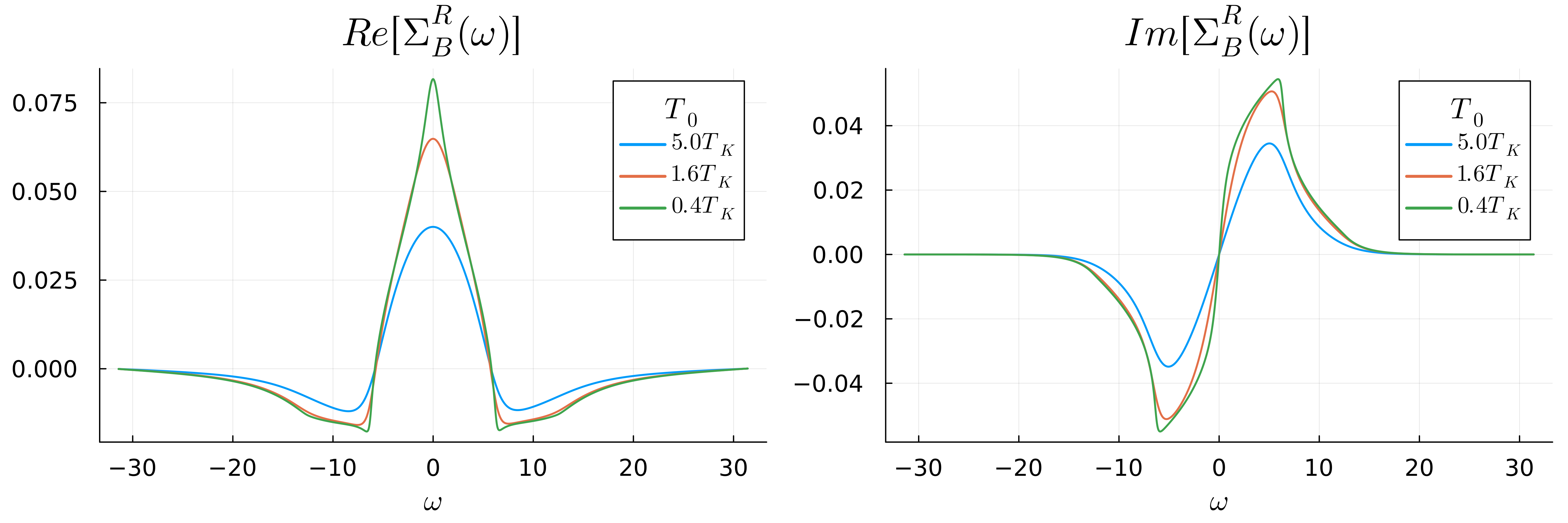}
\caption{Self-energies of the boson $\phi$ }\label{fig:InitialStateBosonSelfEnergy}
\end{subfigure}
\caption{Spectral functions and self energies of the fermion $f$ and the boson $\phi$ with the temperatures $5T_K$, $1.6T_K$, and $0.4T_K$.}
\end{figure}

We discuss the spectral functions of $f$-fermion and $\phi$-boson with different temperatures as shown in 
Fig.~\ref{fig:InitialStateSpectralFunctions}. Note that the Kondo peak develops around the $\omega=0$ point when the temperature becomes lower than the $T_K$. The Kondo transition can be seen more clearly in the self energies as shown in Fig.~\ref{fig:InitialStateFermionSelfEnergy} and~\ref{fig:InitialStateBosonSelfEnergy}.

\subsection{Quantum quench set-up \& Non-equilibrium properties}
With the several different initial states, we consider the quantum quench of the coupling constant $J(t)$ at $t=0$ from the initial value $J_0$ to $J_1$. Therefore 
\begin{gather}
J(t)=J_0(1-\Theta(t))+J_1\Theta(t).
\end{gather}

To numerically solve the Kadanoff-Baym equations, we use the 3rd order Adams-Bashforth-Molton method~\cite{NumericalRecipes} with the Simpson rule for the numerical integration. We have checked the consistency of the numerical results with the sum-rules in Appendix.~\ref{Appendix:SumRules} and some trivial cases as sanity checks. Convergence and independence of the discrete path grid are also examined. For a flow chart of the numerical simulation of the Kadanoff-Baym equations, please see Appendix~\ref{Appendix:KBsolving}. In our simulations, the two-dimensional time lattice with the size $600\times 600$ and $\Delta t=0.1$ is used. As numerical values of $J_1$, we consider a total of nine values given by 
\begin{gather}
J_1=\Big\{\begin{array}{cc}6,7,8,9 & <J_0(=10)\\
11,12,13,14,15 & >J_0(=10)
\end{array}
.
\end{gather}

In the analysis, we used transformed coordinates $\mathcal{T}=\frac{t+t’}{2}$ and $t_r=t-t’$ rather than the original coordinates $t$ and $t’$. In the remaining text, we use Green’s functions with the arguments $\mathcal{T}$ and $t_r$ or $\omega_r$ where $\omega_r$ is a frequency defined in the following Fourier transformation with respect to $t_r$: $
G(\mathcal{T},\omega_r)=\int dt_r e^{-i\omega_r t_r}G(\mathcal{T},t_r)$. 

\subsubsection{Kondo order parameter}\label{sec:EvolutionOfKondoOrder}

\begin{figure}[h]
\centering
\begin{subfigure}{0.45\textwidth}
\centering
\includegraphics[scale=0.065]{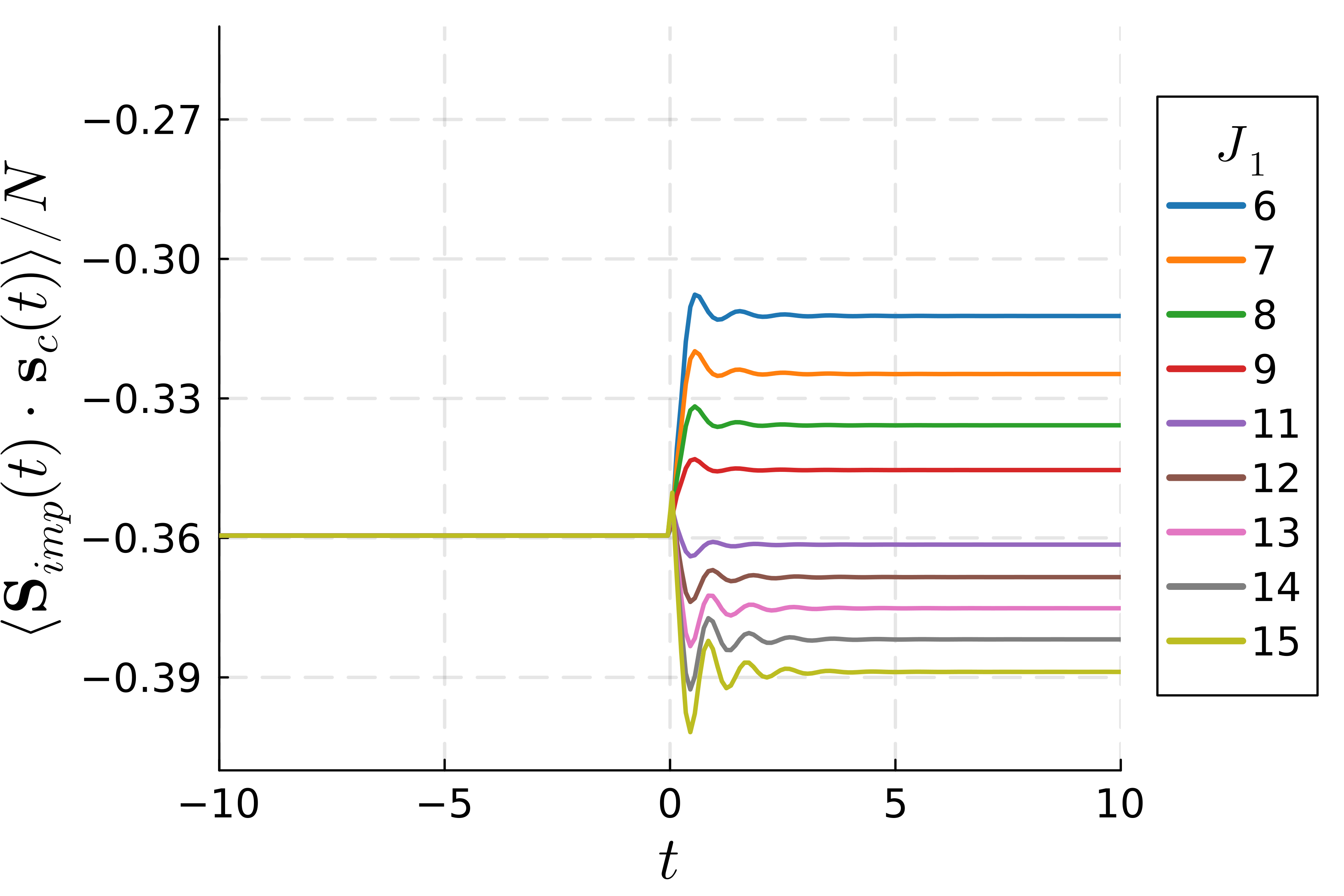}
\caption{$T_0=0.4T_K$ case}\label{fig:EvolutionOfKondoOrderT004TKCase}
\end{subfigure}
~
\begin{subfigure}{0.45\textwidth}
\centering
\includegraphics[scale=0.065]{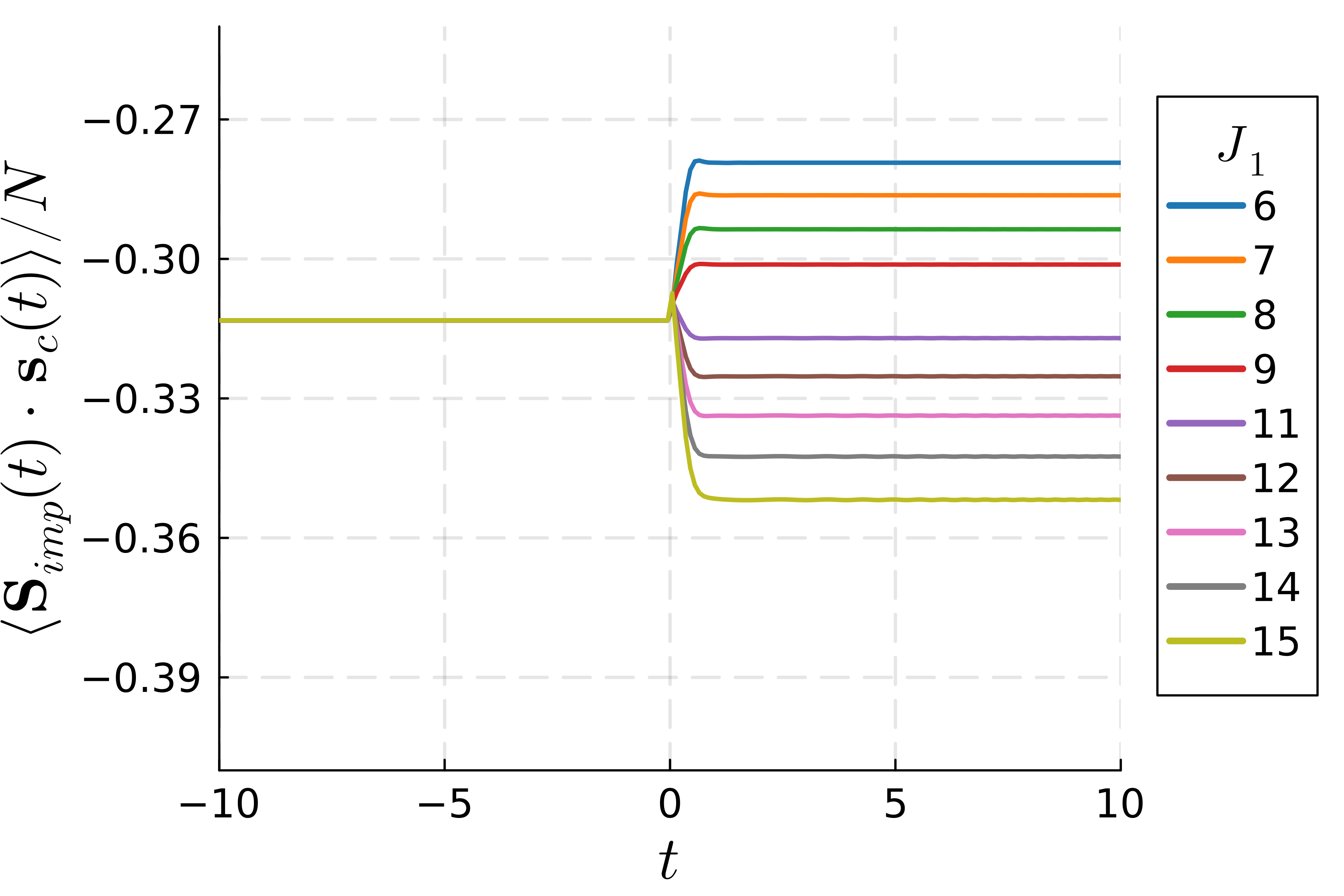}
\caption{$T_0=5T_K$ case}\label{fig:EvolutionOfKondoOrderT05TKCase}
\end{subfigure}
\caption{Evolutions of the $\langle \vec{S}_{imp}(t)\cdot\vec{s}_c(t)\rangle /N$ for (a) $T_0=0.4T_K$ and (b) $T_0=5T_K$ cases with different $J_1$ values. Here the value of $\gamma$ is 1.}\label{fig:EvolutionOfKondoOrder}
\end{figure}

\begin{figure*}
\centering
\begin{subfigure}{0.49\textwidth}
\centering
\includegraphics[scale=0.25]{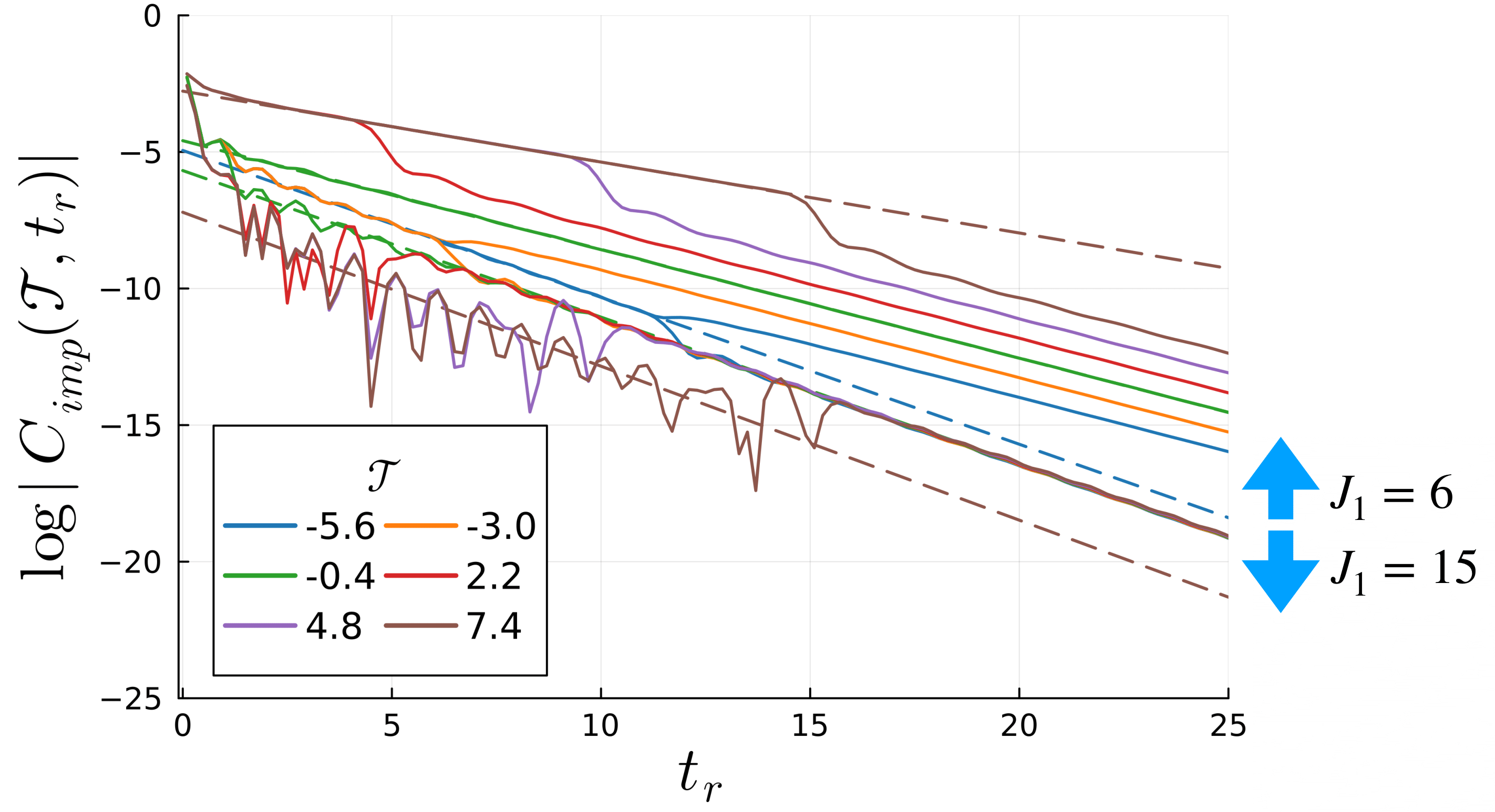}
\caption{$t$ vs $\log|C_{imp}(\mathcal{T},t_r)|$}
\end{subfigure}
~
\begin{subfigure}{0.49\textwidth}
\centering
\includegraphics[scale=0.25]{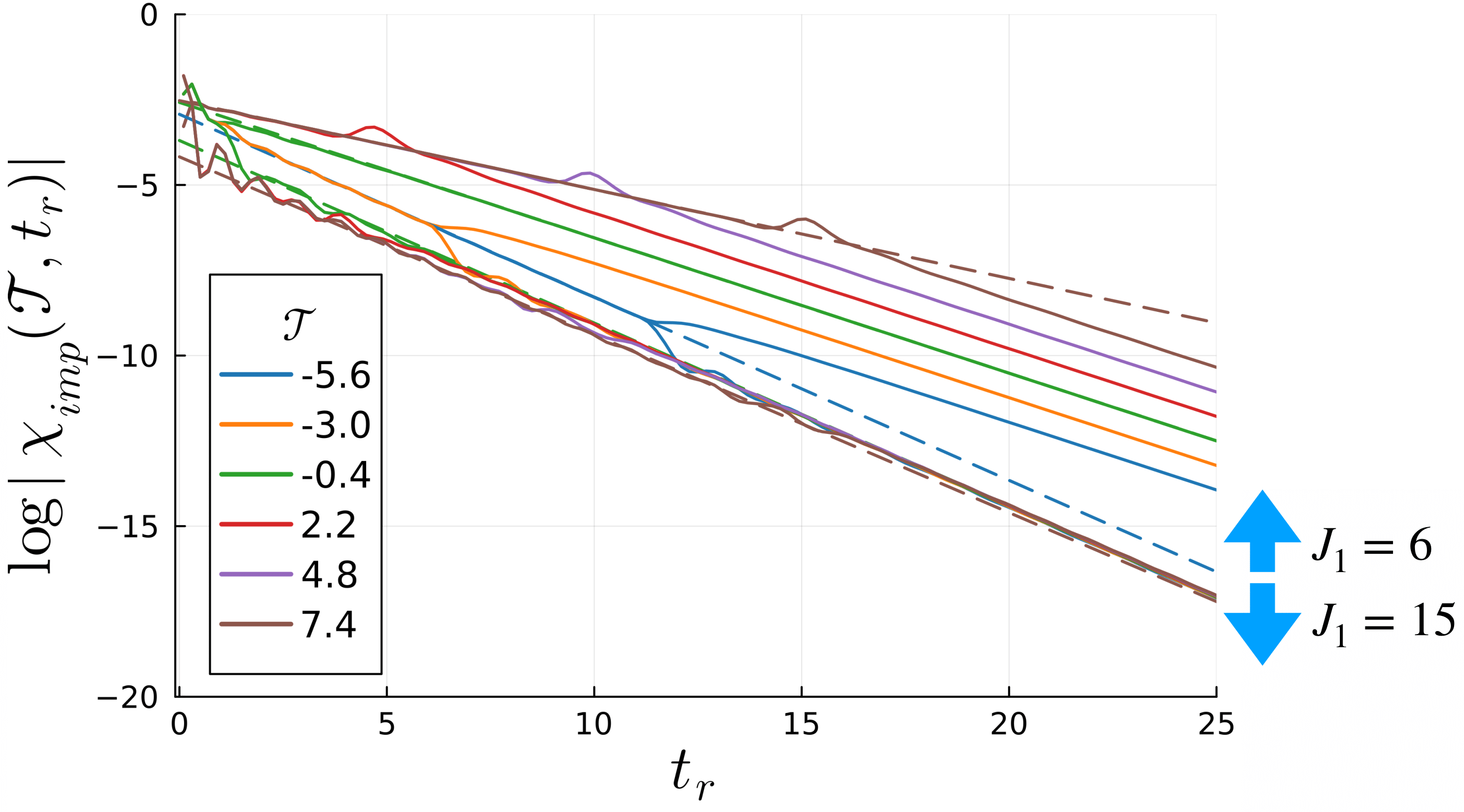}
\caption{$t$ vs $\log|\chi_{imp}(\mathcal{T},t_r)|$}
\end{subfigure}
\caption{Log plots of $\chi_{imp}(\mathcal{T},t_r)$ and $C_{imp}(\mathcal{T},t_r)$ for $T_0=0.4T_K$ case. The color of the graph denotes the values of $\mathcal{T}$. Dashed lines with blue, green and brown colors for both $J_1=6$ and $J_1=15$ cases are the linear fitting lines of the graphs at $\mathcal{T}=-5.6,$ $-0.4$ and $7.4$ respectively. Slopes of dashed lines in time order are given by  $(-0.54,-0.40,-0.26)$ for $C_{imp}$ with $J_1=6$, $(-0.54,-0.54,-0.-0.56)$ for $C_{imp}$ with $J_1=15$, $(-0.54, -0.40, -0.26)$ for $\chi_{imp}$ with $J_1=6$, and $(-0.54,-0.54,-0.52)$ for $\chi_{imp}$ with $J_1=15$. }\label{fig:CollectionOfLogPlotLowT0}
\end{figure*}

\begin{figure*}
\centering
\begin{subfigure}{0.49\textwidth}
\centering
\includegraphics[scale=0.25]{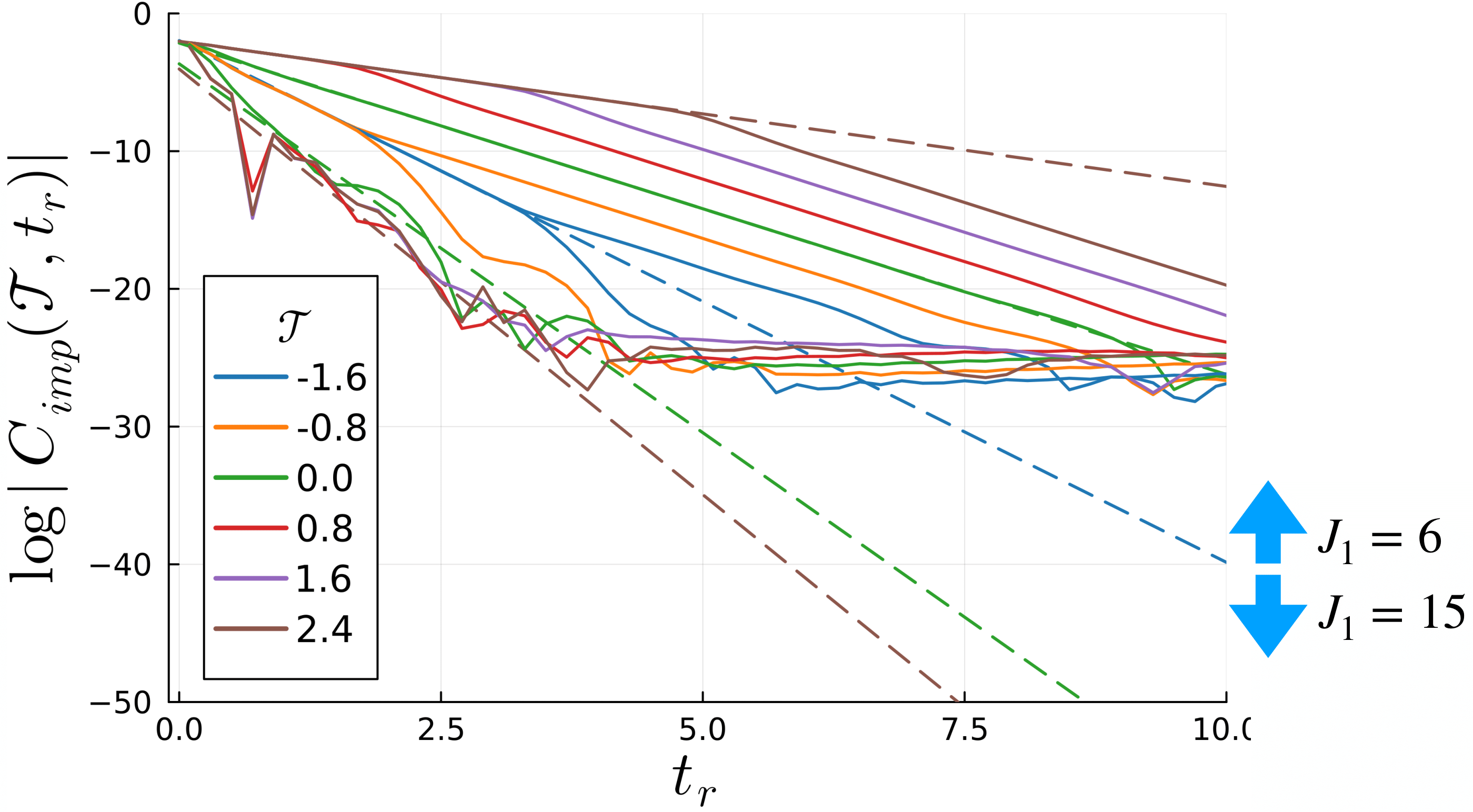}
\caption{$t$ vs $\log|C_{imp}(\mathcal{T},t_r)|$}
\end{subfigure}
~
\begin{subfigure}{0.49\textwidth}
\centering
\includegraphics[scale=0.25]{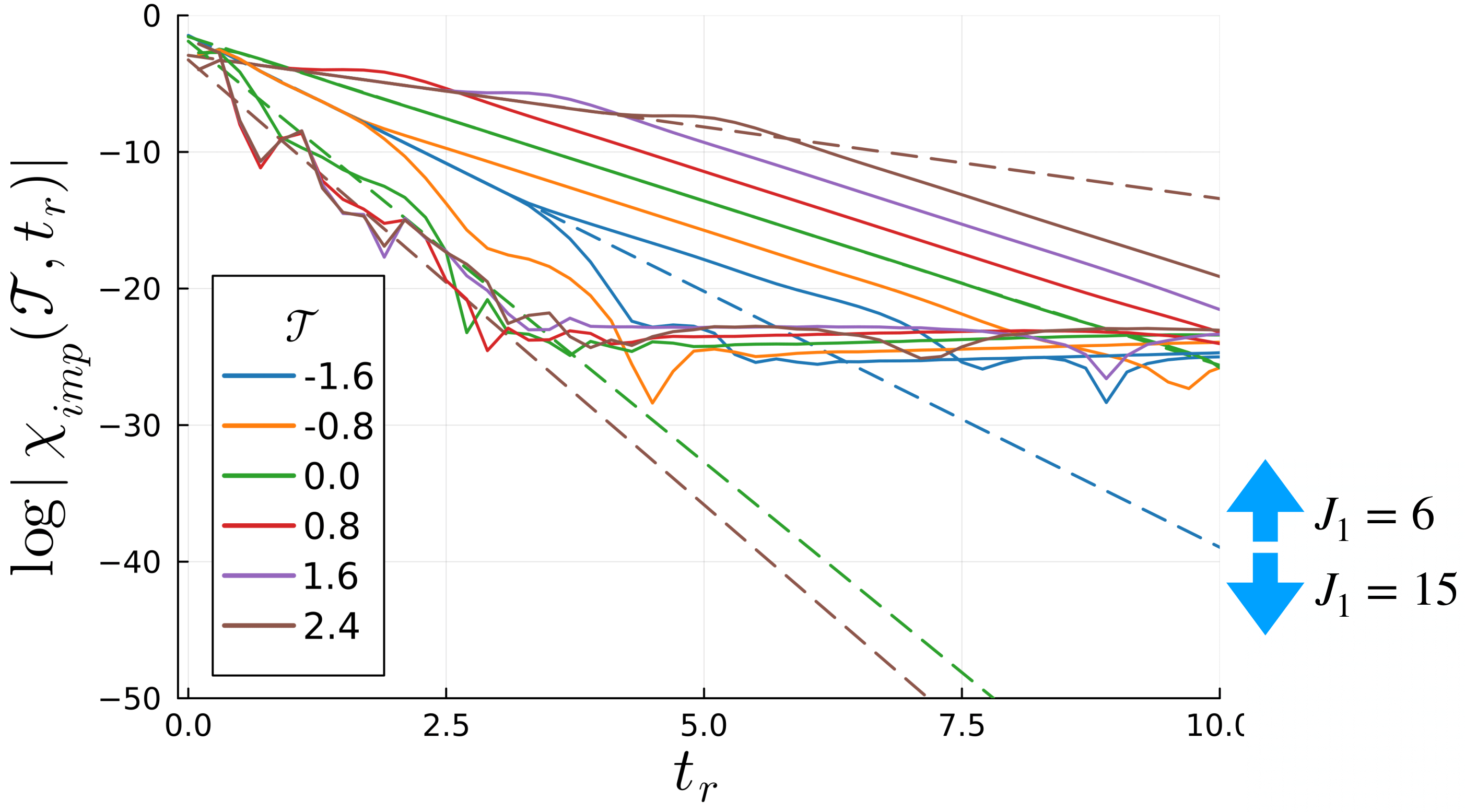}
\caption{$t$ vs $\log|\chi_{imp}(\mathcal{T},t_r)|$}
\end{subfigure}
\caption{Log plots of $\chi_{imp}(\mathcal{T},t_r)$ and $C_{imp}(\mathcal{T},t_r)$ for $T_0=5T_K$ case. The color of the graph denotes the values of $\mathcal{T}$. Dashed lines with blue, green and brown colors are the linear fitting lines of the graphs at $\mathcal{T}=-1.6,$ $0.0$ and $2.4$ respectively. Slopes of dashed lines in time order are given by $(-3.8,-2.4,-1.1)$ for $C_{imp}$ with $J_1=6$, $(-3.8,-5.3,-6.2)$ for $C_{imp}$ with $J_1=15$, $(-3.7,-2.4,-1.0)$ for $\chi_{imp}$ with $J_1=6$, and $(-3.7,-6.2,-6.5)$ for $\chi_{imp}$ with $J_1=15$.}\label{fig:CollectionOfLogPlotHighT0}
\end{figure*}

In this section, we consider the evolution of the Kondo order parameter (Eq.~\eqref{eq:KondoOrder}) under the quantum quench. 
Fig.~\ref{fig:EvolutionOfKondoOrder} shows the evolution of the Kondo order parameter for $T_0=0.4T_K$ and $T_0=5T_K$ respectively. 
In both cases, the absolute values of the Kondo order parameter decrease when the coupling constant $J$ is reduced while increasing when $J$ increases. It is physically natural and expected results. 
However, there is one major distinction between these two cases in the transient time scale. In the case of $T_0=5T_K$ (Fig.~\ref{fig:EvolutionOfKondoOrderT05TKCase}), there are no oscillations in a transient time scale. In contrast to the $T_0=5T_K$ case, strong oscillating patterns are observed in the $T_0=0.4T_K$ case (Fig.~\ref{fig:EvolutionOfKondoOrderT004TKCase}) for all values of $J_1$. Besides, it turns out that all frequencies of these oscillating patterns are the same regardless of the values of the $J_1$ and $J_0$. 
We find that these oscillations only occur when the initial state is closed to the over-screened Kondo state which is the strongly entangled state. 
It is expected that these oscillations are closely related to the revival phenomena in the quench dynamics of entangled states~\cite{Michailidis}.
Similar oscillation behaviors of the spin-spin correlation functions will be discussed in Sec.~\ref{sec:SpinSpinCorrel}. 
More interestingly, the value of the frequency is given by $\Lambda$ which is the energy cut-off of the conduction electron. This universal frequency can be understood from the Friedel oscillation in the time domain. An impurity in a metal can cause density oscillation in space, where the corresponding wavelength is $k_F^{-1}$ with $k_F$ being the Fermi momentum of the conduction electrons~\cite{Affleck2008}.
In the quench setup, the sudden change of the Kondo coupling can induce oscillations in the time domain, and the corresponding frequency is controlled by the energy cut-off of the conduction electrons. Similar universal frequency is also observed in the spin-chain emulator of the two-impurity Kondo model~\cite{Bayat}

\subsubsection{Spin-spin correlation function: $C_{imp}(\mathcal{T},t_r)$ and $\chi_{imp}(\mathcal{T},t_r)$.}\label{sec:SpinSpinCorrel}

\begin{figure*}
\centering
\includegraphics[scale=0.45]{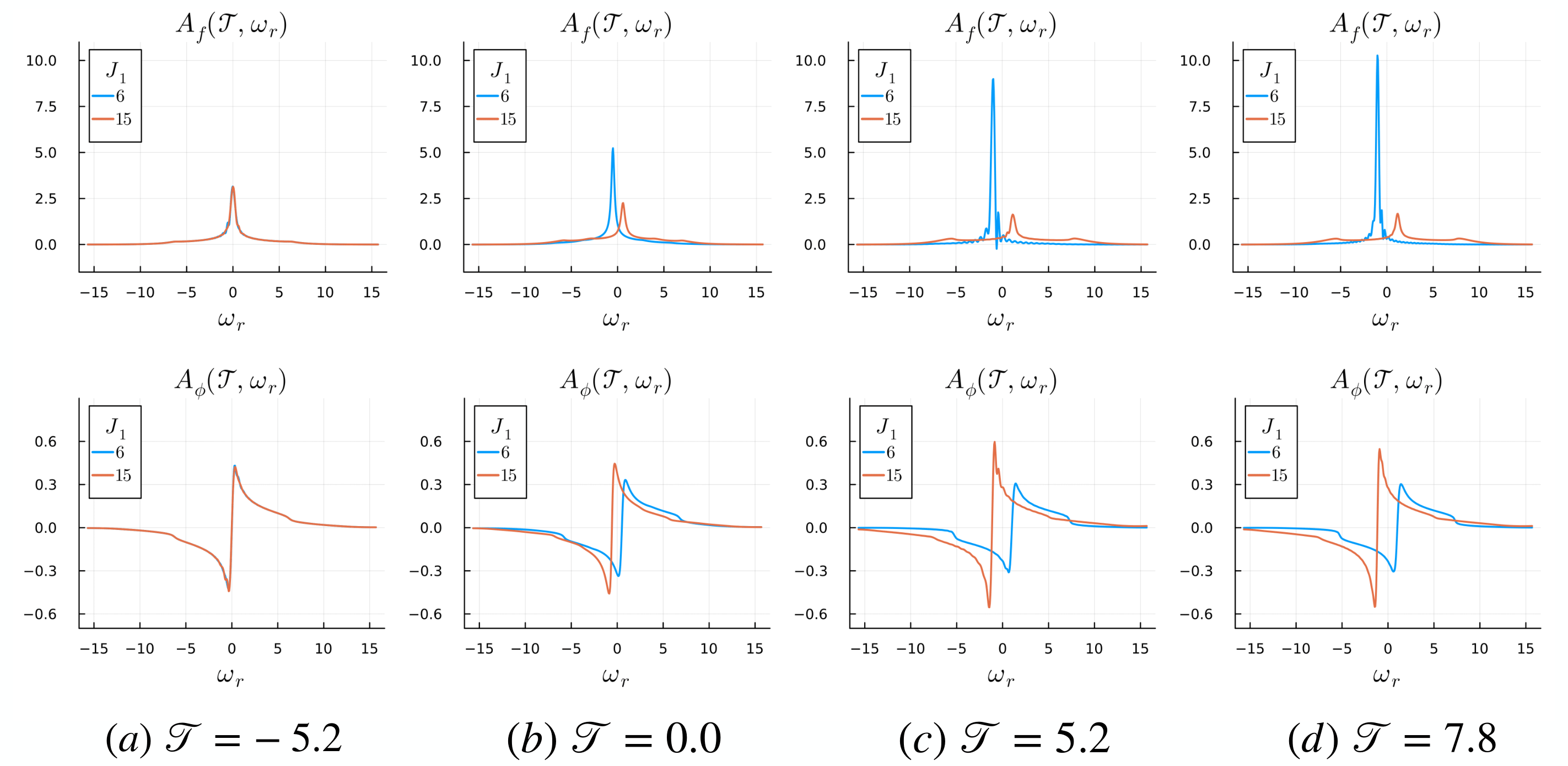}\caption{Evolution of the spectral functions of $f$-fermion and $\phi$-boson for $J_1=6$ and $J_1=15$ cases with the initial temperature $T_0=0.4T_K$.}\label{fig:EvolutionOfSpectralFunctionsT004Case}
\end{figure*}

In this section, we investigate the effects of the quantum quench on the impurity spin by considering the spin-spin correlation functions ($C_{imp}(\mathcal{T},t_r)$ and $\chi_{imp}(\mathcal{T},t_r)$). Due to the limitation of the numerical power, it is difficult to investigate the long time limit compared to the equilibrium case. Therefore we only focus on the transient dynamics of the $C_{imp}(\mathcal{T},t_r)$ and $\chi_{imp}(\mathcal{T},t_r)$. In the long time limit, we expect that the $C_{imp}(\mathcal{T},t_r)$ and $\chi_{imp}(\mathcal{T},t_r)$ show power law behaviors same to that observed in the equilibrium case~\cite{Heyl2010}.

Fig.~\ref{fig:CollectionOfLogPlotLowT0} and~\ref{fig:CollectionOfLogPlotHighT0} show the log plots of $C_{imp}(\mathcal{T},t_r)$ and $\chi_{imp}(\mathcal{T},t_r)$ with different values of $\mathcal{T}$ all together to see the change of exponential decaying rates of them in transient time scale more clearly.

When the $J_1=6$, the transient behavours of $C_{imp}(\mathcal{T},t_r)$ and $\chi_{imp}(\mathcal{T},t_r)$ with $T_0=0.4T_K$ are similar to that of $T_0=5T_K$. The log plots for $J_1=6$ case in Fig.~\ref{fig:CollectionOfLogPlotLowT0} and~\ref{fig:CollectionOfLogPlotHighT0} show that the exponential decay rates of both the $C_{imp}(\mathcal{T},t_r)$ and $\chi_{imp}(\mathcal{T},t_r)$ decrease as $\mathcal{T}$ increases. We expect that this is because the decay rates are proportional to the imaginary part of the self-energy of $f$-fermion $\tilde{\Sigma}_f(t,t’)$ (Eq.~\eqref{eq:Sigmaf}) which is proportional to the $J^2$. As a result, the decay rate is decreased by reducing the value of $J$ from $10$ to $6$. It is supported by the fact that ratios between slopes of initial time (Dashed blue line in Fig.~\ref{fig:CollectionOfLogPlotLowT0},~\ref{fig:CollectionOfLogPlotHighT0}) and later time (Dashed brown line in Fig.~\ref{fig:CollectionOfLogPlotLowT0},~\ref{fig:CollectionOfLogPlotHighT0}) is close to the value $J_1^2/J_0^2 (=0.36)$. We have checked that the ratio between slopes becomes closer to the value of $J_1^2/J_0^2$ as the initial state with a higher initial temperature is considered.

However when $J_1=15$, transient behaviours of $C_{imp}(\mathcal{T},t_r)$ and $\chi_{imp}(\mathcal{T},t_r)$ with $T_0=5T_K$ are different from that with $T_0=0.4T_K$. In the $T_0=5T_K$ case, the exponential decay rates are increased as $\mathcal{T}$ is increased and the ratios between decay rates of initial time and that of later time are close to the value of $J_1^2/J_0^2(=2.25)$ as shown in Fig.~\ref{fig:CollectionOfLogPlotHighT0} (For the initial state with higher temperature, the ratio become closer to the value of $J_1^2/J_0^2$). Therefore it can be also understood with the previous argument based on the fact that self-energy $\tilde{\Sigma}_f(t,t’)$ is proportional to $J^2$. Unlike the $T_0=5T_K$ case, the exponential decay rates remain almost the same value in the $T_0=0.4T_K$ case (Fig.~\ref{fig:CollectionOfLogPlotHighT0}). It means that dissipation between the spin and conduction electron do not get stronger even the coupling between them increases.
  
 These distinct features of the $T_0=0.4T_K$ case with $J_1=15$ can be understood because the impurity spin, in this case, is strongly entangled with conduction electrons, forming the over-screened Kondo state. Since this state is close to the eigenstate of the Kondo Hamiltonian, it does not react to the change of the Kondo coupling constant much. This is a contrasting feature compared to the $T_0=5T_K$ case where the state is closer to the eigenstate of the free fermion Hamiltonian. 
 
 Additionally, rapid oscillations with the same frequency discussed in Sec.~\ref{sec:EvolutionOfKondoOrder} are observed in $C_{imp}(\mathcal{T},t_r)$ and $\chi_{imp}(\mathcal{T},t_r)$ only for the $T_0=0.4T_K$ case with $J_1=15$ as shown in Fig.~\ref{fig:CimpAndChiImpLowT0} and~\ref{fig:CimpAndChiImpHighT0} in Appendix.~\ref{Appendix:EvolutionsOfCandChi}. As discussed in the Sec.~\ref{sec:EvolutionOfKondoOrder}, the oscillating pattern can be considered as one feature that the quench dynamics when the initial state is in a strongly entangled state~\cite{Michailidis}.

\subsubsection{Spectral functions}
From the spectral functions, we can obtain the effective temperature and the thermalization time under the quench. Since there are not many qualitative differences in the evolutions of the spectral function itself between the cases with different initial temperatures, we show the evolutions of the spectral functions for the initial temperature $T_0=0.4T_K$ case only with the largest $J_1(=15)$ and smallest $J_1(=6)$ values in Fig.~\ref{fig:EvolutionOfSpectralFunctionsT004Case}. 

Since the initial temperature is lower than the Kondo temperature $T_K$, the spectral function $A_f(\mathcal{T},\omega_r)$ with $\mathcal{T}=-5.2$ (Fig.~\ref{fig:EvolutionOfSpectralFunctionsT004Case} (a)) shows a  sharp Kondo peak.  In the $J_1=6$ case, the spectral function $A_f(\mathcal{T},\omega_r)$ shows an oscillating pattern during the evolution. Even when $\mathcal{T}=7.8$ (Fig.~\ref{fig:EvolutionOfSpectralFunctionsT004Case} (d)), it still shows a small oscillating near the peak. It is checked that the oscillating pattern disappears at a later time around 
$\mathcal{T}\approx 15$. 
 However, there is no such oscillating pattern that appears in the $A_f(\mathcal{T},\omega_r)$ for $J_1=15$ case. In every cases, the center of the $A_f(\mathcal{T},\omega_r)$ and $A_\phi(\mathcal{T},\omega_r)$ are moved. This is due to the time-dependent self-energy correction. However the change of the center or chemical potential does not affect the physically observable quantities based on the argument given in Appendix~\ref{app:U1symmetry}.

\paragraph{Effective temperature \& quasi-equilibration time}\label{sec:EffTempAndThermalTime}

\begin{figure*}[!htbp]
\centering
\begin{subfigure}{0.32\textwidth}
\centering
\includegraphics[scale=0.05]{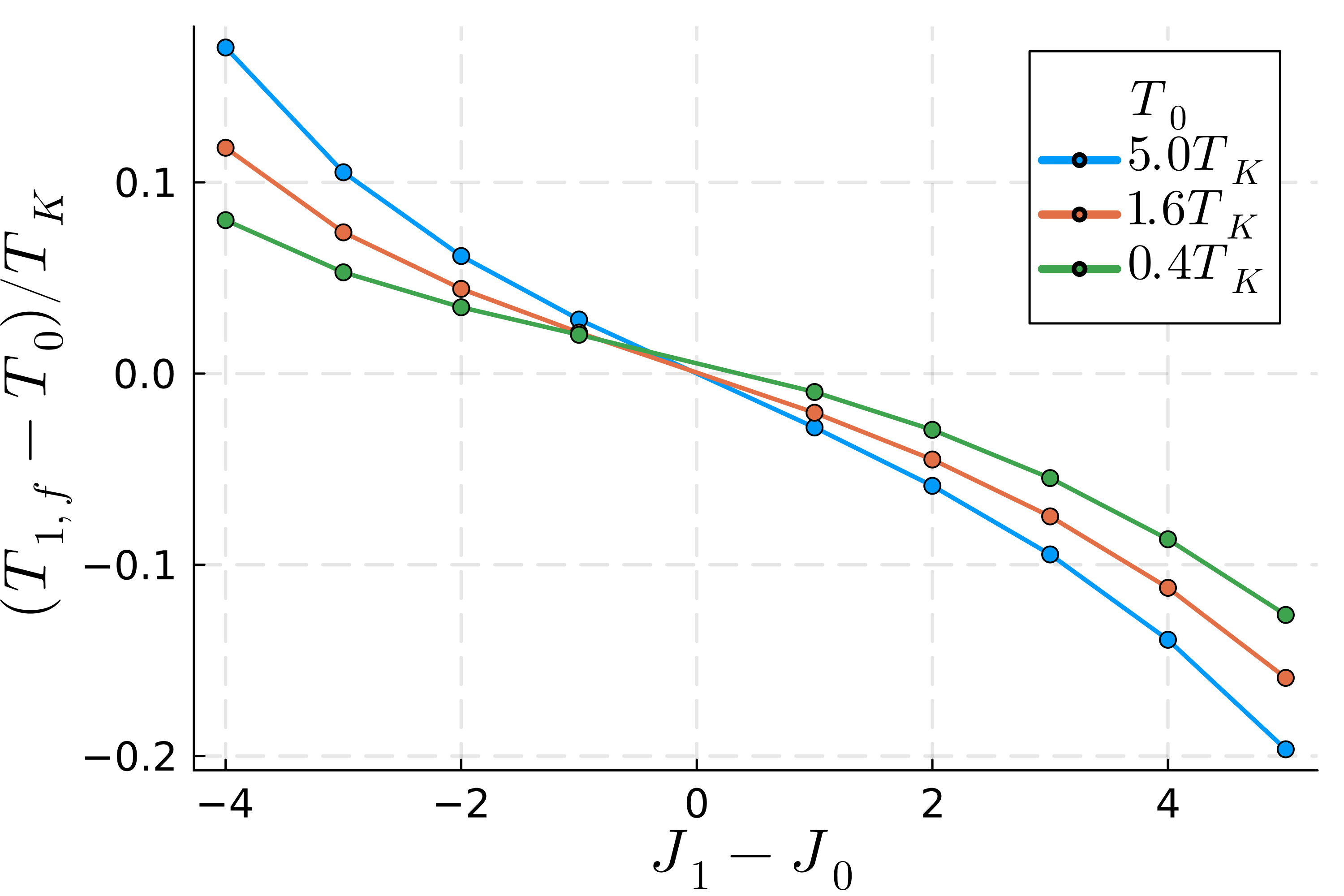}
\caption{Final temperature of $f$-fermion}\label{fig:FinalTemperaturesOfFermion}
\end{subfigure}
~
\begin{subfigure}{0.32\textwidth}
\centering
\includegraphics[scale=0.05]{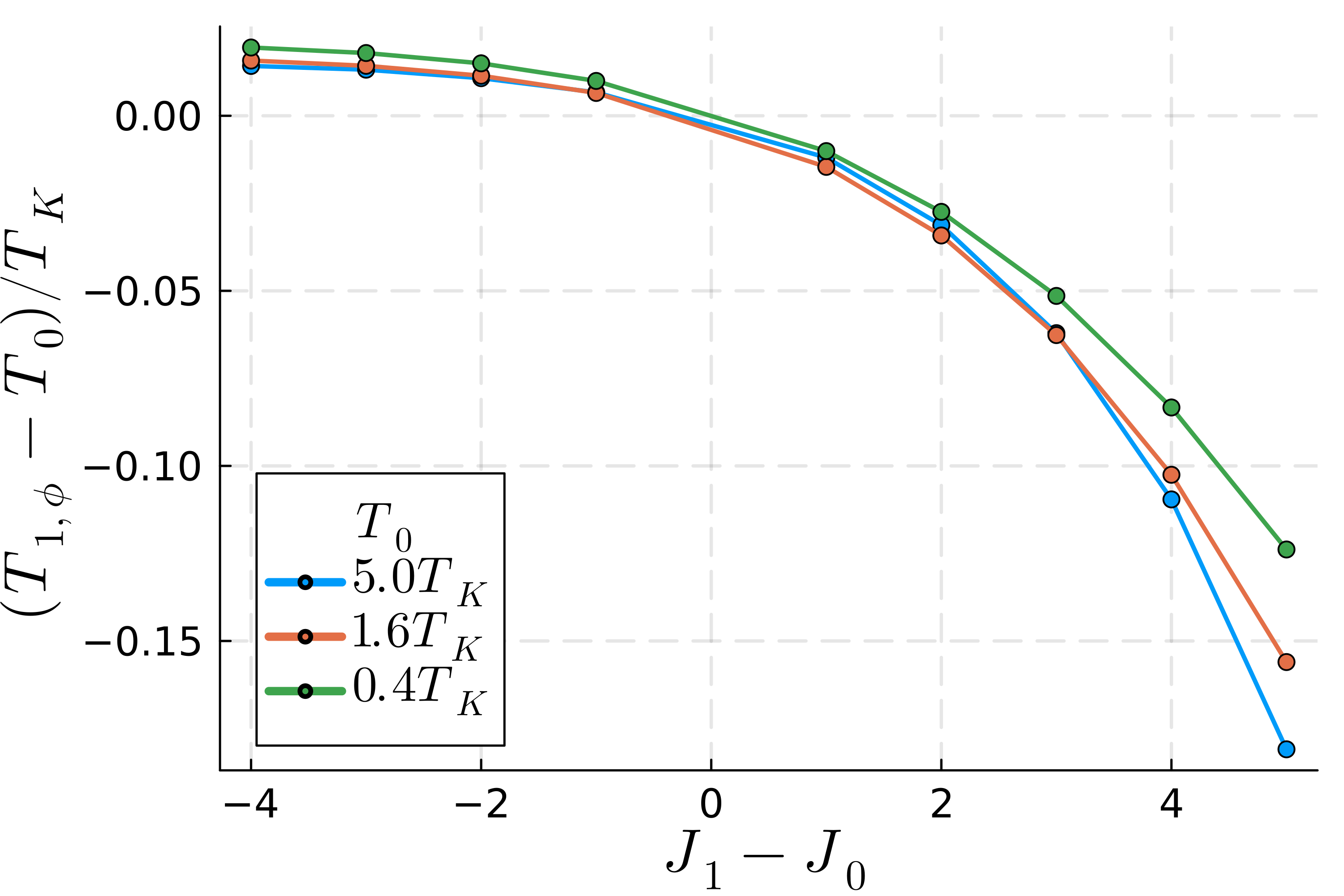}
\caption{Final temperature of $\phi$-boson}\label{fig:FinalTemperaturesOfBoson}
\end{subfigure}
~
\begin{subfigure}{0.32\textwidth}
\centering
\includegraphics[scale=0.05]{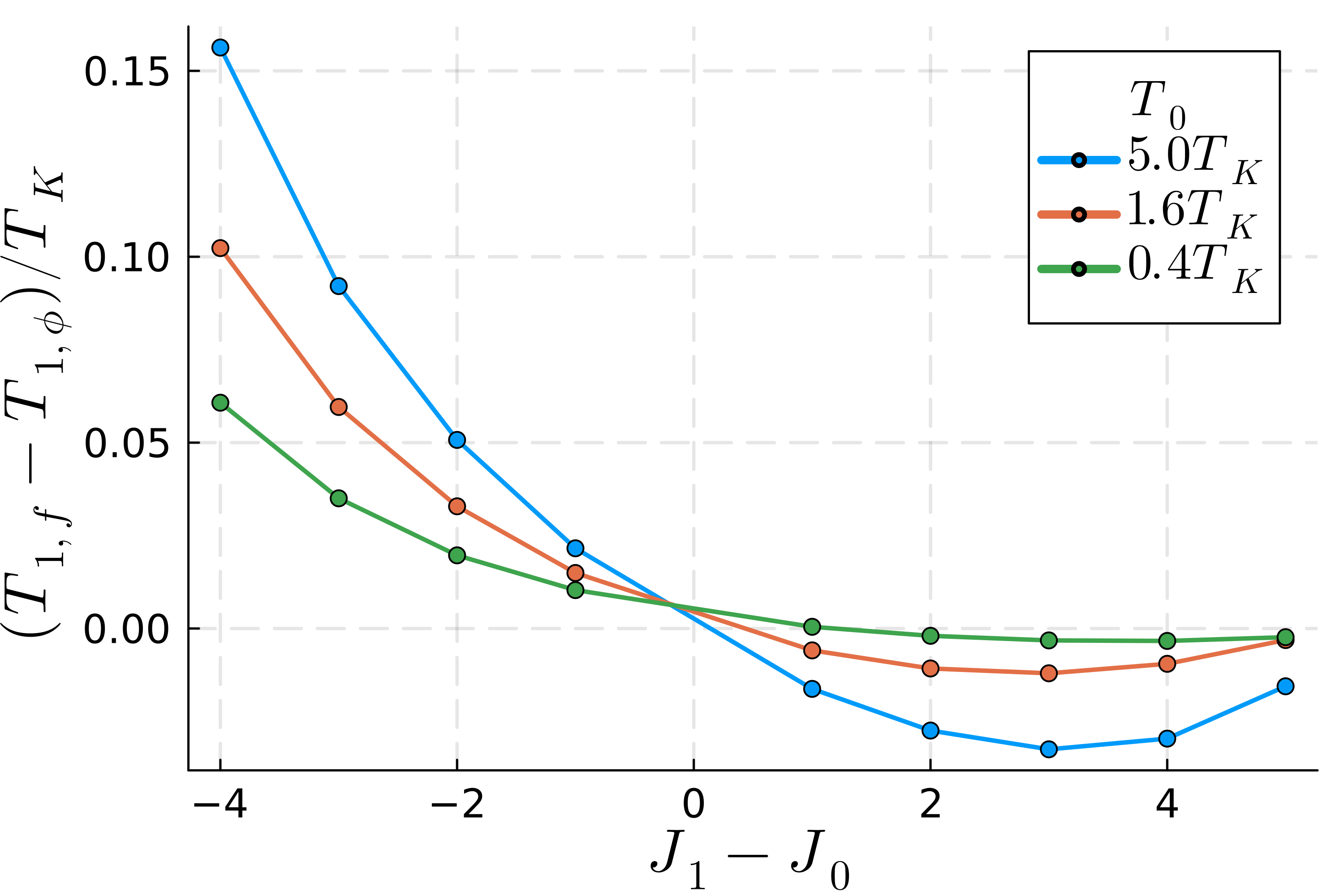}
\caption{Difference between final temperatures of $f$-fermion and $\phi$-boson.}\label{fig:FinalTemperaturesDifference}
\end{subfigure}
\caption{Final temperatures after quenching as a function of $J_1-J_0$ for the cases with the different initial temperatures $0.4T_K$, $1.6T_K$, and $5T_K$.}\label{fig:FinalTemperatures}
\end{figure*}

For more quantitative analysis, we obtain the effective temperatures of the $f$-fermion and $\phi$-boson using the Fluctuation-Dissipation relations
\begin{subequations}
\begin{gather}
\frac{iG_f^K(\mathcal{T},\omega_r)}{A_f(\mathcal{T},\omega_r)}=\tanh\Big(\frac{\omega_r}{2T_{f}(\mathcal{T})}\Big),\\
\frac{iG_\phi^K(\mathcal{T},\omega_r)}{A_\phi(\mathcal{T},\omega_r)}=\coth\Big(\frac{\omega_r}{2T_{\phi}(\mathcal{T})}\Big)
\end{gather}
\end{subequations}
where $T_{f}(\mathcal{T})$ and $T_{\phi}(\mathcal{T})$ are the effective temperatures of the $f$-fermion and $\phi$-boson respectively at the time $\mathcal{T}$. We found that the values of $T_{f}(\mathcal{T})$ and $T_{\phi}(\mathcal{T})$ converge as $\mathcal{T}$ is increased. Let us set converged values of $T_{f}(\mathcal{T})$ and $T_{\phi}(\mathcal{T})$ to $T_{1,f}$ and $T_{1,\phi}$ respectively. Additionally, we also define the thermalization time as a difference between two times $\mathcal{T}_{0,f/\phi}$ and $\mathcal{T}_{1,f/\phi}$ where $\mathcal{T}_{0,f/\phi}$ is a time when the effective temperature $T_{f/\phi}(\mathcal{T})$ first deviates from the initial temperature $T_0$ in the forward time direction while $\mathcal{T}_{1,f/\phi}$ is a time when the effective temperature $T_{f/\phi}(\mathcal{T})$ first deviates from the final temperature $T_{1,f/\phi}$ in the backward time direction. As a qualitative criterion for the deviation, we used the following criteria:  $\frac{|T_{f/\phi}(\mathcal{T}_{0/1,f/\phi})-T_{0/1,f/\phi}|}{T_{0/1,f/\phi}}>0.001$.

\begin{figure}[h]
\centering
\begin{subfigure}{0.5\textwidth}
\centering
\includegraphics[scale=0.055]{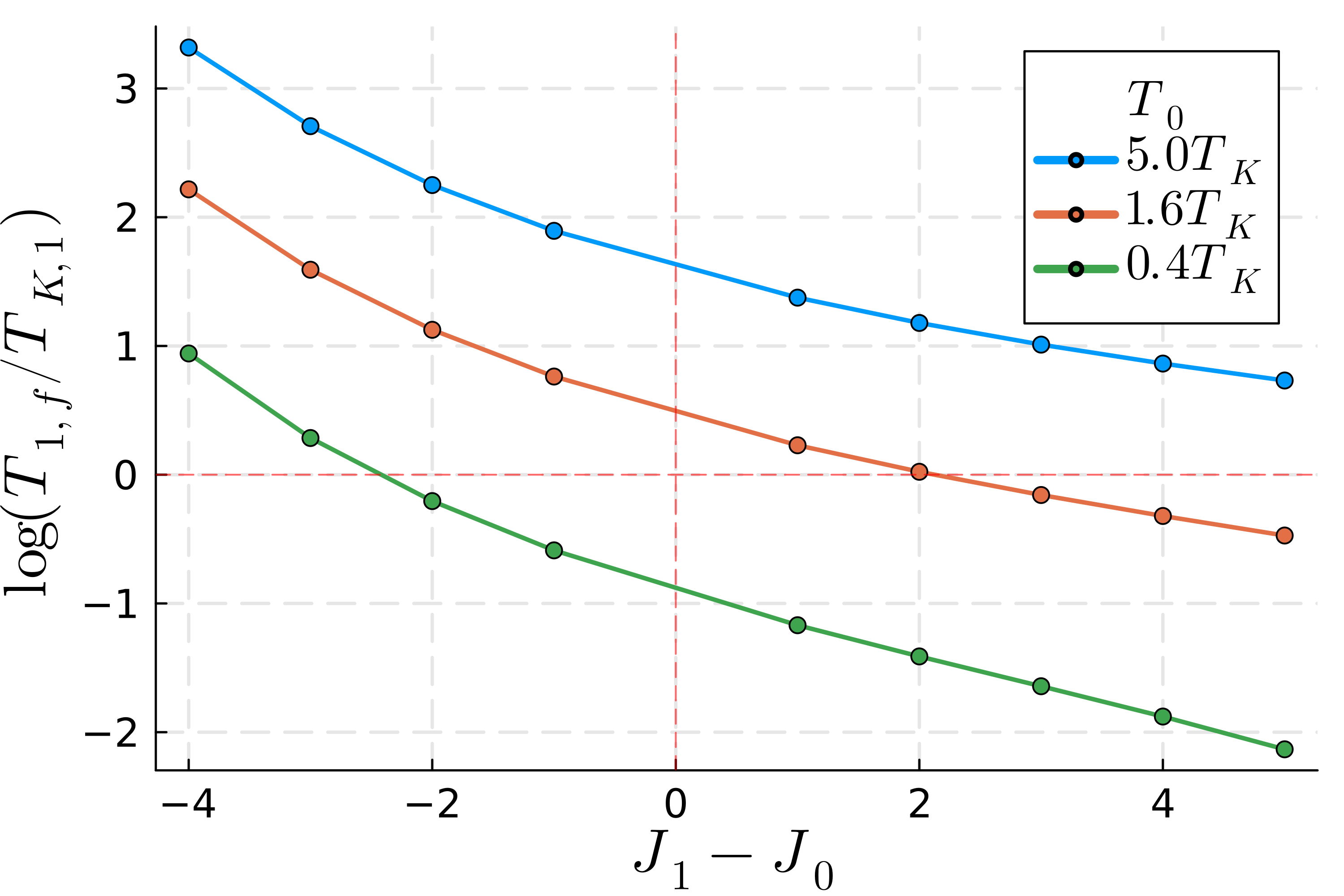}
\caption{$f$-fermion case}
\end{subfigure}
~
\begin{subfigure}{0.5\textwidth}
\centering
\includegraphics[scale=0.055]{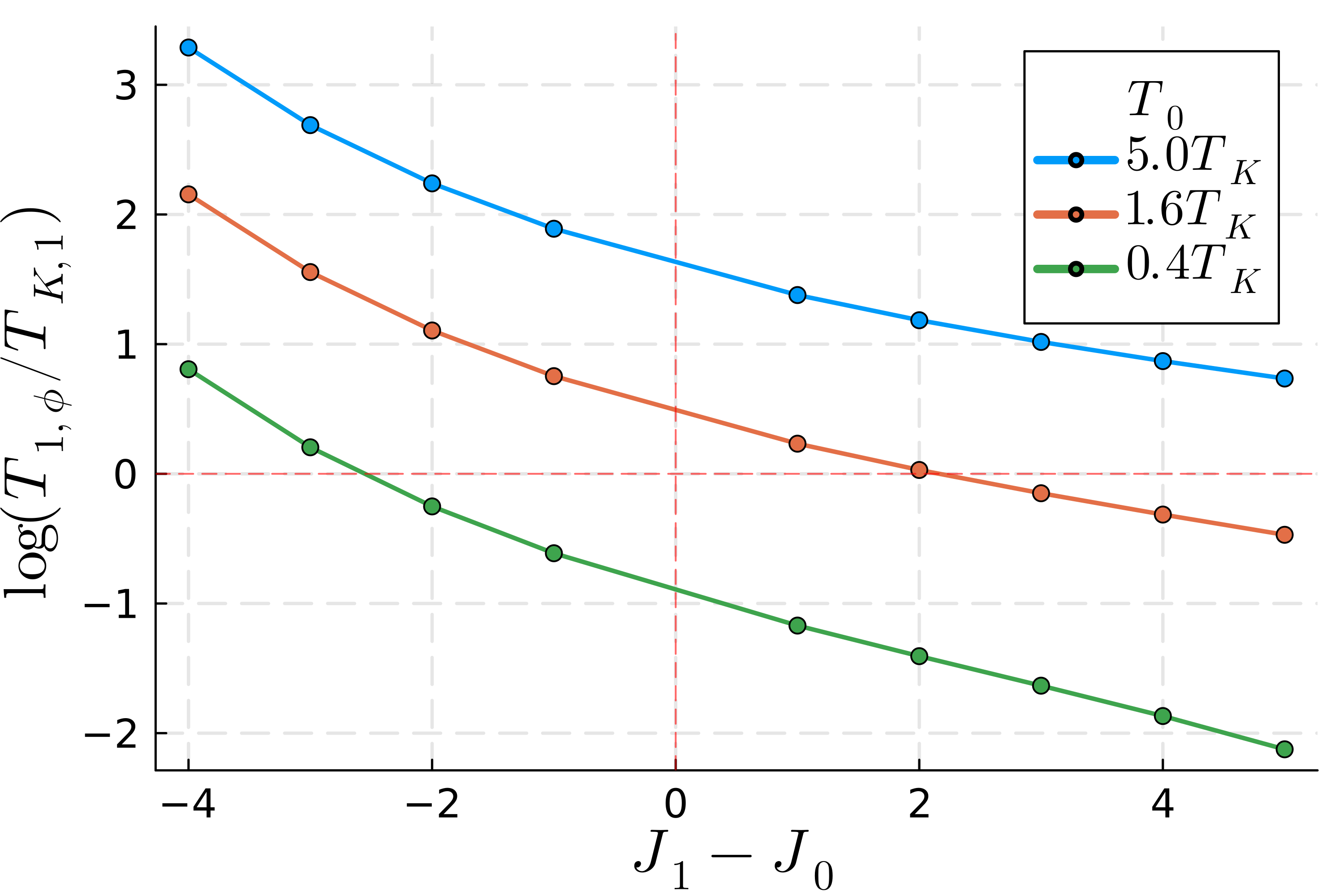}
\caption{$\phi$-boson case}
\end{subfigure}
\caption{Logarithmic ratio between final temperatures and the Kondo temperatures ($T_{K,1}=\Lambda e^{-\frac{1}{J_1N_F}}$) after quench as a function of $J_1-J_0$ for the cases with the different initial temperatures $0.4T_K$, $T_K$, $1.6T_K$, $5T_K$ and $5T_K$. The dashed red line denote the zero values of the logarithmic ratio and $J_1-J_0$ for clarity.}\label{fig:FinalTempVsKondo}
\end{figure}

Figure \ref{fig:FinalTemperatures} shows the final temperatures of the $f$-fermion (Figure \ref{fig:FinalTemperaturesOfFermion}), $\phi$-boson (Figure \ref{fig:FinalTemperaturesOfBoson}), and the temperature difference between them (Figure \ref{fig:FinalTemperaturesDifference}) for different initial states. The figures indicate that the $f$-fermion and $\phi$-boson experience an increase in temperature as the coupling constant $J_1$ decreases below $J_0$, whereas they experience a decrease in temperature when $J_1$ exceeds $J_0$. This cooling effect arises due to the formation of the over-screened state, which occurs when the coupling between the impurity spin and the conduction electrons strengthens.

\begin{figure*}
\centering
\begin{subfigure}{0.4\textwidth}
\includegraphics[scale=0.043]{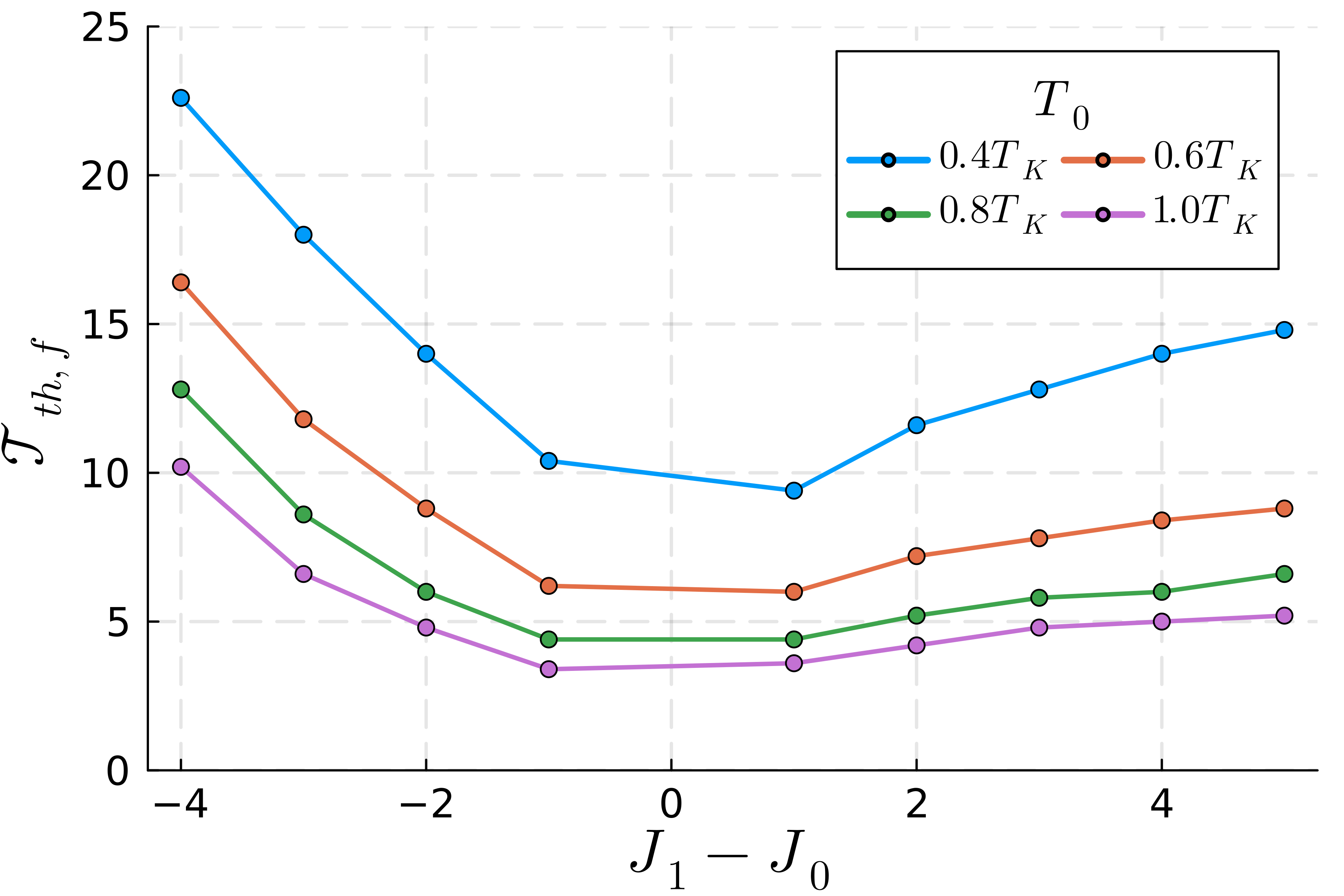}
\caption{Thermalization time of $f$-fermion}
\label{fig:ThermalTimefLowTemp}
\end{subfigure}
~
\begin{subfigure}{0.4\textwidth}
\includegraphics[scale=0.043]{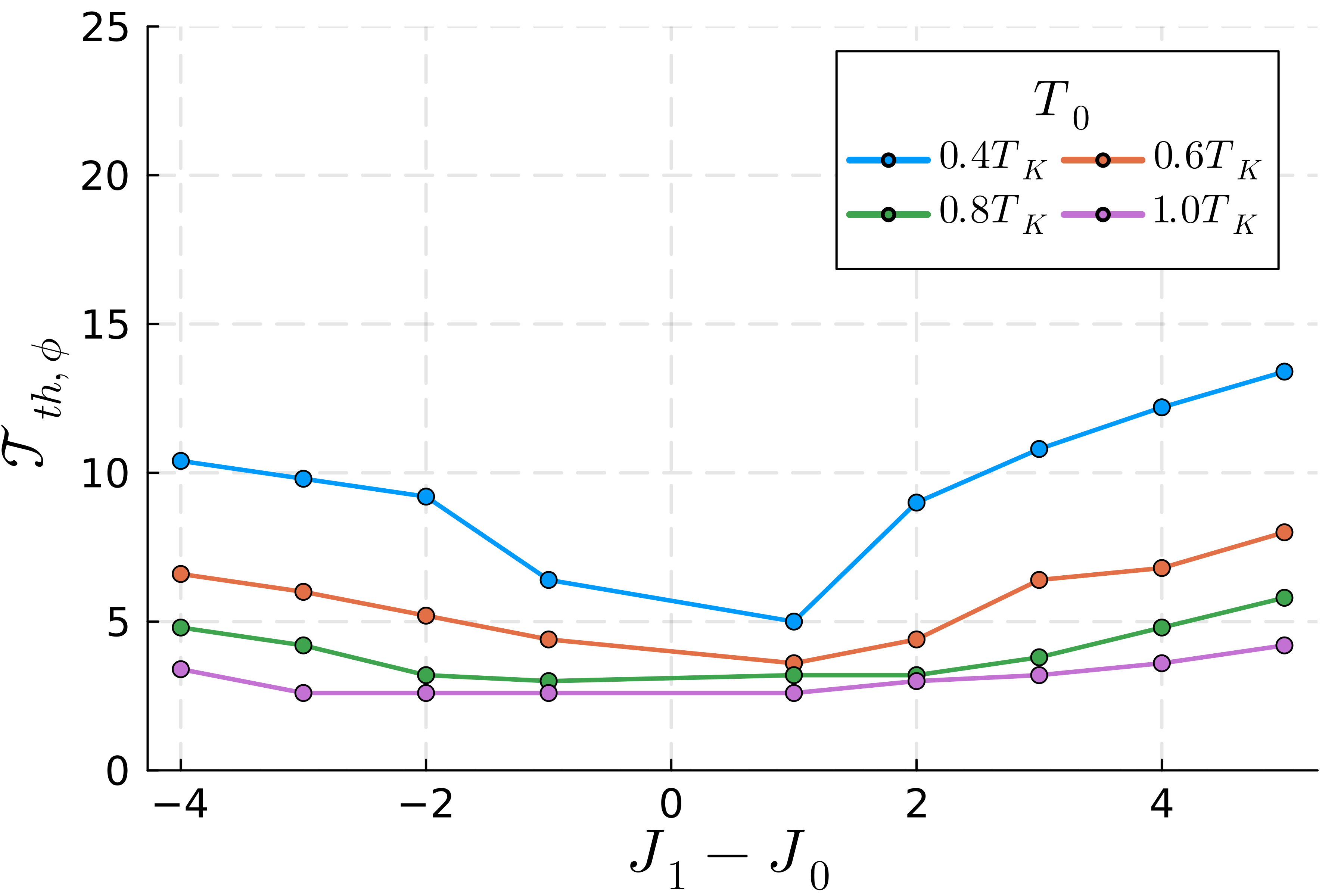}
\caption{Thermalization time of $\phi$-boson}
\label{fig:ThermalTimephiLowTemp}
\end{subfigure}
\caption{Thermalization time as a function of $J_1-J_0$ for thermalized cases with the initial temperatures $0.4T_K$, $0.6T_K$, $0.8T_K$, and $1.0T_K$. }\label{fig:ThermalizationTimeForLowDiffTemp}
\end{figure*}

\begin{figure*}
\centering
\begin{subfigure}{0.4\textwidth}
\includegraphics[scale=0.043]{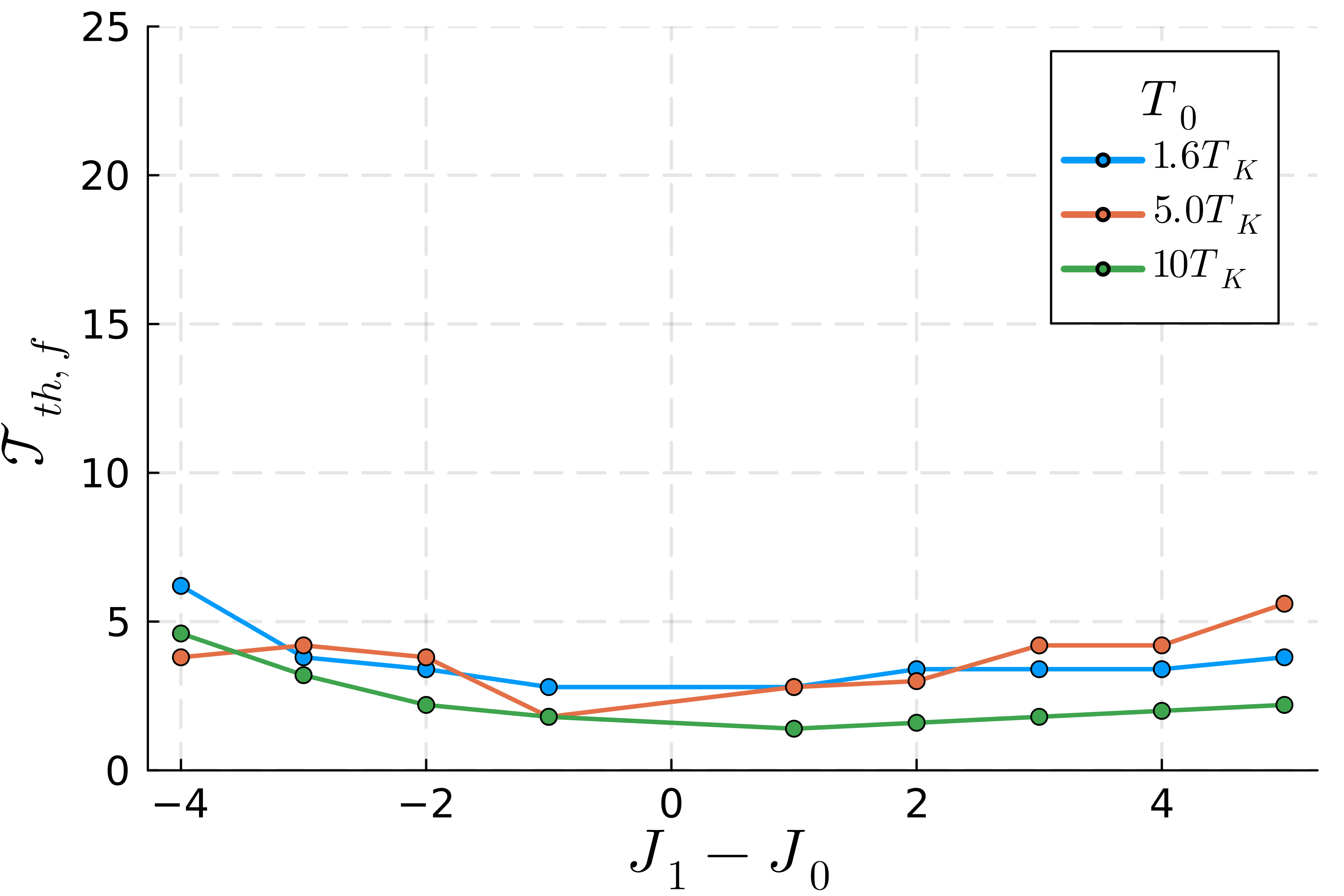}
\caption{Thermalization time of $f$-fermion}
\end{subfigure}
~
\begin{subfigure}{0.4\textwidth}
\includegraphics[scale=0.043]{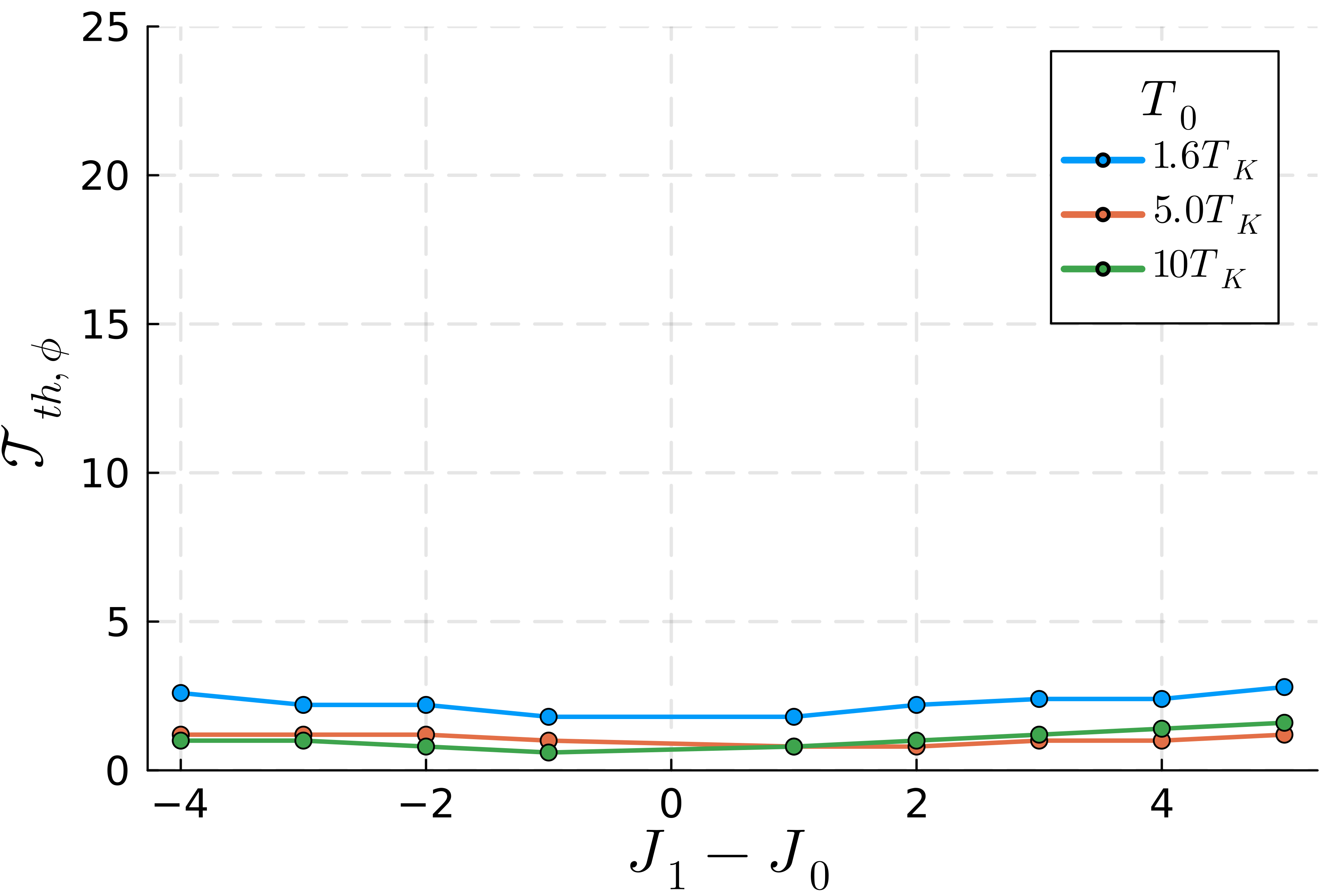}
\caption{Thermalization time of $\phi$-boson}
\end{subfigure}
\caption{Thermalization time as a function of $J_1-J_0$ for thermalized cases with the initial temperatures $1.6T_K$, $5.0T_K$, and $10T_K$. }\label{fig:ThermalizationTimeForHighDiffTemp}
\end{figure*}

In all the cases we studied, the effective temperature differences between the $f$-fermion and the $\phi$-boson
are always finite as shown in Figure \ref{fig:FinalTemperaturesDifference}. 
Notably, we observe the effective temperature differences can be small when the initial state is the over-screened state and the post-quench Kondo coupling is greater than the pre-quench Kondo coupling $J_0<J_1$. This property can also be understood from a phenomenological description employing the quantum Boltzmann equations. We will discuss this in Sec.~\ref{sec:QBEapproach}.

On the other hand, when the Kondo coupling constant is decreased $(J_1<J_0)$, 
the effective temperature differences between the $f$-fermion and the $\phi$-boson
is significantly large as shown in Fig.~\ref{fig:FinalTemperaturesDifference}. 
We can understand the large temperature difference from the decoupling between the fermion and boson due to the higher final temperature compared to the Kondo temperature $T_{K,1}$ of the quenched system. $T_{K,1}$ is the Kondo temperature obtained from the post-quench Kondo coupling constant $J_1$. Fig.~\ref{fig:FinalTempVsKondo} shows that the final effective temperatures of both $f$-fermion and the $\phi$-boson are higher than the Kondo temperatures of the post-quench system for $J_1<J_0$. As a result, $f$-fermion and $\phi$-boson are decoupled and can only be thermalized within their own Hilbert spaces separately, resulting in two very different temperatures. Our observation of this incoherent thermalization in the MCKI model is quite similar to the thermalization discussed in the mixed SYK model~\cite{Larzul2022} containing both the SYK$_2$ and SYK$_4$ terms. 
The incoherently thermalized states in the MCKI model can be interpreted as pre-thermalized states specified with the slow thermalization in the SYK model. The analogy between the incoherently thermalized states in the MCKI model and the pre-thermalized state in the SYK model becomes clear when we consider the re-feedback effect to the $f$-fermion and $\phi$-boson from the conduction electron with the $1/N$-correction. We will discuss it in the next section.

Now we discuss the quasi-equilibration time. Fig.~\ref{fig:ThermalizationTimeForLowDiffTemp} and~\ref{fig:ThermalizationTimeForHighDiffTemp} show the quasi-equilibration times of $f$-fermion and $\phi$-boson for the cases with initial temperatures lower than the $T_K$ and higher than the $T_K$ respectively. 
We observe the quasi-equilibration times for all the cases we studied are about an $\mathcal{O}(1)$ timescale.
In case of the low initial temperatures ($\leq T_K$), both thermalization times $\mathcal{T}_{th,f}$ (Fig.~\ref{fig:ThermalTimefLowTemp}) and $\mathcal{T}_{th,\phi}$ (Fig.~\ref{fig:ThermalTimephiLowTemp}) increase as the initial temperature is decreased. We expect that this is related to the non-Fermi liquid state formation in the low-temperature limit as discussed in~\cite{Ratiani2010}. Compared to the low initial temperature cases ($T_0\leq T_K$), the quasi-equilibration times of high initial temperature cases ($T_0\gg T_K$) do not depend on the initial temperature much as shown in Fig.~\ref{fig:ThermalizationTimeForHighDiffTemp}. Additionally, their quasi-equilibration times are much shorter than that of the low initial temperature case. However, we should keep in mind that $f$-fermion and $\phi$-boson are incoherently thermalized to two different temperatures. To form a full thermalized state with the same temperature, it is necessary to consider the $1/N$ correction or re-feedback effect from the conduction electron. 

\paragraph{Feedback effect to the conduction electron by the self-energy correction of order $\frac{1}{N}$}

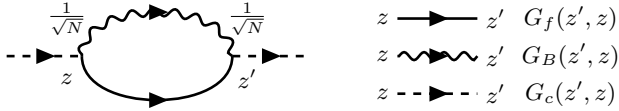
\begin{figure}[h]
\centering
\begin{tikzpicture}[baseline=-0.1cm]
\begin{feynhand}
\vertex (a) at (0,0); \vertex (b) at (1,0); \vertex (c) at (3,0); \vertex (d) at (4,0); 
\propag[chasca] (a) to (b); \propag[chabos] (b) to [in=90, out=90](c); \propag[chasca] (c) to (d);
\propag[fer] (b) to [in=-90,out=-90](c); 
\node at (0.8,0.5) {$\frac{1}{\sqrt{N}}$};
\node at (3.2,0.5) {$\frac{1}{\sqrt{N}}$};
\node at (0.8,-0.3) {$z$};
\node at (3.2,-0.3) {$z'$};
\end{feynhand}
~
	\begin{feynhand}
\vertex (a)  at (5,0.5) {$z$}; \vertex (b) at (6.5,0.5) {$z'$};
\propag[fer] (a) to (b);
\node at (7.5,0.5) {$G_f(z’,z)$};
\end{feynhand}
~
	\begin{feynhand}
\vertex (a)  at (5,0) {$z$}; \vertex (b) at (6.5,0) {$z'$};
\propag[chabos] (a) to (b);
\node at (7.5,0.) {$G_B(z’,z)$};
\end{feynhand}
~
	\begin{feynhand}
\vertex (a)  at (5,-0.5) {$z$}; \vertex (b) at (6.5,-0.5) {$z'$};
\propag[chasca] (a) to (b);
\node at (7.5,-0.5) {$G_c(z’,z)$};
\end{feynhand}
\end{tikzpicture}
\caption{One-loop self-energy correction of the order $\mathcal{O}(1/N)$ to the conduction electron.}
\label{fig:ConductionOneLoopSelf}
\end{figure}

 \begin{figure}[h]
\centering
\includegraphics[scale=0.06]{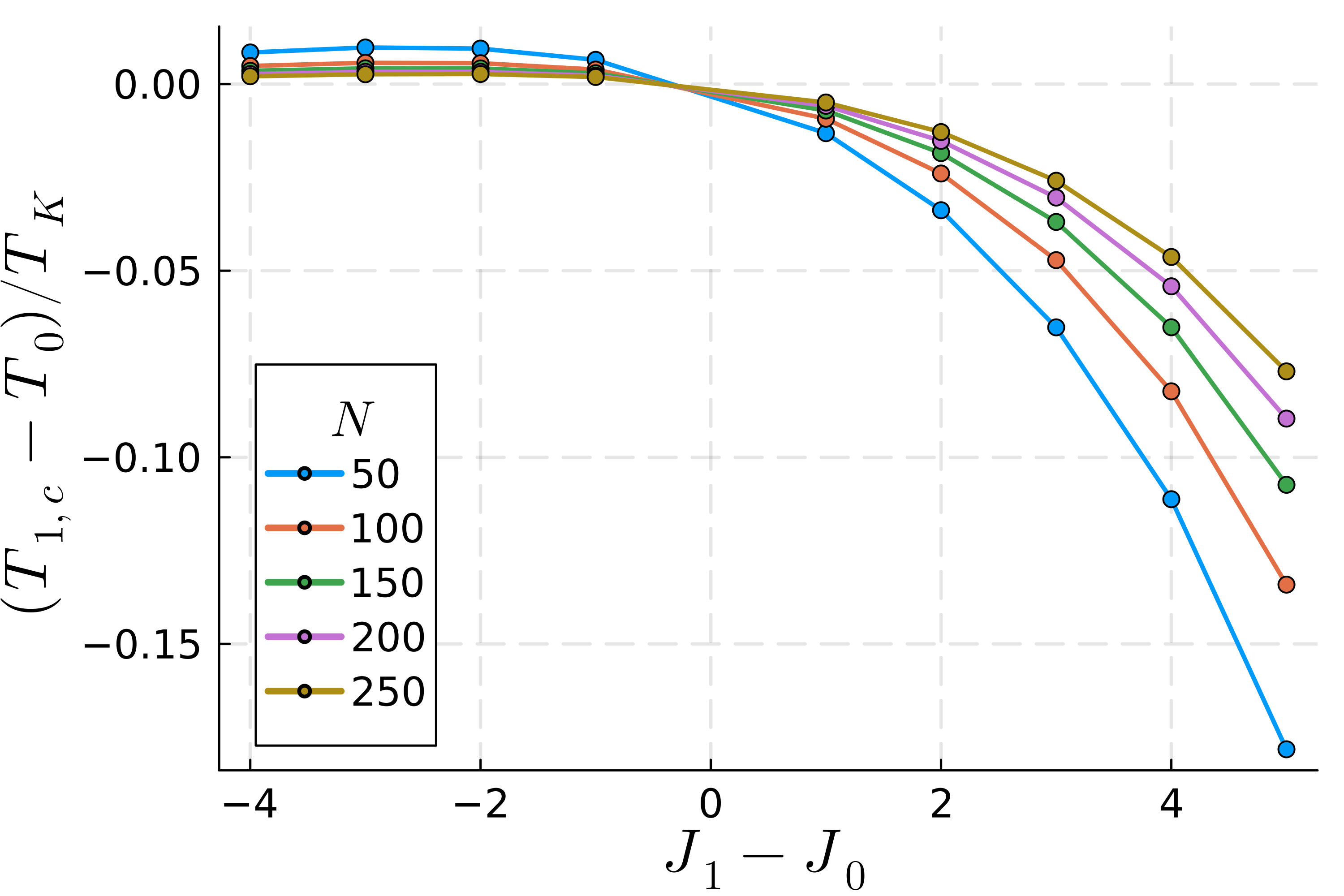}
\caption{Temperature difference of the conduction electron with the self-energy correction of  $\mathcal{O}(N^{-1})$. Here $T_0=5T_K$ case is considered.}\label{fig:ConductionElargeNFinalTemp}
\end{figure}

In the large-$N$ limit, there is no self-energy correction to the conduction electron. Therefore the temperature of the conduction electron remains at the initial temperature $T_0$. Here by considering the feedback effect of the order $\frac{1}{N}$ from the $f$-fermion and $\phi$-boson to the conduction electron, we argue the possibility of the full-thermalization by the re-feedback effect from the conduction electron to the $f$-fermion and $\phi$-boson.

 To investigate the feedback effect to the conduction electron from $f$-fermion and $\phi$-boson, we consider the one-loop self-energy correction, of order $1/N$ in Fig.~\ref{fig:ConductionOneLoopSelf}, to the conduction electron and calculate the change of the conduction electron’s effective temperature. For details of the Feynman rule, see Appendix.~\ref{Appendix:SelfConsFeynmanRules}. For simplicity, we calculate the $1/N$ self-energy correction in Fig.~\ref{fig:ConductionOneLoopSelf} using the solutions of Kadanoff-Baym equations (zeroth order in $1/N$) instead of solving the self-consistent equations including every $1/N$-corrections. In the large-$N$ limit, we expect that there would be not much difference between the solutions of the $ 1/N$ correction obtained by the approximated way and exact way. For details of calculating the effective temperature with the $1/N$-correction, see Appendix.~\ref{Appendix:EffectiveTempOfConductionElectron}. Fig.~\ref{fig:ConductionElargeNFinalTemp} shows the change of the effective temperature of the conduction electron as a function of $J_1-J_0$ for different $N$ values. The larger the value of $N$, we can see that the temperature change decreases. With this feedback effect, the effective temperature of the conduction electron becomes closer to that of $f$-fermion and $\phi$-boson. Therefore the dressed conduction electron, with the $1/N$ correction, will give a re-feedback effect to the $f$-fermion and $\phi$-boson and make them fully thermalize into the final state with the same temperature. Since the relaxation time of the conduction electron is proportional to the inverse of the imaginary part of the self-energy $N$, we expect that the lower bound for the thermalization time is of order $\mathcal{O}(N)$. The fact that it takes a time scale with the lower bound of order $\mathcal{O}(N)$ for the partial thermalized state to be the fully thermalized state makes the analogy we made between the partial thermalized states in the MCKI model and the pre-thermalized state in the SYK model more clear.

\section{Quantum Boltzmann equations}
\label{sec:QBEapproach}
To further verify the incoherent thermalization property from the previous analysis, we consider the quantum Boltzmann approach~\cite{AncelLarzul,kamenev_2011}. 
From Kadanoff-Baym equations (Eqs.~\eqref{eq:KBequation}) and the parametrization of $G_{f,\phi}^K$ using the $F_{f,\phi}$ defined as 
\begin{gather}
G_{f/\phi}^K=G_{f/\phi}^R\circ F_{f/\phi}-F_{f/\phi}\circ G_{f/\phi}^A,
\end{gather}
we obtain the following quantum Boltzmann equations
\begin{subequations}
\begin{align}
&\partial_{\mathcal{T}}F_f(\mathcal{T},\omega_r)=-\frac{\gamma J^2}{2}\int \frac{d\Omega}{2\pi}A_c(\omega_r+\Omega)A_\phi(\mathcal{T},\Omega)\nonumber\\
&\times \Big[1-F_c(\omega_r+\Omega)F_{\phi}(\mathcal{T},\Omega)\nonumber\\
&+\Big(F_\phi(\mathcal{T},\Omega)-F_c(\omega_r+\Omega)\Big)F_f(\mathcal{T},\omega_r)\Big],\\
&F_\phi(\mathcal{T},\omega_r)=\Big[\int \frac{d\Omega}{2\pi}A_c(\omega_r+\Omega)A_f(\mathcal{T},\Omega)\nonumber\\
&\times [1-F_c(\omega_r+\Omega)F_f(\mathcal{T},\Omega)]\Big]\Big[\int\frac{d\Omega}{2\pi}A_c(\omega_r+\Omega)\nonumber\\
&\times A_f(\mathcal{T},\Omega)[F_c(\omega_r+\Omega)-F_f(\mathcal{T},\Omega)]\Big]^{-1}.\label{eq:QBFphiEq}
\end{align}
\end{subequations}
Here we use identities:
\begin{subequations}
\begin{align}
&(i)\; f=g\circ h\nonumber\\
&\Rightarrow f(\mathcal{T},\omega_r) = g(\mathcal{T},\omega_r)e^{\frac{i}{2}\Big(\overleftarrow{\partial}_{\mathcal{T}}\overrightarrow{\partial}_{\omega_r}-\overleftarrow{\partial}_{\omega_r}\overrightarrow{\partial}_{\mathcal{T}}\Big)}
 h(\mathcal{T},\omega_r)\label{eq:WignerTransformationId1},\\
&(ii)\; f(t,t’)=g(t,t’)h(t’,t)\nonumber\\
&\Rightarrow  f(\mathcal{T},\omega_r)=\int\frac{d\omega_1}{2\pi}g(\mathcal{T},\omega_r+\omega_1)h(\mathcal{T},\omega_1).
\end{align}
\end{subequations}
We ignore all derivative terms in Eq.~\eqref{eq:WignerTransformationId1}. We will justify this approximation below.

Note that $F_\phi(\mathcal{T},\omega_r)$ is determined only by $A_f(\mathcal{T},\Omega)$ and $F_f(\mathcal{T},\Omega)$ from Eq.~\eqref{eq:QBFphiEq}. Although the above quantum Boltzmann equations are much simpler than Kadanoff-Baym equations, it is still quite difficult to obtain the analytical solutions. Therefore instead of solving the full quantum Boltzmann equations, we use only Eq.~\eqref{eq:QBFphiEq} to discuss how the final temperature of the $\phi$-boson depends on a form of the spectral function and the final temperature of the $f$-fermion phenomenologically. Since we assume that the final state is in equilibrium, the assumption of ignoring all derivative terms can be justified. 

Let us assume the temperature of the final state of $f$-fermion is given by $T_{1,f}$ then $F_f(\omega)=\tanh\Big(\frac{\omega}{2T_{1,f}}\Big)$. Additionally using the fact that $F_c(\omega)=\tanh\Big(\frac{\omega}{2 T_0}\Big)$, $A_c(\omega)\propto \theta(\Lambda-|\omega|)$ where $T_0$ is the temperature of the conduction bath, Eq.~\eqref{eq:QBFphiEq} is simplified into
\begin{align}
F_\phi(\omega_r)&= \Bigg[\int_{-\Lambda}^{\Lambda} \frac{d\Omega}{2\pi}A_f(\Omega-\omega_r)\coth\Big(\frac{\Omega}{2T_0}-\frac{\Omega-\omega_r}{2T_{1,f}}\Big)\nonumber\\
&\times \Big[\tanh\Big(\frac{\Omega}{2 T_0}\Big)-\tanh\Big(\frac{\Omega-\omega_r}{2T_{1,f}}\Big)\Big]\Bigg]\nonumber\\
&\times \Bigg[\int_{-\Lambda}^{\Lambda}\frac{d\Omega}{2\pi}A_f(\Omega-\omega_r)\Big[\tanh\Big(\frac{\Omega}{2 T_0}\Big)\nonumber\\
&-\tanh\Big(\frac{\Omega-\omega_r}{2T_{1,f}}\Big)\Big]\Bigg]^{-1}\label{eq:FphiFinalState}
\end{align}
where we have used the identity: $\tanh(x-y)=\frac{\tanh x-\tanh y}{1-\tanh x\tanh y}$. 

As a general form of $A_f(\Omega)$, we use
\begin{align}
A_f(\Omega)&=2\frac{\Gamma+ |\Omega|^{1-\alpha}\sin\Big(\frac{\alpha \pi}{2}\Big)}{|\Omega|^{2(1-\alpha)}\cos^2\Big(\frac{\alpha \pi}{2}\Big)+\Big(\Gamma+|\Omega|^{1-\alpha}\sin\Big(\frac{\alpha\pi}{2}\Big)\Big)^2},\nonumber\\
&(0\leq \alpha<1)
\end{align}
where $\Gamma$ is a constant scattering rate and $\alpha$ is an anomalous dimension originating from the Kondo interaction. The above spectral function is obtained from the Matsubara Green’s function of the form $G_f(i\omega_n)=\frac{1}{i\omega_n|\omega_n|^{-\alpha}+isgn(\omega_n)\Gamma }$ by an analytic continuation.

$A_f(\Omega)$ with $\alpha=0$ is a usual spectral function of the Fermi liquid state with the scattering rate $\Gamma$ while $A_f(\Omega)$ with the non-zero $\alpha$ describes non-Fermi liquid state. In the Multi-Channel Kondo system, the high-temperature state of $f$-fermion can be considered as a Fermi liquid state with the finite scattering rate due to the conduction electron while the low-temperature state is a non-Fermi liquid state so-called over-screened state with $\alpha$ is given by $\frac{1}{1+\gamma}$~\cite{Parcollet}. 

\begin{figure}[h]
\centering
\begin{subfigure}{0.45\textwidth}
\centering
\includegraphics[scale=0.25]{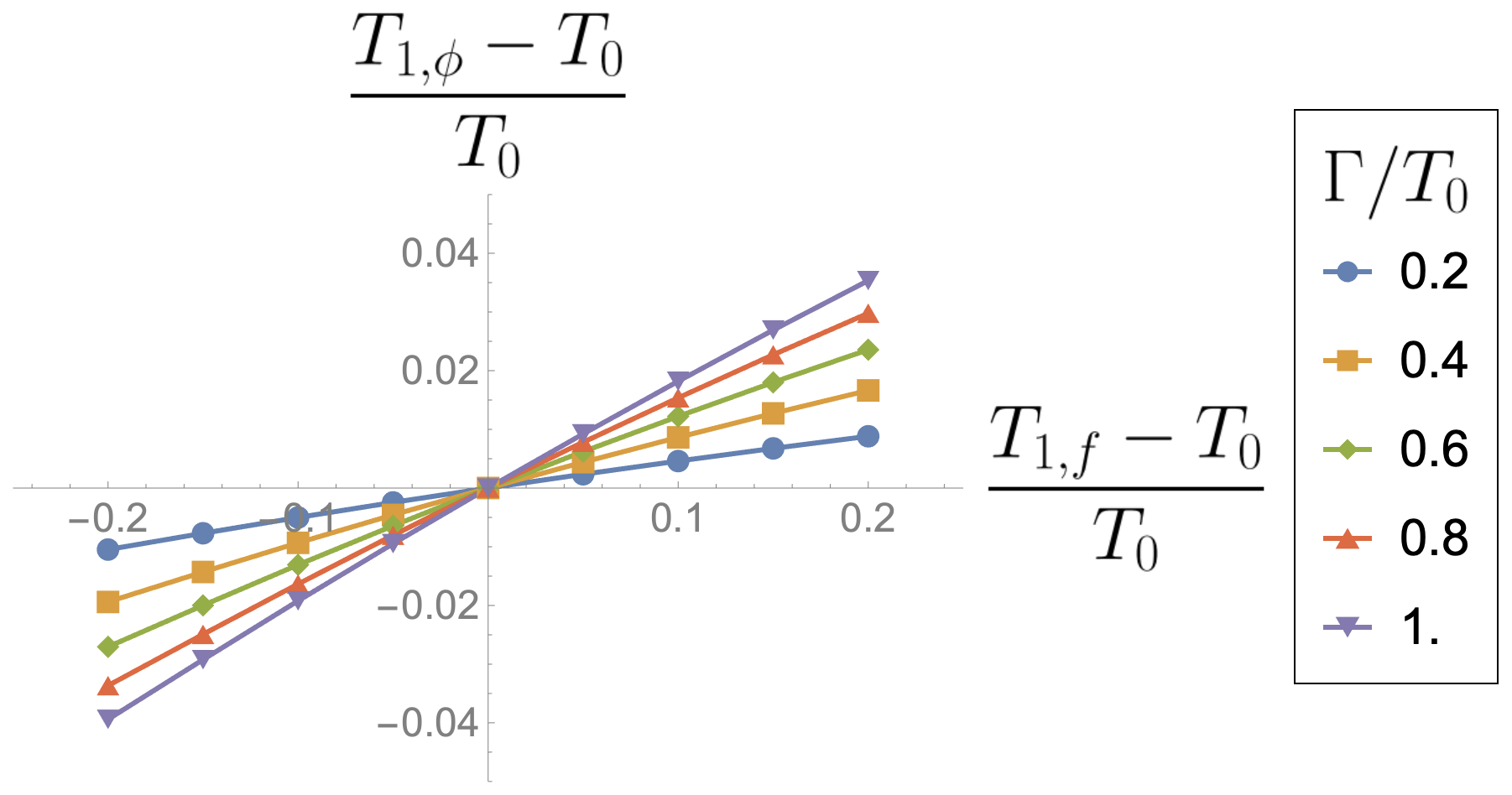}
\caption{$T_{1,\phi}$ as a function of $T_{1,f}$ for different values $\Gamma$ with $\alpha=0$.}\label{fig:T1phiForDiffGamma}
\end{subfigure}
~
\begin{subfigure}{0.45\textwidth}
\centering
\includegraphics[scale=0.25]{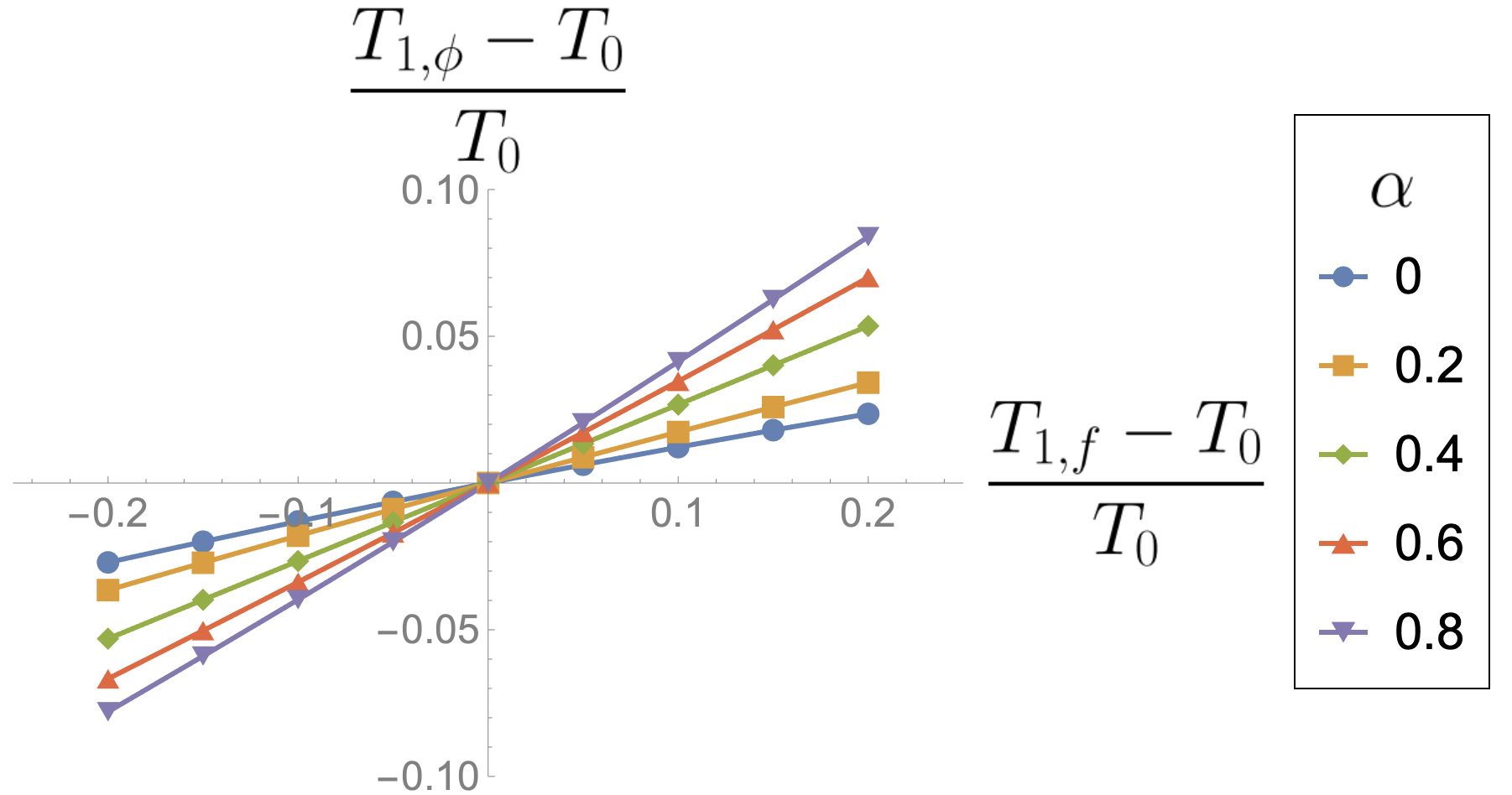}
\caption{$T_{1,\phi}$ as a function of $T_{1,f}$ for different values $\alpha$ with $\Gamma/T_0=0.4$.}\label{fig:T1phiForDiffAlpha}
\end{subfigure}
\caption{$T_{1,\phi}$ as a function of $T_{1,f}$ with different values of $\Gamma$ and $\alpha$.}
\label{fig:T1phiAsFunctionT1f}
\end{figure}

As a phenomenological approach to understand the difference of final effective temperatures of $f$-fermion ($T_{1,f}$) and $\phi$-boson ($T_{1,\phi}$) for different quench protocols, we obtain the final effective temperature of $\phi$-boson for a given value of $T_{1,f}$ with different forms of $A_{f}(\Omega)$ by varying $\alpha$ and $\Gamma$. 

If we consider the $T_{1,f}=T_0$ case as a sanity check, one can easily check that the $F_{\phi}(\omega_r)=\coth\Big(\frac{\omega_r}{2T_0}\Big)$ from Eq.~\eqref{eq:FphiFinalState} regardless of the form of $A_f(\Omega)$. It is consistent with the fact that the system is in a full equilibrium state. 

Now let us consider when $T_{1,f}$ is different from $T_0$. By fitting the numerically obtained $F_{\phi}(\omega_r)$ with $\coth\Big(\frac{\omega_r}{2T_{1,\phi}}\Big)$, we obtain values of $T_{1,\phi}$. Fig.~\ref{fig:T1phiAsFunctionT1f} shows the values of $(T_{1,\phi}-T_{0})/T_0$ as a function of $(T_{1,f}-T_{0})/T_0$ for different $\Gamma$ values with $\alpha=0$ (Fig.~\ref{fig:T1phiForDiffGamma}) and different $\alpha$ values with the fixed $\Gamma$ (Fig~\ref{fig:T1phiForDiffAlpha}). 

From Fig.~\ref{fig:T1phiAsFunctionT1f}, we can see that the final temperature of $\phi$-boson becomes closer to that of $f$-fermion as the scattering rate $\Gamma$ and the anomalous dimension $\alpha$ increase. To some extent, it indicates that the temperature difference between the $f$-fermion and the $\phi$-boson becomes smaller when we consider the quench with the larger $J$ values and the initial temperature lower than $T_K$ where the initial state is given by the non-Fermi liquid state. Additionally, the analysis here shows that the magnitude of difference between the $T_{1,\phi}$ and $T_0$ ($=(T_{1,\phi}-T_0)/T_0$) is always smaller than that between $T_{1,f}$ and $T_{0}$ ($=(T_{1,f}-T_0)/T_0$) as shown in Fig.~\ref{fig:T1phiAsFunctionT1f}. It is consistent with the numerical results discussed in Sec.~\ref{sec:EffTempAndThermalTime}.

\section{Conclusion}\label{sec:Conclusion}

In this paper, we have investigated the transient behavior and quasi-equilibrium properties in the Multi-Channel Kondo system following a quantum quench of the Kondo coupling constant. We use the $SU(N)$ MCKI model and Scwinger-Keldysh formalism with the large-$N$ method. 
Through numerical analysis of the transient dynamics, we observe several universal properties in the evolutions of the spin-spin correlation functions and the Kondo order parameter. The transient properties of the spin-spin correlation functions exhibit an exponential decay. 
Interestingly, the exponential decay rate is invariant when the initial state corresponds to the over-screened state and the post-quench Kondo coupling $J_1$ is larger than the pre-quench Kondo coupling $J_0$. Conversely, the exponential decay rates strongly depend on the quench protocols
and the ratio of the decay rates between the pre-quench and the post-quench states is closed to the value of $J_1^2/J_0^2$ in other cases.
Furthermore, we observe that the oscillation pattern of the Kondo order parameter exhibits the universal frequency given by the energy cut-off of the conduction electrons, during the transient timescale for the initial state being the over-screened state. 
On the other hand, for the initial states corresponding to high-temperature Fermi liquid states, no such oscillations are observed during the transient time. The observed oscillation patterns, which occur exclusively in the over-screened initial state, are closely linked to the revival phenomenon commonly observed in entangled states under a quench~\cite{Michailidis}. These oscillations exhibit a frequency determined by the energy cut-off of the conduction electron and bear similarity to the Friedel oscillation in the time domain.

In terms of the long-time behaviors following quantum quenches, 
we find that the MCKI model reaches a quasi-equilibrium state, characterized by a thermal distribution.  
However, the temperatures of the $f$-fermion and $\phi$-boson are not equilibrated into the same value.
To validate our numerical observations, we have performed a careful analysis and corroborated our findings using
quantum Boltzmann equations. 
Despite the quasi-equilibrium state exhibiting different effective temperatures of $f$ and $\phi$,
we have determined that the system 
 can still evolve into a fully thermalized final state by
considering the $1/N$ correction of the self-energy in the conduction electrons.
We find that the dressed conduction electrons would be able to eventually thermalize the total system with the thermalization time proportional to $N$, 
which is estimated from the relaxation time of the conduction electrons. Furthermore, we find the quantum cooling effect can happen when the Kondo coupling constant is suddenly increased. This quantum cooling effect is attributed to the reduction of the impurity entropy and can be thought of as an algorithmic cooling.

Our study systematically analyzes the non-equilibrium properties of the MCKI model under sudden quantum quenches. We offer complementary results to the previous studies~\cite{Ratiani2010,Heyl2010,Dalum2020,Dalum20202,Karki2020,Kirchner2009,Ribeiro2013}, which have mostly focused on the steady states. 
Leveraging the large-N structure of the model, we anticipate that our findings can provide valuable guidance for investigating non-equilibrium phenomena in the MCKI model using methods such as Conformal Field Theory (CFT)~\cite{Parcollet} and holography~\cite{Erdmenger2017}. 
Moreover, by employing the same framework, it is feasible to explore non-equilibrium setups in diverse physical systems.
One intriguing system worthy of consideration is the pseudo-gap Kondo Impurity model~\cite{Withoff1990,Kan1995,Fritz2004}. 
Given that the pseudo-gap Kondo system exhibits a quantum phase transition~\cite{Fritz2004}, it would be fascinating to study and investigate non-equilibrium phenomena, such as the dynamical quantum phase transition (DQPT)~\cite{Heyl2018} in this system. Another interesting direction involves exploring the transient non-equilibrium current through a multi-channel Kondo impurity subjected to sudden voltage quenches. This quench protocol has primarily been studied using computationally intensive methods like quantum Monte Carlo (QMC)~\cite{Werner2010,Schiro2010,Antipov2016,Krivenko2019}, density matrix renormalization group (DMRG)~\cite{Cazalilla2002,Kirino2008,Heidrich2009,Kirino2011}, and time-dependent numerical renormalzation group (TDNRG)~\cite{Anders2006,Eidelstein2012,Nghiem2017} methods. 
Our findings offer valuable insights into the non-equilibrium dynamics of the MCKI model in the transient timescale, revealing several previously undiscovered features in the many-body quantum dynamics.




\section*{Acknowledgements}
I.J. acknowledge the discussion with Francesco Piazza. We thank Stefan Kirchner, Chung-Hou Chung, Miguel A. Cazalilla, and Ethan Lake for the discussion,
and the National Center for Theoretical Sciences (NCTS) for the support. This work is supported by the National Science and Technology Council of Taiwan under Grants No. NSTC 112- 2636-M-007-007.



\begin{appendix}


\section{Derivation of the large-$N$ saddle point self-consistent equations using the Feynman diagrams}\label{Appendix:SelfConsFeynmanRules}

Here we provide a derivation of the saddle point self-consistent equations (Eq.~\eqref{eq:SKEquations}) using Feynman diagrams. Based on the Hubbard-Stratonovich transformed action Eq.~\eqref{eq:LargeNEffectiveAction1}, we obtain following Feynman rules:

\begin{gather*}
\begin{tikzpicture}[baseline=-0.1cm]
\begin{feynhand}
\vertex (a)  at (0,0) {$z$}; \vertex (b) at (2,0) {$z'$};
\propag[fer] (a) to (b);
\end{feynhand}
\end{tikzpicture}
=-i\langle f_{\alpha}(z')f_{\beta}^\dagger(z)\rangle=\delta_{\alpha\beta}G_{f}(z',z)\\
\begin{tikzpicture}[baseline=-0.1cm]
\begin{feynhand}
\vertex (a)  at (0,0) {$z$}; \vertex (b) at (2,0) {$z'$};
\propag[chasca] (a) to (b);
\end{feynhand}
\end{tikzpicture}
=-i\langle c_{i\alpha}(z')c_{j\beta}^\dagger(z)\rangle=\delta_{ij}\delta_{\alpha\beta}G_{c}(z',z)\\
\begin{tikzpicture}[baseline=-0.1cm]
\begin{feynhand}
\vertex (a)  at (0,0) {$z$}; \vertex (b) at (2,0) {$z'$};
\propag[chabos] (a) to (b);
\end{feynhand}
\end{tikzpicture}
=-i\langle B_{i}(z')B_{j}^\dagger(z)\rangle=\delta_{ij}G_{B}(z',z)\\
\begin{tikzpicture}[scale=0.7,baseline=-0.1cm]
\begin{feynhand}
\vertex (a)  at (-1.5,0) {$i,\alpha$}; \vertex (b) at (0,0); \vertex (c) at (1,1){$\alpha$}; \vertex (d) at (1,-1){$i$}; 
\propag[chasca] (a) to (b);
\propag[fermion] (b) to (c);
\propag[chabos] (b) to (d);
\end{feynhand}
\end{tikzpicture}
=-\frac{i}{\sqrt{N}},
~
\begin{tikzpicture}[scale=0.7,baseline=-0.1cm]
\begin{feynhand}
\vertex (a)  at (-1.5,0) {$i,\alpha$}; \vertex (b) at (0,0); \vertex (c) at (1,1){$\alpha$}; \vertex (d) at (1,-1){$i$}; 
\propag[antsca] (a) to (b);
\propag[fermion] (c) to (b);
\propag[antbos] (b) to (d);
\end{feynhand}
\end{tikzpicture}
=-\frac{i}{\sqrt{N}}
\end{gather*}

Using the above Feynman rules, one-loop self energies of the $f$ fermion and the $B$ boson are given by
\begin{gather}
\Sigma_{f}(z',z)=
\begin{tikzpicture}[scale=0.7,baseline=-0.1cm]
\begin{feynhand}
\vertex (a) at (0,0); \vertex (b) at (1,0); \vertex (c) at (3,0); \vertex (d) at (4,0); 
\propag[fer] (a) to (b); \propag[chasca] (b) to [in=90, out=90](c); \propag[fer] (c) to (d);
\propag[chabos] (c) to [in=-90,out=-90](b); 
\node at (0.8,-0.3) {$z$};
\node at (3.2,-0.3) {$z'$};
\end{feynhand}
\end{tikzpicture}
= i\gamma G_c(z',z)G_B(z,z'),\label{eq:fSelfEnergyFeynmanDiagram}
\\
\Sigma_{B}(z',z)=
\begin{tikzpicture}[scale=0.7,baseline=-0.1cm]
\begin{feynhand}
\vertex (a) at (0,0); \vertex (b) at (1,0); \vertex (c) at (3,0); \vertex (d) at (4,0); 
\propag[chabos] (a) to (b); \propag[chasca] (b) to [in=90, out=90](c); \propag[chabos] (c) to (d);
\propag[fer] (c) to [in=-90,out=-90](b); 
\node at (0.8,-0.3) {$z$};
\node at (3.2,-0.3) {$z'$};
\end{feynhand}
\end{tikzpicture}
=i G_c(z’,z)G_f(z,z’)\label{eq:BSelfEnergyFeynmanDiagram}
\end{gather}
where 
\begin{gather}
G_f^{-1}=G_{f,0}^{-1}-\Sigma_f,\; G_B^{-1}=-J^{-1}+\Sigma_B.
\end{gather}

It can be easily checked that the self-energies of the $f$ and $B$ with more than one loop and the self-energy of the conduction electron $c$ are suppressed by the $\frac{1}{N}$ factor. As a result, Eqs.~\eqref{eq:fSelfEnergyFeynmanDiagram} and~\eqref{eq:BSelfEnergyFeynmanDiagram} become exact in the $N\rightarrow \infty$ limit.

\section{$U(1)$ symmetry \& Ward identity}\label{app:U1symmetry}
The action Eq.~\eqref{eq:MCKImpurityAction} is invariant under the following $U(1)$ transformation of the impurity fermion $f_\sigma$:
\begin{gather}
f_\sigma(z)\rightarrow e^{i\alpha }f_\sigma(z),\; f_\sigma^\dagger(z)\rightarrow f_\sigma^\dagger(z)e^{-i\alpha}
\end{gather}
where $\alpha$ is constant. 

From the Noether theorem, $Q=\sum_{\alpha=1}^Nf_{\alpha}^\dagger f_{\alpha}$ is conserved classically. To derive the quantum version of the Noether theorem and Ward identity, let us consider the local $U(1)$ transformation by considering a non-constant $\alpha(z)$. Under the local $U(1)$ transformation, the action Eq.~\eqref{eq:MCKImpurityAction} transforms as follows
\begin{align}
S_{MCKI,r=0}&\rightarrow S_{MCKI,r=0}\nonumber\\
&-\sum_{\sigma}\int_\mathcal{C} dz f_\sigma^\dagger(z)\partial_z\alpha(z)f_\sigma(z)
\end{align} 
while the Jacobian of the integral measure $\mathcal{D}(f^\dagger,f)$ remains invariant. From the fact that the partition function $Z$ is invariant under the above changes, we can derive the following identity:
\begin{align}
Z&=\int\mathcal{D}(f^\dagger,f)\mathcal{D}(\cdots)e^{iS}\nonumber\\
&\rightarrow \int\mathcal{D}(f^\dagger,f)\mathcal{D}(\cdots )e^{iS}\Bigg(1-i\int_{\mathcal{C}}dz f_\sigma^\dagger(z)\partial_z\alpha(z)f_\sigma (z)\Bigg)\nonumber\\
&\Rightarrow \partial_z\sum_{\sigma}\Big\langle f_\sigma^\dagger(z)f_\sigma(z)\Big\rangle=0
\end{align}
which is a quantum version of the Noether theorem. 
 
Not only the partition function $Z$ but also expectation values of the correlation functions should be invariant under the local $U(1)$ transformation. Considering the two-point correlation function $\langle f_{\alpha}(z_1)f_\alpha^\dagger(z_2)$, we can derive the following identity:
\begin{align}
&\Big\langle f_\sigma(z_1)f^\dagger_\sigma(z_2)\Big\rangle \nonumber\\
&\rightarrow \Big\langle f_\sigma(z_1)f_\sigma^\dagger(z_2)\Big\rangle+ i \Big([\alpha(z_1)-\alpha(z_2)]\langle f_\sigma(z_1)f_\sigma^\dagger(z_2)\rangle
\nonumber\\
&-\sum_{\sigma’}\int_{\mathcal{C}}dz \partial_z\alpha(z)\langle f^\dagger_{\sigma’}(z)f_{\sigma’}(z)f_\sigma(z_1)f_\sigma(z_2)\rangle_c \Big)\nonumber\\
&\Rightarrow \partial_z\Big(G_f(z_1,z)G_f(z,z_2)\Big)\nonumber\\
&=i\Big(\delta(z-z_1)-\delta(z-z_2)\Big) G_f(z_1,z_2)\label{eq:WardIdentity1}
\end{align}
which is Ward identity. 

In addition to the above identities, we can check that the change of the $\lambda(z)$ on the real axis of the contour $\mathcal{C}$ ($\mathcal{C}_1\cup \mathcal{C}_2$ in Fig.~\ref{fig:KeldyshContour}) does not affect to the value of physical quantities using the local $U(1)$ transformation. To see that, suppose an arbitrary physical observable operator $\mathcal{O}(z_1,z_2,\cdots)$ which is a composite operator consisting of the impurity spin operator $\vec{S}$ and the conduction electron operator $c$, $c^\dagger$. This physical observable is invariant under the local $U(1)$ transformation of the $f$ fermions since the impurity spin operator $\vec{S}$ is invariant under the transformation. Now let us consider the expectation value of the physical observable $\mathcal{O}(z_1,z_2,\cdots)$:  
\begin{align}
&\langle \mathcal{O}(z_1,z_2,\cdots)\rangle_{\lambda_0}=\frac{1}{Z}\int\mathcal{D}(f^\dagger,f)\mathcal{D}(\cdots )e^{iS}\mathcal{O}(z_1,z_2,\cdots)\nonumber\\
&\rightarrow\frac{1}{Z}\int \mathcal{D}(f^\dagger,f)e^{iS-i\sum_\sigma \int_\mathcal{C}dz (\partial_z)f^\dagger_\sigma(z)f_\sigma(z)}\mathcal{O}(z_1,z_2,\cdots)\nonumber\\
&=\langle \mathcal{O}(z_1,z_2,\cdots)\rangle_{\lambda_0+\partial_z\alpha(z)}\label{eq:ChangeOfLambda}
\end{align}
where $\lambda_0$ is a constant value of the $\lambda(z)$ on the imaginary axis of the contour $\mathcal{C}$ ($\mathcal{C}_3$ in Fig.~\ref{fig:KeldyshContour}). The value of the $\lambda_0$ is determined by the initial state. 

From the relation Eq.~\eqref{eq:ChangeOfLambda}, we can see that the change of the $\lambda(z)$ on the real axis of the contour $\mathcal{C}$ ($\mathcal{C}_1\cup\mathcal{C}_2$) does not affect to the physical quantities. As a result, the value of $\lambda(z)$ can be fixed to $\lambda_0$ which is determined from the initial state. In the remaining text, we consider $\lambda(z)$ is given by the constant value $\lambda_0$.

\section{Preparation of the Initial state}\label{Appendix:PreparationOfInitialState}

\subsection{Kadanoff-Baym equation in the equilibrium}
Consider an initial state with a Kondo coupling $J_0$ at temperature $T_0$. Since the initial state is assumed to be in equilibrium, we change all bi-local fields $\mathcal{O}(t,t')$ into $\mathcal{O}(t-t')$. Then the resulting self-consistent equations of the Green's functions from the Kadanoff-Baym equations~\eqref{eq:KBequation} are given as follows:
\begin{subequations}
\begin{gather}
G_f^{R}(\omega)=\frac{1}{\omega\pm i\eta -\lambda_0+i\gamma JG_c^<(t=0)-\tilde{\Sigma}_f^{R}(\omega)},\\
G_\phi^{R}(\omega)=\frac{\Sigma_B^{R}(\omega)}{J\Sigma_B^{R}(\omega)-1}
\end{gather}
\end{subequations}
where 
\begin{subequations}
	\begin{gather}
		\tilde{\Sigma}_f^{>/<}(t)=i\gamma J^2 G_{\phi}^{</>}(-t)G_c^{>/<}(t),\\
		\Sigma_B^{>/<}(t)=iG_{f}^{</>}(-t)G_c^{>/<}(t),\\ 
		G_f^<(t=0)=iq_0\;,G_f^>(t=0)=-i(1-q_0)\\
G_f^{R}(t)=\Theta(t)[G_f^>(t)-G_f^<(t)],\\
\tilde{\Sigma}_f^{R}(t)=\Theta(t)[\tilde{\Sigma}_f^>(t)-\tilde{\Sigma}_f^<(t)],\\
G_\phi^{R}(t)=\Theta(t)[G_\phi^>(t)-G_\phi^<(t)].
\end{gather}
\end{subequations}

\subsection{Kubo-Martin-Schwinger (KMS) condition}

To take account of the initial temperature $T_0$, the KMS condition needs to be considered. For the Green's function $G_f$, $G_c$ and $G_\phi$, the following KMS conditions are used:
\begin{gather}
	G_{f/c}^<(\omega)=-e^{-\beta \omega}G_{f/c}^{>}(\omega)\nonumber\\
	\Rightarrow \Bigg\{\begin{array}{l}G_{f/c}^>(\omega)=-i(1-n_F(\omega))A_{f/c}(\omega) \\ G_{f/c}^{<}(\omega)=in_F(\omega)A_{f/c}(\omega)\end{array},\label{eq:KMSfermion}\\
	G_\phi^<(\omega)=e^{-\beta\omega}G_\phi^>(\omega)\nonumber\\
	\Rightarrow \Bigg\{\begin{array}{l}G_{\phi}^>(\omega)=-i(1+n_B(\omega))A_{\phi}(\omega) \\ G_{\phi}^{<}(\omega)=-in_B(\omega)A_{\phi}(\omega)\end{array}
\end{gather} 
where $A_{f/c/\phi}=-2ImG_{f/c/\phi}^R$ and $n_{F/B}(\omega)=\frac{1}{e^{\beta \omega}\pm 1}$.

\subsection{Analytic form of the Green's functions of the conduction electrons}\label{Appendix:AnalyticFormOfConduction}

Since there is no self-energy correction to the conduction electron, it is a just non-interacting case. Therefore we can obtain an analytic form of the $A_c(\omega)$ and $G_c^{>/<}(\omega)$ using Eq.~\eqref{eq:Gc0} and the KMS condition Eq.~\eqref{eq:KMSfermion} as follows:
\begin{align}
A_c(\omega)&=-2ImG_c^R(\omega)\approx \frac{2\pi VD_F}{N_{site}} \int_{-\Lambda}^\Lambda d\epsilon \delta(\omega-\epsilon)\nonumber\\
&=\frac{2\pi VD_F}{N_{site}}\Theta(\Lambda-|\omega|),\\
&\Big\{\begin{array}{l} G_c^>(\omega)=-i\Big(1-n_F(\omega)\Big)2\pi \frac{V D_F}{N_{site}}\Theta(\Lambda-|\omega|),\\ G_c^<(\omega)=in_F(\omega)2\pi \frac{VD_F}{N_{site}}\Theta(\Lambda-|\omega|)\end{array}
\end{align}
where $V$ and $D_F$ are a volume of the system and a density of states near the Fermi energy per volume respectively and $\Lambda$ is an energy cut-off near the Fermi energy point. Self-consistency of the above results can be easily shown. Here we consider only conduction electrons near the Fermi energy based on the physical assumption that only conduction electrons near Fermi energy play an important role. 

If we do not make the approximation that only conduction electrons near Fermi energy are considered then integration of the $A_c(\omega)$ is given by value 1 as follows: $\int\frac{d\omega}{2\pi}A_c(\omega)=\frac{1}{N_{site}}\sum_p =1$.
However, due to the approximation, the sum-rule of $A_c(\omega)$ is changed and the integration of the $A_c(\omega)$ should be smaller than the value 1. Therefore 
\begin{align}
	&\int\frac{d\omega}{2\pi}A_{c}(\omega)=\frac{VD_F}{N_{site}}\int d\omega \Theta(\Lambda-|\omega|)\nonumber\\
	&=\frac{2\Lambda V D_F}{N_{site}}<1	
\end{align}
Here $2\Lambda V D_F$ is a number of conduction electrons within the energy regime $-\Lambda<\epsilon<\Lambda$ near the Fermi energy. We will use a notation $\nu_c=2\Lambda V D_F/N_{site}$ as a conduction electron filling in the remaining context. Then,
\begin{gather}
	A_c(\omega)=\frac{\nu_c\pi}{\Lambda} \Theta(\Lambda-|\omega|),	\\
	G_c^>(\omega)=-i\Big(1-n_F(\omega)\Big)\frac{\nu_c\pi}{\Lambda}  \Theta(\Lambda-|\omega|),\\
	G_c^<(\omega)=in_F(\omega)\frac{\nu_c\pi}{\Lambda} \Theta(\Lambda-|\omega|).
\end{gather}

For the density of states of the conduction electron near the Fermi energy, we use the notation $N_f$ which is given by $\frac{\nu_c}{2\Lambda}$.

\subsection{Flow chart of programming}
The numerical process to obtain the initial states is given as follows:
\begin{gather*}
	\textit{1. Prepare initial guess: }G_f^{>/<,(0)}(t),\; G_\phi^{>/<,(0)}(t) \\
	\textit{2. Calculate the self energies: }\\
	\tilde{\Sigma}_{f}^{>/<,(0)}(t),\; \Sigma_B^{>/<,(0)}(t)\rightarrow \tilde{\Sigma}_f^{R,(0)}(t),\; \Sigma_B^{R,(0)}(t)\\
	\textit{3. Find $\lambda_0$ value satisfying the condition: }\\
	G_f^<(t=0)=iq_0,\; G_f^>(t=0)=-i(1-q_0)\\
	\textit{4. Define new retarded Green's function: }\\
	G_f^{R,(0)}=x G_{f,1}^{R}\Big[G_f^{>/<,(0)}\Big]+(1-x) G_{f,2}^{R}\Big[\tilde{\Sigma}_f^{R,(0)},\lambda\Big],\\
	G_\phi^{R,(0)}=x G_{\phi,1}^{R}\Big[G_\phi^{>/<,(0)})\Big]+(1-x) G_{\phi,2}^{R}\Big[\Sigma_\phi^{R,(0)}\Big]
	\end{gather*}
	\begin{gather*}
	\textit{5. Obtain new greater and lesser Green's function }\\
	\textit{using KMS conditions:}\\
	G_f^{>,(1)}(\omega)=-i(1-n_F(\omega))A_f\Big[G_f^{R,(0)}\Big],\\
	G_f^{<,(1)}(\omega)=in_F(\omega)A_f\Big[G_f^{R,(0)}\Big]\\
	G_\phi^{>,(1)}(\omega)=-i(1+n_B(\omega))A_\phi\Big[G_\phi^{R,(0)}\Big],\\
	G_\phi^{<,(1)}(\omega)=-in_B(\omega)A_\phi\Big[G_\phi^{R,(0)}\Big]\\
	\textit{6. Go to step 2 and continue the same step}\\
	\textit{until following convergence conditions hold:}\\
	|G_{f/\phi}^{>/<,(n+1)}-G_{f/\phi}^{>/<,(n)}|/|G_{f/\phi}^{>/<,(n)}|\ll \epsilon
\end{gather*}
where 
\begin{gather*}
	G^{R}_{f/\phi,1}\Big[G_{f/\phi}^{>/<,(n)}\Big]=\Theta(t)[G_{f/\phi}^{>,(n)}-G_{f/\phi}^{<,(n)}],\\
	G^{R}_{f,2}\Big[\tilde{\Sigma}_{f}^{R,(n)},\lambda\Big]=\Big[\omega+i\eta-\lambda+i\gamma J G_c^<(0)-\tilde{\Sigma}_f^{R,(n)}(\omega)\Big]^{-1},\\
	G^{R}_{\phi,2}\Big[\Sigma_{\phi}^{R,(n)}\Big]=\frac{\Sigma_{\phi}^{R,(n)}(\omega)}{J\Sigma_{\phi}^{R,(n)}(\omega)-1}.
\end{gather*}

\section{Flow chart of numerical solving of the Kadanoff-Baym equations}\label{Appendix:KBsolving}

\begin{figure*}
\centering
\includegraphics[scale=0.4]{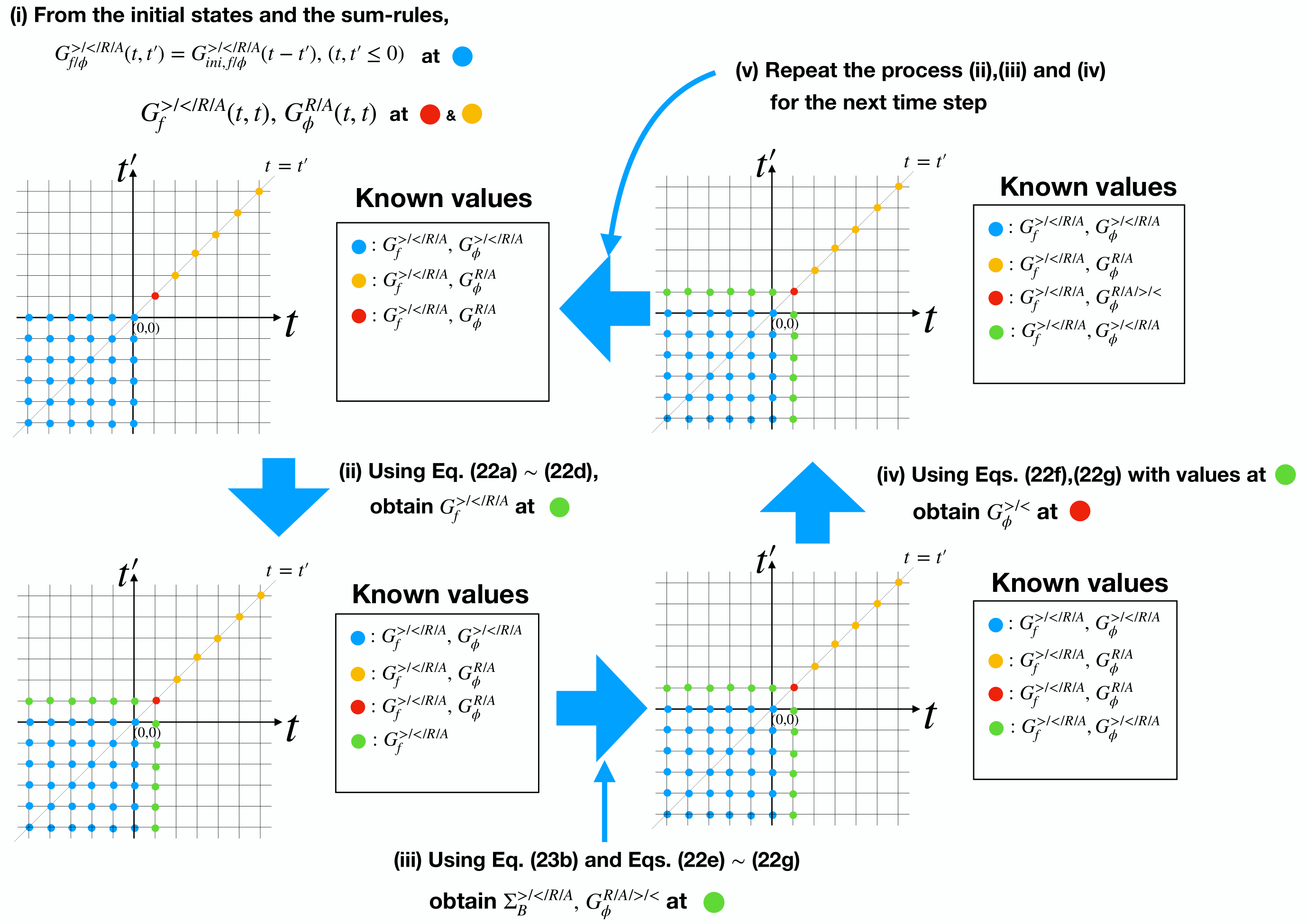}
\caption{Flow chart of the Kadanoff-Baym equation solver.}\label{fig:FlowChartForKBSolver}
\end{figure*}

Fig. \ref{fig:FlowChartForKBSolver} shows the schematic flow chart for solving the Kadanoff-Baym equations step by step.

\section{Effective temperature of the conduction electron with $1/N$-self energy correction}\label{Appendix:EffectiveTempOfConductionElectron}

The Keldysh-Schwinger equation for the conduction electrons with the one-loop self energy correction of order $1/N$ in Fig.~\ref{fig:ConductionOneLoopSelf} is given by 
\begin{subequations}
\begin{gather}
\int_{\mathcal{C}}d\bar{z}[G_{c}^{-1}(z,\bar{z})-\Sigma_c(z,\bar{z})]\bar{G}_c(z,z’)=\delta(z-z’),\\
\Sigma_c(z,z’)=\frac{i}{N}G_f(z,z’)G_B(z,z’)
\end{gather}
\end{subequations}
where $G_c^{-1}(z,z’)=G_{c,0}^{-1}(z,z’)+\frac{q_0}{N}J(z)\delta(z-z’)$ and $\bar{G}_c(z,z’)$ is a Green’s function of conduction electron with the one-loop self energy. 

Using Eq.~\eqref{eq:RelationGBGphi} and $G_f^<(t,t)=iq_0$, the above equations become
\begin{subequations}
\begin{gather}
\int_{\mathcal{C}}d\bar{z}[G_{c,0}^{-1}(z,\bar{z})-\tilde{\Sigma}_c(z,\bar{z})]\bar{G}_c(z,z’)=\delta(z-z’),\\
\tilde{\Sigma}_c(z,z’)=\frac{i}{N}J(z)J(z’)G_f(z,z’)G_\phi(z,z’).
\end{gather}
\end{subequations}

Applying the Keldysh rotation, we obtain the following Kadanoff-Baym equations of conduction electrons consisting of retarded, advanced, and Keldysh Green’s function with real-time arguments:
\begin{align}
[G_{c,0}^{-1}-\tilde{\Sigma}_c]\circ \bar{G}_c=\mathbf{1}\label{eq:barGcKBeq}
\end{align}
where $\circ$ notation means the matrix product in the Keldysh and time space. Here $G_{c,0}^{-1}$, $\tilde{\Sigma}_c$ and $\bar{G}_c$ are matrices given by 
\begin{align*}
G_{c,0}^{-1}&=\left(\begin{array}{cc}(G_{c,0}^R)^{-1} & - (G_{c,0}^R)^{-1}G_{c,0}^K(G_{c,0}^A)^{-1} \\ 0 & (G_{c,0}^A)^{-1}  \end{array}\right), \\
\tilde{\Sigma}_c&=\left(\begin{array}{cc}\tilde{\Sigma}_c^R & \tilde{\Sigma}_c^K \\ 0 & \tilde{\Sigma}_c^A  \end{array}\right),\; \bar{G}_c=\left(\begin{array}{cc}\bar{G}_c^R & \bar{G}_c^K \\ 0 & \bar{G}_c^A\end{array}\right)
\end{align*}

Using the above matrix expressions and Eq.~\eqref{eq:barGcKBeq}, we obtain the $\bar{G}_c^{R/A}$ and $\tilde{G}_c^K$ in terms of the $G_{c,0}^{R/A/K}$ and $\tilde{\Sigma}_c^{R/A/K}$ as follows:
\begin{align}
\bar{G}_c^{R/A}&=\Big[(G^{R/A}_{c,0})^{-1}-\tilde{\Sigma}_c^{R/A}\Big]^{-1},\\
\bar{G}_c^K&=\Big[(G_{c,0}^R)^{-1}-\tilde{\Sigma}_c^R\Big]^{-1}\circ \Bigg([G_{c,0}^R]^{-1}G_{c,0}^K[G_{c,0}^A]^{-1}\nonumber\\
&+\tilde{\Sigma}_{c}^K\Bigg)\circ \Big[[G_{c,0}^A]^{-1}-\tilde{\Sigma}_c^A\Big]^{-1}\nonumber\\
&=\bar{G}_c^R\circ  \Bigg([G_{c,0}^R]^{-1}G_{c,0}^K[G_{c,0}^A]^{-1}+\tilde{\Sigma}_{c}^K\Bigg) \circ \bar{G}_{c}^A
\end{align}

From the Fluctuation-Dissipation relation: $
\lim_{\mathcal{T}\rightarrow \infty}\frac{i\bar{G}_c^K(\mathcal{T},\omega_r)}{-2Im \bar{G}_c^R(\mathcal{T},\omega_r)}\approx\tanh\Big(\frac{\omega_r}{2 T_{1,c}}\Big)$, we obtain the effective temperatures of the conduction electron shown in Fig.~\ref{fig:ConductionElargeNFinalTemp}.

\section{Derivation of sum-rules} 
\label{Appendix:SumRules}

Here we obtain sum-rules which serve as good tools to check numerical results. In expressing the sum-rules, we use the Greens functions in terms of total time $\mathcal{T}=\frac{t+t’}{2}$ and relative time $t_r=t-t’$.

For the $f$-fermions, we can obtain the following trivial sum-rules from Eq.~\eqref{eq:GfConstraint} and the commutation relation of $f$ fermion $\{f_i,f_i^\dagger\}=1$:
\begin{subequations}
\begin{gather}
\int \frac{d\omega_r}{2\pi}G_f^<(\mathcal{T},\omega_r)=iq_0,\\
\int \frac{d\omega_r}{2\pi}G_f^>(\mathcal{T},\omega_r)=-i(1-q_0),\\
\int \frac{d\omega_r}{2\pi}A_f(\mathcal{T},\omega_r)=1
\label{eq:fSumRules}
\end{gather}
\end{subequations}
where 
\begin{gather*}
G_f^{</>/R}(\mathcal{T},\omega_r)=\int dt_r e^{-i\omega_r t_r} G_f^{f</>/R}(\mathcal{T},t_r),\\
A_f(\mathcal{T},t_r)=-2Im G_f^R(\mathcal{T},t_r).
\end{gather*}

Now let us consider the sum-rule for the $\phi$-boson. The sum-rule for the boson $\phi$ is subtle compared to the $f$ fermion. Using Eq.~\eqref{eq:DefOfPhi}, we obtain the following sum-rule for the spectral function of the boson $\phi$:
\begin{align}
	&\int\frac{d\omega_r}{2\pi}A_\phi(\mathcal{T},\omega_r)=\frac{1}{K}\sum_{i=1}^K\Big\langle [\phi_i,\phi_i^\dagger]\Big\rangle\nonumber\\
	&=\frac{1}{NK}\sum_{i=1}^K\sum_{\alpha=1}^N\Big\langle \Big[f_\alpha^\dagger f_\alpha \{c_{i,\alpha},c_{i,\alpha}^\dagger\}-c_{i,\alpha}^\dagger c_{i,\alpha}\Big]\Big\rangle \nonumber\\
	&=\frac{1}{NK}\sum_{i=1}^K\sum_{\alpha=1}^N\Big[\langle f_\alpha^\dagger f_\alpha\rangle \Big(\int\frac{d\omega_r}{2\pi}A_c(\omega)\Big)-\langle c_{i,\alpha}^\dagger c_{i,\alpha}\rangle \Big]\nonumber\\
	&=q_0\nu_c+iG_c^<(t=0)\label{eq:phiSumRule}
\end{align}
where $A_\phi(\mathcal{T},\omega_r)=\int dt_r e^{-i\omega_r t_r}A_{\phi}(\mathcal{T},t_r)$, $A_{\phi}(\mathcal{T},t_r)=-2ImG_\phi^R(\mathcal{T},t_r)$ and $\nu_c$ is the filling of the conduction electron defined in Appendix~\ref{Appendix:AnalyticFormOfConduction}.

Lastly, combining the Ward identity Eq.~\eqref{eq:WardIdentity1} with Eqs.~\eqref{eq:IntegroDiffGfEq1} and~\eqref{eq:IntegroDiffGfEq2} gives the following relation:
\begin{gather}
\int_{\mathcal{C}}d\bar{z}\Big[G_f(z_1,\bar{z})\tilde{\Sigma}
_f(\bar{z},z)G_f(z,z_2)\nonumber\\
-G_f(z_1,z)\tilde{\Sigma}_f(z,\bar{z})G_f(\bar{z},z_2)\Big]=0.\label{eq:WardIdentity2}
\end{gather}
This relation holds for the general values of $z$, $z_1$ and $z_2$. When $z=z_1=z_2$, Eq.~\eqref{eq:WardIdentity2} is reduced to a following relation: 
\begin{align}
&\partial_t G_f^<(t,t')\Big|_{t'=t}+\partial_{t'} G_f^<(t,t')\Big|_{t'=t}\nonumber\\
&=\int d\bar{t}\Big[\tilde{\Sigma}_f^<(t,\bar{t})G_f^A(\bar{t},t)+\tilde{\Sigma}_f^R(t,\bar{t})G_f^<(\bar{t},t)\nonumber\\
&-G_f^R(t,\bar{t})\tilde{\Sigma}_f^<(\bar{t},t)-G_f^<(t,\bar{t})\tilde{\Sigma}_f^A(\bar{t},t)\Big]=0
\end{align}
which can be also obtained from the Kadanoff-Baym equations (Eq.~\eqref{eq:GfKBEq1},~\eqref{eq:GfKBEq2}) and the fact that $G_f^<(t,t)=iq_0$.  

Note that these sum-rules hold for the general non-equilibrium setting. We have used these sum-rules as a sanity check for the numerical simulation.  

\section{Evolutions of $\mathcal{C}_{imp}(\mathcal{T},t_r)$ and $\chi_{imp}(\mathcal{T},t_r)$ after the quenching for $T_0=10T_K$ and $T_0=0.4T_K$ cases with $J_1=6,15$.}\label{Appendix:EvolutionsOfCandChi}

Here we present the numerical solutions of $\mathcal{C}_{imp}(\mathcal{T},t_r)$ and $\chi_{imp}(\mathcal{T},t_r)$ for $T_0=10T_K$ and $T_0=0.4T_K$ cases with $J_1=6,15$.

\begin{figure*}
\centering
\includegraphics[scale=0.4]{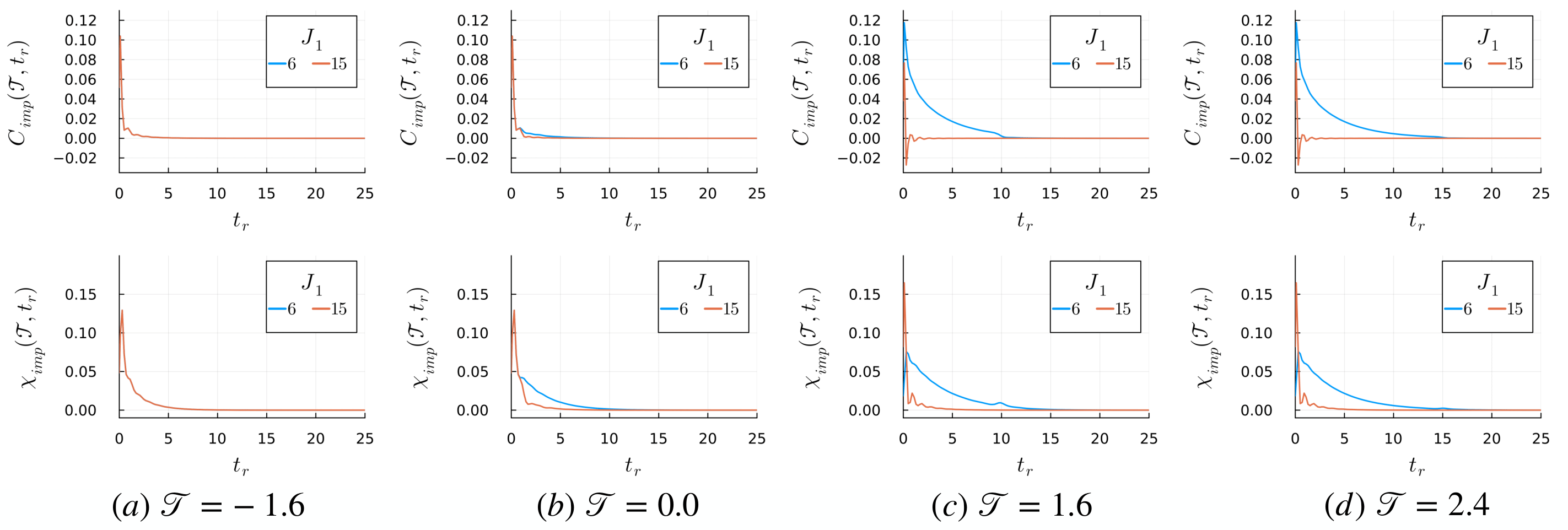}
\caption{Evolution of $C_{imp}(\mathcal{T},t_r)$ and $\chi_{imp}(\mathcal{T},t_r)$ for $T_0=0.4T_K$ case with $J_1$ are given by $6$ and $15$. }\label{fig:CimpAndChiImpLowT0}
\end{figure*}

\begin{figure*}
\centering
\includegraphics[scale=0.4]{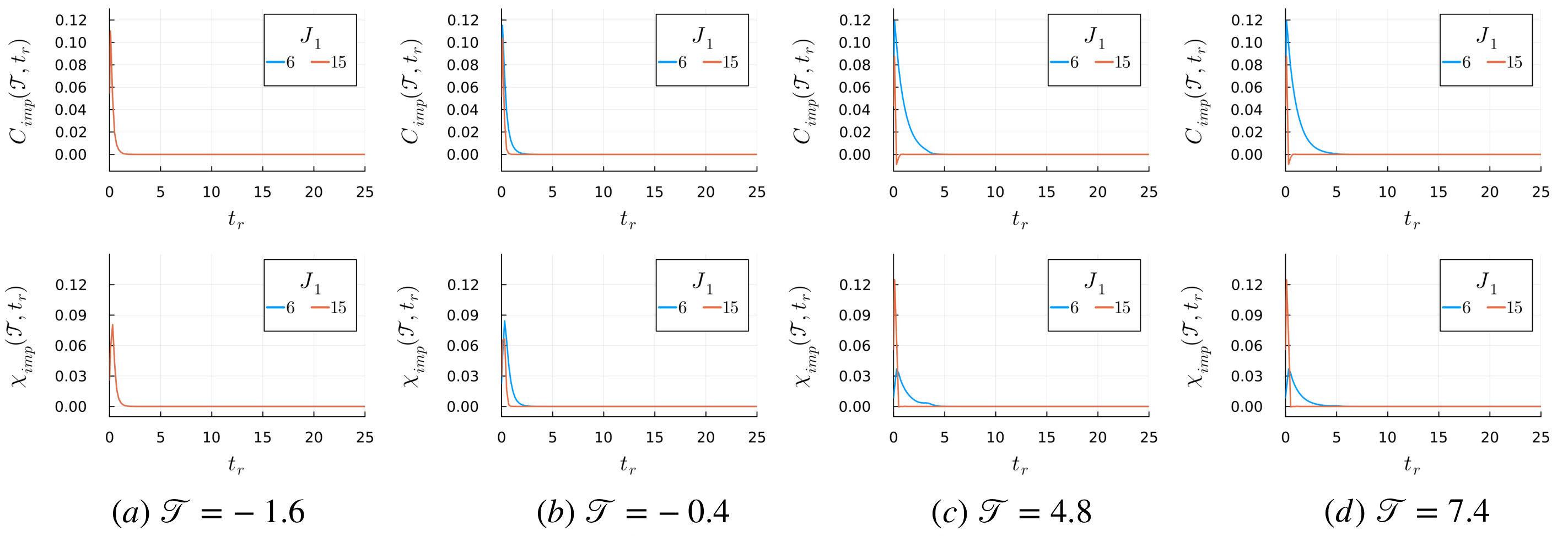}
\caption{Evolution of $C_{imp}(\mathcal{T},t_r)$ and $\chi_{imp}(\mathcal{T},t_r)$ for $T_0=5T_K$ case with $J_1$ are given by $6$ and $15$. }\label{fig:CimpAndChiImpHighT0}
\end{figure*}

\end{appendix}
\bibliography{references.bib}

\end{document}